\documentclass[a4paper,fleqn]{cas-dc}

\usepackage[authoryear,longnamesfirst]{natbib}

\usepackage{amsmath,amssymb,amsfonts}
\usepackage{graphicx}
\usepackage{subfigure} 
\usepackage{textcomp}
\usepackage{xcolor}
\usepackage{tcolorbox}
\usepackage{float}
\usepackage{makecell}
\usepackage{colortbl}
\usepackage{diagbox}
\usepackage{multirow}
\usepackage{booktabs}
\usepackage{mdframed}
\usepackage{utfsym}
\newmdenv[skipabove=2mm]{findingbox}

\usepackage{indentfirst}
\usepackage{amsfonts}
\usepackage{multirow}
\usepackage{fixltx2e}
\usepackage{amsmath, amsfonts}
\usepackage{textcomp}
\usepackage{xcolor}
\usepackage{amsmath}
\usepackage{adjustbox}
\usepackage{float}
\usepackage{array,diagbox,siunitx}
\usepackage{xspace}
\usepackage{algorithm}
\usepackage{algorithmicx}
\usepackage{algpseudocode}
\usepackage{amsmath,amsfonts}
\usepackage{amssymb}
\usepackage{array}
\usepackage{balance}
\usepackage{booktabs}
\usepackage[skip=1pt,labelfont=bf]{caption}
\usepackage{calligra}
\usepackage{color}
\usepackage{colortbl}
\usepackage{courier}
\usepackage{csvsimple}
\usepackage{enumitem}
\usepackage{fancybox}
\usepackage{fontenc}
\usepackage{graphicx}
\usepackage{listings}
\usepackage{longtable}
\usepackage{lscape}
\usepackage{makecell}
\usepackage{moreverb}
\usepackage{multicol}
\usepackage{multirow}
\usepackage{pifont}
\usepackage{rotating}
\usepackage{setspace}
\usepackage{subfigure}
\definecolor{myred}{RGB}{200, 50, 50}
\definecolor{darkpastelred}{rgb}{0.76, 0.23, 0.13}
\definecolor{ao(english)}{rgb}{0.0, 0.5, 0.0}
\definecolor{darkpastelred}{rgb}{0.76, 0.23, 0.13}
\definecolor{ao(english)}{rgb}{0.0, 0.5, 0.0}
\definecolor{yellow}{RGB}{255,255,153}
\definecolor{grey}{RGB}{224,224,224}

\newboolean{showcomments}
\setboolean{showcomments}{true}
\ifthenelse{\boolean{showcomments}}
 { \newcommand{\mynote}[2]{
      \fbox{\bfseries\sffamily\scriptsize#1}
        {\small$\blacktriangleright$\textsf{\emph{#2}}$\blacktriangleleft$}}}
        { \newcommand{\mynote}[2]{}}

\definecolor{DarkOrange}{rgb}{0.8,0.3,0.0}
\definecolor{DarkCyan}{rgb}{0.0, 0.55, 0.55}
\definecolor{DarkCyel}{rgb}{1.0, 0.49, 0.0}
\definecolor{yellow-green}{rgb}{0.6, 0.8, 0.2}

\newcolumntype{?}{!{\vrule width 1pt}}

\def\tsc#1{\csdef{#1}{\textsc{\lowercase{#1}}\xspace}}
\tsc{WGM}
\tsc{QE}
\begin{document}
\let\WriteBookmarks\relax
\def\floatpagepagefraction{1}
\def\textpagefraction{.001}

\shorttitle{An Empirical Study of AI Techniques in Mobile Applications }

\title [mode = title]{An Empirical Study of AI Techniques in Mobile Applications } 

\shortauthors{Li et al.}

\author[1]{Yinghua Li}
\ead{yinghuali@acm.org}

\author[1]{Xueqi Dang}
\cormark[1]
\ead{xueqi.dang@uni.lu}

\author[2]{Haoye Tian}
\ead{haoye.tian@unimelb.edu.au}

\author[1]{Tiezhu Sun}
\ead{tiezhu.sun@uni.lu}

\author[3]{Zhijie Wang}
\ead{zhijie.wang@ualberta.ca}

\author[3,4]{Lei Ma}
\ead{ma.lei@acm.org}

\author[1]{Jacques Klein}
\ead{jacques.klein@uni.lu}

\author[1]{Tegawendé F. Bissyandé}
\ead{tegawende.bissyande@uni.lu}

\affiliation[1]{
    organization={University of Luxembourg},
    city={Luxembourg},
    country={Luxembourg}
}

\affiliation[2]{
    organization={The University of Melbourne},
    city={Melbourne},
    country={Australia}
}

\affiliation[3]{
    organization={University of Alberta},
    city={Edmonton},
    country={Canada}
}

\affiliation[4]{
    organization={University of Tokyo},
    city={Tokyo},
    country={Japan}
}

\cortext[1]{Corresponding author.}

\begin{abstract}
The integration of artificial intelligence (AI) into mobile applications has significantly transformed various domains, enhancing user experiences and providing personalized services through advanced machine learning (ML) and deep learning (DL) technologies. AI-driven mobile apps typically refer to applications that leverage ML/DL technologies to perform key tasks such as image recognition and natural language processing. Despite existing research exploring how mobile apps exploit AI techniques, they have the following main limitations: 1) Most existing studies focus on DL-based apps, with limited research on ML-based apps. 2) Existing research typically focuses on investigating the apps and the technologies utilized in the apps, lacking user-level analysis. 3) The number of apps studied is limited, with only 1,000 to 2,000 ML/DL apps identified after filtering.
To fill the gap, in this paper, we conducted the most extensive empirical study on AI applications, exploring on-device ML apps, on-device DL apps, and AI service-supported (cloud-based) apps. Our study encompasses 56,682 real-world AI applications, focusing on three crucial perspectives: \textbf{1) Application analysis}, where we analyze the popularity of AI apps and investigate the update states of AI apps; \textbf{2) Framework and model analysis}, where we analyze AI framework usage and AI model protection; 
\textbf{3) User analysis}, where we examine user privacy protection and user review attitudes. 
Our study has strong implications for AI app developers, users, and AI R\&D. On one hand, our findings highlight the growing trend of AI integration in mobile applications, demonstrating the widespread adoption of various AI frameworks and models. On the other hand, our findings emphasize the need for robust model protection to enhance app security. Additionally, our study highlights the importance of user privacy and presents user attitudes towards the AI technologies utilized in current AI apps. We provide our AI app dataset (currently the most extensive AI app dataset) as an open-source resource for future research on AI technologies utilized in mobile applications. 
\end{abstract}


% make the title area
\maketitle

\section{Introduction}

The advent of artificial intelligence (AI) has revolutionized numerous fields, with mobile applications being a significant beneficiary. AI-driven mobile apps leverage machine learning algorithms, natural language processing, and computer vision to enhance user experiences, improve functionalities, and provide personalized services. These advancements are particularly evident in areas such as recommender systems (\cite{lu2015recommender, wang2015collaborative}) handwriting recognition (\cite{pham2014dropout}) and face detection (\cite{hjelmaas2001face, dospinescu2016face}).  

From the perspective of the underlying technology, AI applications can be broadly categorized into two main categories: machine learning-based apps (\cite{sun2021mind}) and deep learning-based apps (\cite{xu2019first}). Machine learning apps rely on traditional machine learning algorithms (e.g., decision trees (\cite{rokach2005decision}), clustering algorithms (\cite{xu2015comprehensive}), and logistic regression (\cite{lavalley2008logistic})) to achieve the apps' functionality. For example, in the field of healthcare, logistic regression can be used for diabetes prediction (\cite{joshi2021predicting}). 

Deep learning apps leverage deep learning techniques to perform various tasks, such as image recognition (\cite{jones2020plant}), speech recognition (\cite{matarneh2017speech}), and natural language processing (\cite{locke2021natural}). For instance, Google Lens (\cite{bilyk2020comparing}) is a prevalent deep learning mobile app that uses image recognition technology to identify objects, landmarks, and text in photos. Users can point their phone's camera at an object, and Google Lens can provide information about what they see, such as identifying the type of flower or offering detailed information about a landmark.  

From the perspective of deployment methods, AI apps can be categorized into two groups: on-cloud inference and on-device inference (\cite{xu2019first}). In cloud-based inference, mobile devices connect to cloud servers via the network and transmit data to the servers. ML/DL models operating on the servers carry out inferences on the data and send the results back to mobile devices. Cloud-based inference benefits from robust computing capabilities and flexibility, making it suitable for dealing with intricate models and large data volumes. Nonetheless, it relies on network connectivity, which can lead to latency and data privacy concerns. In contrast, on-device inference enables ML/DL models to run directly on the mobile device. On-device inference offers low latency and offline capabilities, ensuring immediate responses and better data privacy protection. However, the drawback is that mobile devices have limited computing resources, posing challenges in handling complex models.
  
In the literature (\cite{xu2019first, sun2021mind}), some studies have focused on investigating the deployment of ML/DL on mobile devices. \cite{xu2019first} present the first empirical study on how real-world Android apps exploit DL techniques. Their research focuses on three aspects: the characteristics of DL apps, what they use DL for, and what their DL models are. \cite{sun2021mind} present the first empirical study of ML model protection on mobile devices. This study explored the extent of model protection usage in apps, the robustness of existing model protection techniques, and the potential impacts of stolen models.  

Although their research is valuable and in-depth, there are the following limitations in the scope of their study. 
\begin{itemize}[leftmargin=*]
    
    \item The majority of existing studies (\cite{xu2019first, sun2021mind}) concentrated on DL-based apps, with a lack of research on classical ML-based apps. While existing studies (\cite{sun2021mind}) explored model protection in ML apps, their focus has primarily been on the protection aspect without delving into the specific characteristics of ML apps or examining the utilization of ML technologies. 
    
    \item Existing studies mainly focused on investigating the apps and the technologies utilized in the apps, lacking analysis at the user level. However, user reviews play a vital role in improving AI applications. Moreover, protecting user privacy in AI apps is crucial for maintaining user trust and ensuring compliance with data protection regulations. Existing studies lack such an analysis from the user's perspective. 
    
    \item The number of apps studied in existing research is limited. Typically, the range of apps investigated in existing studies is between 10,000 to 50,000. Researchers then filter out the ML/DL apps. Finally, the number of ML/DL apps obtained from this process ranges from 1,000 to 2,000. 

\end{itemize} 

To fill this gap, we conducted the most extensive empirical study on AI applications, comprehensively exploring on-device machine learning (ML) apps, on-device deep learning (DL) apps, and AI service-supported apps (also referred to as cloud-based apps). Our study encompasses 56,682 real-world AI applications. To this end, we designed an automated AI app identification tool named AI Discriminator. On the server, AI Discriminator runs concurrently on 96 threads for approximately 1440 hours (two months), extracting 56,682 real-world AI applications from a pool of 7,259,232 mobile apps in the AndroZoo large-scale application repository (\cite{allix2016androzoo}). The number of AI apps we collected is currently the largest AI app dataset. 

Specifically, our research focuses on three main perspectives: \textbf{1) Application analysis}, which includes AI app popularity analysis and AI app update status analysis. \textbf{2) Framework and model analysis}, which includes AI framework usage analysis and AI model protection analysis. \textbf{3) User analysis}, which includes user privacy protection analysis and user review analysis. Below, we provide a detailed explanation of each aspect of our empirical analysis. 

\begin{itemize} [leftmargin=*]

\item[\ding{182}] \textbf{AI app popularity analysis (Application analysis)} 
From the perspective of app popularity analysis, we investigate the following four aspects: \textbf{1) The annual development volume of AI apps} Understanding the annual development volume of AI apps can provide insight into the growth and adoption rate of AI technologies, contributing to identifying trends in AI investment. \textbf{2) The popular markets for AI apps} Identifying the popular markets for AI apps allows developers to understand where AI technology is being most widely adopted and integrated, providing insights into market entry strategies and investment decisions. 3) \textbf{The prevalent categories of AI apps in the industry} Studying the prevalent types of AI apps in the industry highlights which categories of AI apps are most commonly developed and used. This information can inform developers and companies about the most in-demand AI functionalities, helping them align their development efforts with industry needs. \textbf{4) The popular categories of AI apps in the market} Understanding the popular categories of AI apps in the market provides a detailed view of consumer preferences. It helps identify which types of AI applications are gaining traction among users, providing developers insights for better-targeted product development and marketing strategies.

\item[\ding{183}] \textbf{AI app update status analysis (Application analysis)} Our research on AI app update status is divided into three aspects: \textbf{1) The update frequency of AI apps} Analyzing the update frequency of AI apps reveals how often these apps are maintained and improved. Regular updates can indicate a positive approach to enhancing app performance. \textbf{2) The correlation between AI app updates and AI model updates} Investigating the correlation between AI app updates and AI model updates helps to understand the efficiency of developers in adopting new AI technologies. \textbf{3) The correlation between AI app updates and AI framework updates.} Investigating the correlation between AI app updates and AI framework updates helps identify the dependency of apps on the latest frameworks. Understanding this relationship can guide developers in choosing suitable frameworks. 

\item[\ding{184}] \textbf{AI framework usage analysis (Framework and model analysis)} The research on AI frameworks is mainly divided into three aspects: \textbf{1) The popularity of different AI frameworks} Understanding which AI frameworks are popular can help developers choose suitable framework among the many available options. \textbf{2) The usage of single-framework and multi-framework AI frameworks} Analyzing the usage patterns of single-framework and multi-framework AI systems can helps to understand different development strategies. This analysis can guide developers in choosing the most effective approach. 3) \textbf{The mainstream AI frameworks} Understanding the popularity of mainstream AI frameworks in recent years can provide insights into the popularity trends of AI frameworks, specifically which AI frameworks are gradually being phased out and which ones are on the rise. Developers can refer to the popularity of these frameworks when selecting AI frameworks. 

\item[\ding{185}] \textbf{AI model protection analysis (Framework and model analysis)} The analysis of AI model protection is conducted from two aspects: \textbf{1) Open-source model usage conditions} Open-source models can potentially pose higher security risks. Investigating the use of open-source models in AI apps can shed light on the state of AI model protection. \textbf{2) AI model encryption conditions} Examining model encryption conditions is crucial for understanding to what extent on-device AI models are protected from unauthorized access. 

\item[\ding{186}] \textbf{User privacy protection analysis (User analysis)} We investigate the state of user privacy protection in published AI apps. Protecting user privacy in AI apps is crucial to maintaining user trust. Analyzing the state of user privacy protection can help identify potential security vulnerabilities where improvements are needed.   

\item[\ding{187}] \textbf{User review analysis (User analysis)} We investigate users' attitudes toward AI techniques in AI apps. Understanding users' attitudes toward AI techniques can help developers identify current issues that users perceive in AI apps and create AI apps that better meet user expectations and preferences.

\end{itemize}

Based on the experimental results of the above empirical analysis, we obtained a total of 23 key findings, which can be found in Section~\ref{subsec:RQ1} to Section~\ref{subsec:RQ6}. Among these findings, Findings 1 to 10 belong to application analysis (cf. Section~\ref{subsec:RQ1} and Section~\ref{subsec:RQ2}). Findings 11 to 18 belong to framework and model analysis (cf. Section~\ref{subsec:RQ3} and Section~\ref{subsec:RQ4}). Finding 19 to 23 belong to user analysis (cf. Section~\ref{subsec:RQ5} and Section~\ref{subsec:RQ6}).

In summary, our contributions are as follows:

\begin{itemize}[leftmargin=*]
    \item \textbf{Collection of a large-Scale AI App Dataset for Research} We design AI Discriminator, an automated AI application identification tool, which successfully identified 56,682 AI apps out of 7,259,232 mobile apps from the AndroZoo large-scale repository. We provide this AI app database for further research on AI apps. 
    \item \textbf{Application Analysis} We conduct an empirical analysis of the collected 56,682 AI apps from the perspective of application popularity and update status, providing insights into the prevalence of AI apps across different markets and categories, as well as their update frequency. This can help understand trends in AI app development and maintenance practices, guiding future AI investment and development strategies.
    \item \textbf{Framework and Model Analysis} We conduct an empirical analysis on the collected AI apps from the perspective of frameworks and models. We analyzed the usage of different AI frameworks, including single-framework and multi-framework approaches, and examined the state of AI model protection. This can provide insights to developers in selecting effective AI frameworks, as well as revealing potential issues in current model protection.
    \item \textbf{User Analysis} We conducted analysis on the collected AI apps from the perspective of users. We investigated the state of user privacy protection in AI apps and analyzed user reviews to understand user attitudes toward AI techniques applied in AI apps. This can provide valuable insights for developers to understand the current issues perceived by users and to create AI apps that better meet user expectations and preferences. 
\end{itemize}

The remainder of the paper is organized as follows. Section~\ref{sec:background} introduces the background knowledge of AI models and frameworks as well as deploying mobile ML/DL. Section~\ref{sec:study_design} presents an overview of our empirical study. Section~\ref{sec:app} shows the objective, experimental design, experimental results, and findings of our Application Analysis on AI apps. Additionally, this section also demonstrates the specific operation process of our designed AI app identification tool, AI Discriminator. Section~\ref{sec:frame} presents the Framework and Model Analysis. Section~\ref{sec:users} presents the User Analysis. Section~\ref{sec:discussion} discusses the challenges and opportunities of deploying AI technologies to mobile applications. Section~\ref{sec:relatedWork} presents the related existing work and research. Section~\ref{sec:conclusion} concludes this paper.
\section{Background}
\label{sec:background}

\begin{figure*}[h]
	\centering
	\includegraphics[width=0.8\textwidth]{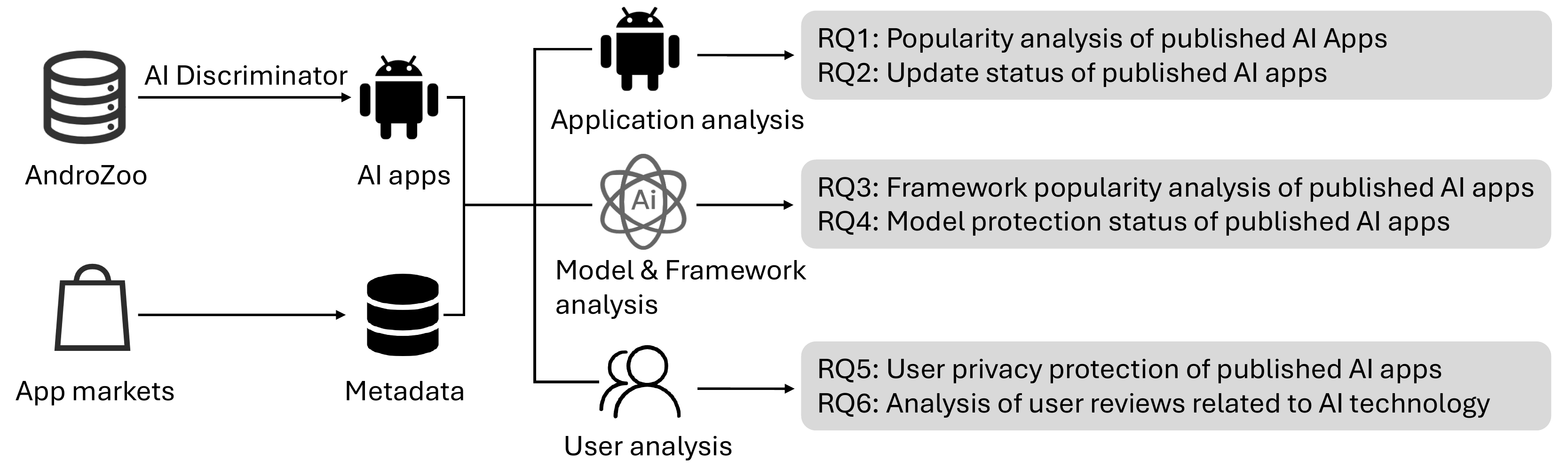}
	\caption{Overview of our explorative study. 
 }
	\label{fig:overview}
\end{figure*}

\subsection{AI models and frameworks}
Artificial Intelligence (AI) (\cite{gamble2020artificial, li2021deeppayload, li2023test, dang2023graphprior}) has become a cornerstone in the development of mobile applications, enabling enhanced functionality, user experience, and efficiency. AI models, including machine learning (ML) (\cite{sun2021mind, dang2024test}) and deep learning (DL) (\cite{huang2021robustness, li2024test, dang2024towards}), are the foundational technologies driving the operation of apps. These models allow mobile applications to perform complex tasks such as image recognition (\cite{jones2020plant}), speech processing (\cite{zhao2017android}), and face recognition (\cite{amos2016openface}) with high accuracy and speed.

The development and deployment of AI-driven mobile apps heavily rely on robust AI frameworks that facilitate the integration and implementation of AI techniques. Prevalent AI frameworks such as TensorFlow (\cite{abadi2016tensorflow}), PyTorch (\cite{paszke2019pytorch}), and Keras (\cite{keras}) provide developers with the tools necessary to build, train, and optimize AI models. These frameworks offer pre-built modules, extensive libraries, and flexible APIs that simplify the complex processes involved in AI development, allowing researchers to concentrate on algorithmic design rather than low-level programming, thus reducing time and labor costs. 

In the context of mobile applications, specialized frameworks like TensorFlow Lite (\cite{tflite}) and Core ML (\cite{thakkar2019introduction}) are designed to optimize AI models for mobile environments. These frameworks ensure that AI models can run efficiently on the limited computational resources available on mobile devices. TensorFlow Lite, for instance, leverages techniques like model quantization and hardware acceleration to enable resource-efficient execution of AI algorithms on mobile devices. Similarly, Core ML is tailored specifically for iOS devices, harnessing the power of Apple's hardware and software integration to provide high-performance and low-latency inference for AI-driven tasks. 

Moreover, in recent years, some companies introduced pre-trained models accompanied by accessible APIs, facilitating remote access to AI services (e.g., Google AI (\cite{googleai}), Baidu NLP (\cite{baidunlp}), and Amazon AI (\cite{amazonai})). Consequently, developers can utilize mobile AI services without necessitating expertise in AI frameworks and model design. These AI services provide ready-to-use functionalities such as image analysis and natural language processing, allowing developers to integrate advanced AI features into their apps with minimal effort. 

Despite the rapid development of AI techniques, there is a growing concern regarding its environmental impact(~\cite{ali2023green}). The energy consumption and carbon emissions associated with AI data centers can be substantial, posing a significant challenge to sustainability efforts. In response to these concerns, the concept of Green AI has emerged, which emphasizes the integration of economic and environmental sustainability into AI systems (\cite{ali2023green}). Green AI aims to leverage machine learning to accelerate the transition to a circular economy, where products and materials are reused and recycled efficiently. The potential applications of Green AI include intelligent production planning, predictive maintenance and reuse marketplaces (\cite{kindylidi2021sustainability}).

\subsection{Deploying mobile ML/DL}
Towards enabling deep learning (DL) on mobile devices, model inference can be broadly classified into two categories: cloud-based inference and on-device inference (\cite{chen2021empirical}). In cloud-based inference, mobile devices connect to cloud servers over the network and send the data to the servers. ML/DL models running on the servers perform inference on the data and return the results to the mobile devices. The cloud-based approach can leverage the powerful computational resources of cloud servers to accelerate the inference process.
However, cloud-based inference also has some disadvantages: \textbf{1) Privacy Risks:} Data needs to be transmitted to the cloud, which can pose privacy leakage risks. \textbf{2) Network Dependence: } It requires a stable network connection. Network latency and bandwidth limitations can affect the inference speed. \textbf{3) Cost:} Using cloud services can incur additional outsourcing costs. 

On the other hand, on-device inference allows for the execution of ML/DL models directly on the mobile device, eliminating the need to send data to remote servers. On-device inference offers several advantages: \textbf{1) Privacy Protection:} Since data is processed locally on the device, it mitigates the risk of data leakage, enhancing user privacy. \textbf{2) Reduced Latency:} Without the need to transmit data over networks, on-device inference can significantly reduce latency, providing faster response times, which are crucial for real-time applications. \textbf{3) Independence from Network Connectivity:} On-device inference does not rely on internet connectivity; thus, it is not limited by network conditions and bandwidth. \textbf{4) Cost Efficiency:} On-device inference can be more cost effective as it avoids the recurring costs associated with data transmission and cloud computing services.

Despite these benefits, on-device inference also poses its own challenges. These include limited computational resources compared to cloud servers, which can restrict the complexity of the models that can be deployed. Additionally, the energy consumption required for local processing can be high, impacting the device's battery life. 

In the literature, several studies have explored deploying deep learning on mobile devices. \cite{xu2019first} investigated how smartphone applications utilize deep learning techniques through the analysis of over 16,500 popular Android apps, aiming to identify and characterize apps that integrate deep learning, understand their purposes, and scrutinize the deep learning models they employ. This research offers insights into current practices and potential optimization areas in mobile deep learning applications.

\section{Analysis Overview}
\label{sec:study_design}

\begin{figure*}[h]
	\centering
	\includegraphics[width=0.6\textwidth]{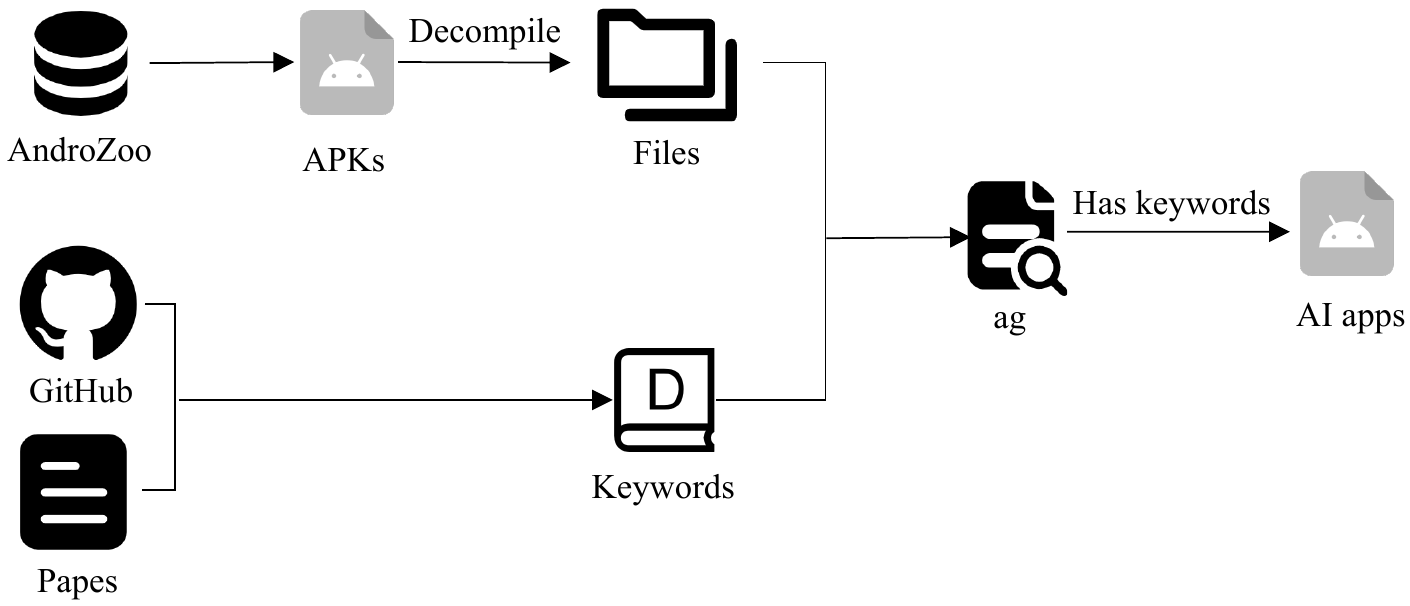}
	\caption{Pipeline of AI app Identification.}
	\label{fig:ai_app_identification}
\end{figure*}

This paper aims to analyze AI-based applications from three perspectives: application analysis (cf. Section~\ref{subsec:app}), model and framework analysis (cf. Section~\ref{sec:frame}), and user analysis (cf. Section~\ref{sec:users}). To this end, we design an automated AI app extraction tool, called AI Discriminator, to recognize AI apps from the AndroZoo application database (\cite{allix2016androzoo}), successfully collecting 56,682 AI apps from 7,259,232 Android apps. 

The workflow of our study is presented in Figure~\ref{fig:overview}. In the initial step, we extract AI apps from Androo using the AI Discriminator. Then, we build a crawler tool to collect important information about these AI apps from the application market. This information includes category, installs, user reviews, etc., for subsequent analysis. Specifically, our analysis focuses on the following three aspects:
\begin{itemize}[leftmargin=*]
\item \textbf{Application analysis}: We study the extracted AI apps from the application level, including an analysis of the popularity of AI apps (RQ1) and an analysis of the update condition of AI apps (RQ2).
\item \textbf{Model and framework analysis}: We analyze the internal techniques of AI apps, including the usage of AI frameworks (RQ3) and the model protection conditions of AI apps (RQ4).
\item \textbf{User analysis}: We analyze AI apps from the user perspective, including privacy protection for users of AI apps (RQ5) and an analysis of user reviews (RQ6).
\end{itemize}

We discuss more details of the AI Discriminator in Section~\ref{subsec:find}.

\section{Applications Analysis}
\label{sec:app}

\subsection{Methodology: finding AI apps}
\label{subsec:find}
We propose \textbf{AI Discriminator} to automatically extract AI apps from the AndroZoo application repository (\cite{allix2016androzoo}). 
In the following, we provide a detailed description of how AI Discriminator works and its accuracy in identifying AI apps (including precision, recall, etc.). 
\subsubsection{Workflow of AI Discriminator}

The input for the AI Discriminator is the SHA-256 hashes (256-bit Secure Hash Algorithm) of apps in AndroZoo. The output is the SHA-256 hashes of the AI apps (apps supported by AI techniques). Here, SHA-256 (256-bit Secure Hash Algorithm) is a hash function used for application encryption. AndroZoo employs it to record a unique ID for each collected app. The overall pipeline of AI Discriminator can be divided into three main steps, as illustrated in Figure \ref{fig:ai_app_identification}:   
    
\begin{itemize}[leftmargin=*]
    \item[\ding{182}] \textbf{Decompilation of APK} In the first step, the AI Discriminator retrieves the APK associated with a given app. It then uses APKTool (\cite{apktool}) to unzip the package and collect its internal files.
    \item[\ding{183}] \textbf{Building an AI Keyword Dictionary}
    AI Discriminator constructs an AI keyword dictionary that includes on-device ML, on-device DL, and AI service keywords. These keywords are collected based on the literature and open-source GitHub codes. Some examples of these keywords are presented in Table \ref{tab:ai_frameworks}. In the following, we describe in detail how we constructed the AI keyword dictionary. Specifically, the construction of the dictionary follows the four steps below:
    \begin{itemize}[leftmargin=*]
    \item \textbf{Collecting AI Keywords from Academic Papers.} In the first step, we collected AI keywords from the existing research (~\cite{xu2019first, sun2021mind}) and their open-source repositories. These keywords have been proven effective for identifying DL/ML-based apps.
    \item \textbf{Collecting AI Keywords from Source Code.} Next, we analyzed the source code of each AI framework and manually collected AI-related keywords. These keywords encompass both algorithm package keywords and model-saving format keywords. To collect the model-saving format keyword, we carefully reviewed the algorithm documentation of each AI framework to identify relevant keywords. For instance, the TensorFlow Lite framework (~\cite{tflite}) generally saves models in the ".tflite" format. As a result, we added ".tflite" as a keyword in our AI keyword dictionary. 
    \item \textbf{Excluding non-AI Keywords.} Since some non-AI terms can also contain simple AI keywords (e.g., LSTM and CNN), the presence of these keywords can reduce the accuracy of the AI Discriminator in identifying AI apps. Therefore, we excluded such terms from the keyword dictionary.
    \item \textbf{Validation and Optimization.} 
    We first evaluate the collected AI keywords on a small dataset. We applied all the collected AI keywords to filter out AI apps and manually checked whether they were truly AI apps. If we found any apps that were incorrectly filtered as AI apps, we identified the keywords that led to the misclassification and removed these non-AI terms from the AI keyword dictionary to optimize Al Discriminator's accuracy.
\end{itemize}

    \item[\ding{184}] \textbf{Keyword Matching} In this step, AI Discriminator performs keyword matching using the code search tool Ag (\cite{ag}), which is a fast and efficient code search tool designed specifically for searching large codebases. For the file packages extracted from all apps, we apply the Ag tool to scan for AI-related keywords that we collected from the previous steps. If any file in an app contains one of the keywords from the pre-defined dictionary, this app is identified as an AI application, and its SHA-256 hash and keyword information will be recorded. For example, if the file of an app contains keywords like ``org.tensorflow.lite'', ``libtensorflowlite.so", and ``N5EigenForTFLite", this app is considered using the TensorFlow Lite framework. Additionally, in terms of the OpenCV framework, since not all OpenCV packages use AI algorithms, with some packages mainly used for image processing, we utilize the keyword "org.opencv.ml" to identify the use of AI algorithms in OpenCV. This keyword is specifically used to identify apps supported by AI-related packages in OpenCV.  
\end{itemize}

\begin{table*}[]
\caption{AI Frameworks given AI Keyword Dictionary}
\label{tab:ai_frameworks}
\centering
\scalebox{0.7}
{
\begin{tabular}{lllcc}
\toprule
\multicolumn{2}{c}{\textbf{AI Frameworks Categories}} & \textbf{AI Frameworks} & \textbf{Platform} & \textbf{Model Format}   \\
\midrule
\multirow{19}{*}{DL Frameworks} & \multirow{9}{*}{Traditional Deep Learning Frameworks} & Tensorflow (\cite{abadi2016tensorflow}) &  Andorid, iOS & .pb, .pbtxt   \\
& & PyTorch (\cite{paszke2019pytorch}) & Andorid, iOS & .pt, .ptl \\
& & MxNet (\cite{paszke2019pytorch}) & Andorid, iOS  & .params \\
& & Caffe (\cite{caffe}) & Andorid, iOS  & .caffemodel, .prototxt \\
& & Caffe2 (\cite{caffe2}) & Andorid, iOS  & .pb  \\
& & Chainer (\cite{chainer})  & / & .chainermodel \\
& & DeepLearning4j (\cite{deeplearning4j}) & Andorid  & .zip \\
& & CNTK (\cite{cntk}) & / & .cntk, .model \\
& & Neuroph (\cite{neuroph}) & / & .model, .nnet \\
  \cmidrule(r){2-5}
& \multirow{8}{*}{Lite Deep Learning Frameworks} & TF Lite (\cite{tflite})  & Andorid, iOS   & .tflite, .lite  \\
& & NCNN (\cite{ncnn}) & Andorid, iOS & .params  \\
& & Paddle Lite (\cite{paddlelite}) & Andorid, iOS & .jar, .so, etc \\
& & MACE (\cite{mace}) & Andorid & .pb, .yml \\
& & FeatherCNN (\cite{feathercnn}) & Andorid, iOS & .feathermodel \\
& & SNPE (\cite{snpe}) & Andorid & / \\
& & CNNDroid (\cite{CNNdroid}) & Andorid & .model \\
& & TVM (\cite{tvm}) & /  & libtvm\_rumtime.so, etc \\
  \cmidrule(r){2-5}
& \multirow{2}{*}{Computer Vision Frameworks} & OpenCV (\cite{opencv}) & Andorid, iOS & TesnorFlow, Caffe, etc \\
& & Baidu OCR (\cite{baiduocr}) & / & / \\
\midrule
\multirow{8}{*}{ML Frameworks}& \multirow{8}{*}{Classical Machine Learning Frameworks} & Xgboost-predictor (ML) & /  & / \\
& & Sklearn-porter (ML, Java) (\cite{skpodamo}) & Andorid & / \\
& & WEKA (ML, Java) (\cite{weka}) & / & .model \\
& & Shogun (ML, C++) (\cite{shogun}) & / & model.prof \\
& & MALLET (ML, Java) (\cite{mallet}) & / & .mallet, .classifier, etc \\
& & Rapid Miner (ML Java) (\cite{rapidminer}) & / & .model \\
& & Datumbox (ML Java) (\cite{datumbox}) & / & .model \\
& & MLPACK (\cite{mlpack}) & / & /  \\
\midrule
\multirow{6}{*}{AI Service} & \multirow{4}{*}{General Cloud AI Frameworks} & Google AI (\cite{googleai}) & /  & / \\
& & Amazon AI (\cite{amazonai}) & /  & / \\
& & Alexa AI (\cite{alexaai}) & /  & / \\
& & Azure AI (\cite{azureai}) & /  & / \\
  \cmidrule(r){2-5}
& \multirow{2}{*}{Natural Language Processing Frameworks} & Baidu NLP (\cite{baidunlp}) & / & / \\
& & Baidu Synthesizer (\cite{synthesizer}) & / & / \\
\bottomrule
\end{tabular}
}
\end{table*}

\subsubsection{Evaluation of AI Discriminator}
To validate the accuracy of the AI Discriminator in identifying AI apps, in other words, to verify whether the AI keywords in the Discriminator's keyword dictionary are truly effective, we used five widely adopted statistical metrics for evaluation: False Positive Rate, False Negative Rate, Precision, Recall, and Accuracy. Among these, 1) False Positive Rate reflects the proportion of all non-AI apps that were incorrectly identified as AI apps. 2) False Negative Rate reflects the proportion of all AI apps that were incorrectly identified as non-AI apps. 3) Precision measures the proportion of true AI apps among all the apps identified as AI apps. 4) Recall measures the proportion of true AI apps that were correctly identified. 5) Accuracy reflects the overall correctness of the AI Discriminator in identifying AI apps.

The evaluation methodology is derived from the existing work (\cite{sun2021mind}). Specifically, we manually collected and verified 450 apps, including 225 non-AI apps and 225 AI apps, as samples to evaluate the AI Discriminator. These apps were collected from existing literature sources (\cite{xu2019first, sun2021mind}). The rationale behind employing a sampling approach for evaluation is that determining whether apps belong to the AI or non-AI category, namely establishing the ground truth of each app, requires manual labeling. Manually labeling all the apps can result in substantial time and labor costs. To tackle this challenge, we draw inspiration from the evaluation methodology employed in a previous study conducted by \cite{sun2021mind}, wherein they evaluated their developed app identification tool by sampling 438 apps. In line with this approach, we employ a similar sampling methodology for assessing the AI Discriminator based on 450 apps.

We conducted the evaluation on these 450 labeled applications. Through calculations, we obtained that the AI Discriminator's precision is 100\%, the recall is 0.84, and the accuracy is 0.92. The specific calculations are as follows. 

Note that in the calculations below, True Positives (TP) refers to the number of correctly predicted positive instances. True Negatives (TN) refers to the number of correctly predicted negative instances. False Positives (FP) refers to the number of instances that were incorrectly predicted as positive when they were actually negative. False Negatives (FN) refers to the number of instances that were incorrectly predicted as negative when they were actually positive. '\#apps' refers to the total number of sampled apps.

\begin{itemize} [leftmargin=*]
\item \textbf{False Positive Rate} AI Discriminator detected a total of 188 AI apps, and 0 app belong to non-AI apps. Therefore, the False Positive Rate is: 
\begin{equation}
\text { FPR }=\frac{FP}{FP + TN}=\frac{0}{0+225}=0
\end{equation}

\item \textbf{False Negative Rate} Out of the 225 AI apps, 37 were not identified by AI Discriminator. Therefore, the FNR is: 
\begin{equation}
\text { FNR }=\frac{FN}{TP + FN}=\frac{37}{188+37}=0.16
\end{equation}

\item \textbf{Precision} AI Discriminator detected a total of 188 AI apps, and all of these 188 apps indeed belonged to the AI app category. Therefore, the precision of AI Discriminator is:  

\begin{equation}
\text { Precision }=\frac{TP}{TP+FP}=\frac{188}{188+0}=100\%
\end{equation}

\item \textbf{Recall} Out of the 225 AI apps, 188 were successfully identified. Therefore, the recall of AI Discriminator is:  
\begin{equation}
\text { Recall }=\frac{TP}{TP+FN}=\frac{188}{188+37}=0.84
\end{equation}

\item \textbf{Accuracy} Among the 450 apps, 413 apps were correctly classified. Therefore, the accuracy of AI Discriminator is: 
\begin{equation}
\text { Accuracy }=\frac{TP+TN}{\#apps}=\frac{188+225}{450}=0.92
\end{equation}

\end{itemize}

The evaluation above shows that out of the 225 AI apps, 37 went undetected. The main reason is that some specific keywords were not added to the keyword dictionary of the AI Discriminator to ensure that all apps identified as AI apps are indeed AI apps. For instance, '.prototxt' files are typically associated with deep learning and machine learning frameworks. However, files with the '.prototxt' format can also serve other purposes, not just AI technology, such as configuration or parameter files for software. To prevent the AI Discriminator from falsely predicting non-AI apps as AI apps, we filtered out such keywords. Consequently, this resulted in some specific AI apps being undetected.

\subsection{Experimental Environment}
We ran experiments on a high-performance computer cluster. Each cluster node runs a 2.6 GHz Intel Xeon Gold 6132 CPU with an NVIDIA Tesla V100 16G SXM2 GPU. For the data visualization, we conducted corresponding experiments on a MacBook Pro laptop with Mac OS Big Sur 11.6, Intel Core i9 CPU, and 64 GB RAM.

\subsection{RQ1: Popularity analysis of published AI Apps}
\label{subsec:RQ1}

\subsubsection{Objectives} 
By analyzing the popularity of AI apps, such as identifying which categories of AI apps are more popular, developers can better understand market demands, thereby improving product development and deciding in which areas to invest. In this research question, we analyzed the popularity of AI apps through 4 sub-questions: 

\begin{itemize}[leftmargin=*] 
    \item \textbf{RQ-1.1} Has the released ratio of AI apps increased rapidly in recent years? 
    \item \textbf{RQ-1.2} In which markets are AI apps mainly released?
    \item \textbf{RQ-1.3} Which AI app categories do the top providers prefer?
    \item \textbf{RQ-1.4} Which categories of AI apps are more prevalent in the market?
\end{itemize}
\par 

\subsubsection{Experimental Methodology} 

First, we utilize AI Discriminator (cf. Section~\ref{subsec:find}) to extract AI apps from the AndroZoo application database (\cite{allix2016androzoo}). Next, we analyze these AI apps using the following methodology. 

\begin{itemize}[leftmargin=*]
    
    \item \textbf{Experiment RQ-1.1 (AI app released ratio)} To calculate the proportion of AI apps among all apps published each year, we record the total number of apps published and the number of AI apps published each year, and use Formula~\ref{con:ratio} for calculation: 
    \begin{equation}
    \label{con:ratio}
    \text { Ratio }=\frac{\text { \# AI apps}^y}{\text { \# published apps}^y}
    \end{equation} 

\noindent where $y$ refers to the specific year. $\text { \#AI apps }$ refers to the number AI apps published in the year $y$. $\text { \#published apps }$ refers to the number of all the apps published in the year $y$.

    \item \textbf{Experiment RQ-1.2 (AI app market distribution)} To investigate the distribution of AI apps across different markets, we first obtained information from AndroZoo about which market each AI app belongs to. Then, we calculated the number of AI apps released by each market. 
    \item \textbf{Experiment RQ-1.3 (App provider's preference for AI apps)} To investigate which categories AI apps providers are more focused on, we crawl the provider information for all the collected AI apps from the application market. The crawled information includes the app provider each app belongs to, categories, number of downloads, ratings, etc. We then calculate the number of AI apps developed by each app provider and identify the top ten providers based on the number of AI apps developed. Additionally, we analyze and report the category information of AI apps developed by each of these top ten providers.
    \item \textbf{Experiment RQ-1.4 (Market's preference for AI apps)} To investigate the popularity of different AI app categories in the market, we use a crawler tool to collect metadata for each app, including its category, number of installs, and rating. We then calculate the total number of apps, average number of installs, and average rating for each category. 

\end{itemize}

\begin{figure}[htpb]
	\centering
	\includegraphics[width=0.5\textwidth]{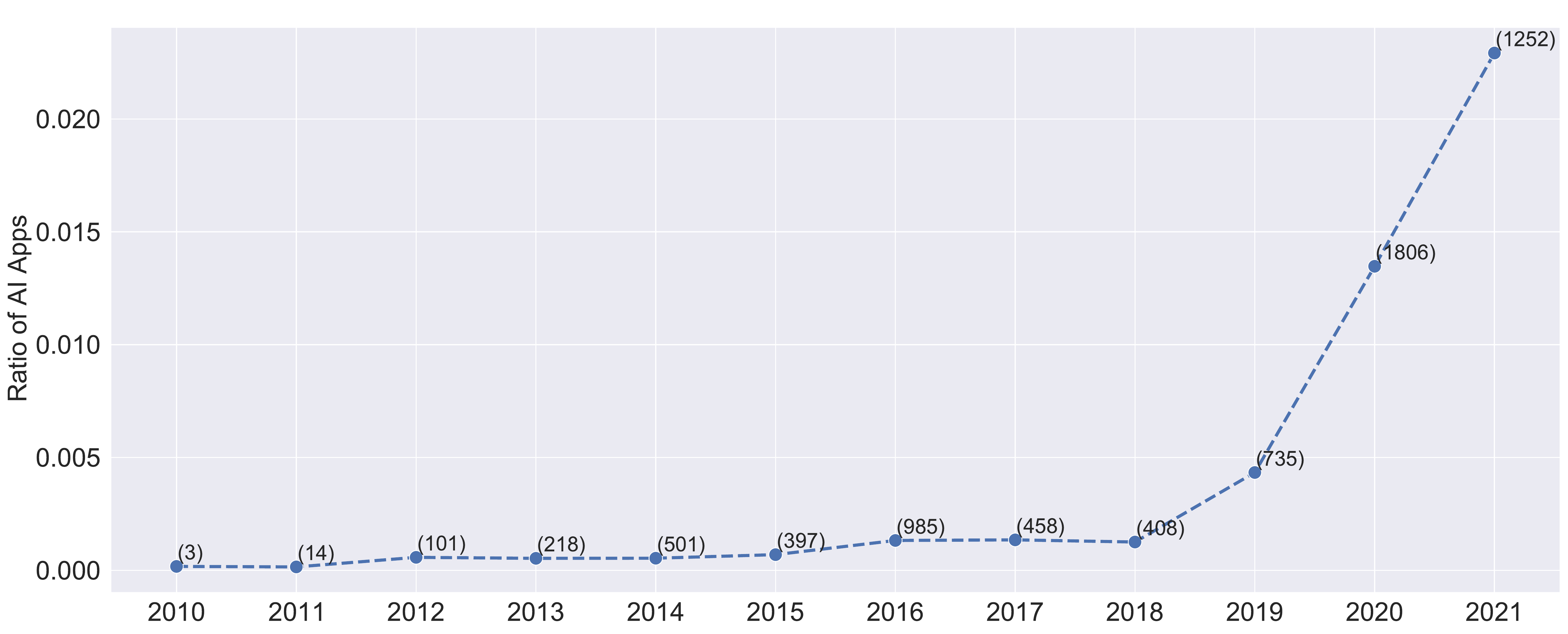}
	\caption{Ratio of AI apps developed in each year.}
	\label{fig:aiapps_years}
\end{figure}

\subsubsection{Results}
Figure \ref{fig:aiapps_years} presents the experimental results of \textbf{RQ-1.1}, showing the proportion of AI apps among the apps released each year. The X-axis exhibits the years, and the Y-axis presents the percentage of AI apps. From Figure \ref{fig:aiapps_years}, we see that from 2010 to 2018, the proportion of AI app development grew relatively slowly, while it rose sharply from 2018 to 2021. The experimental results indicate that from 2018, incorporating AI technology into applications has become a growing trend in mobile app development.

\vspace{0.1cm}
\noindent\colorbox{gray!20}{{\parbox{1.0\linewidth}{
\textbf{Finding 1:} Since 2018, incorporating AI technology into applications has become a growing trend in mobile app development.}}}
\vspace{0.1cm}

\begin{figure}[htpb]
	\centering
	\includegraphics[width=0.5\textwidth]{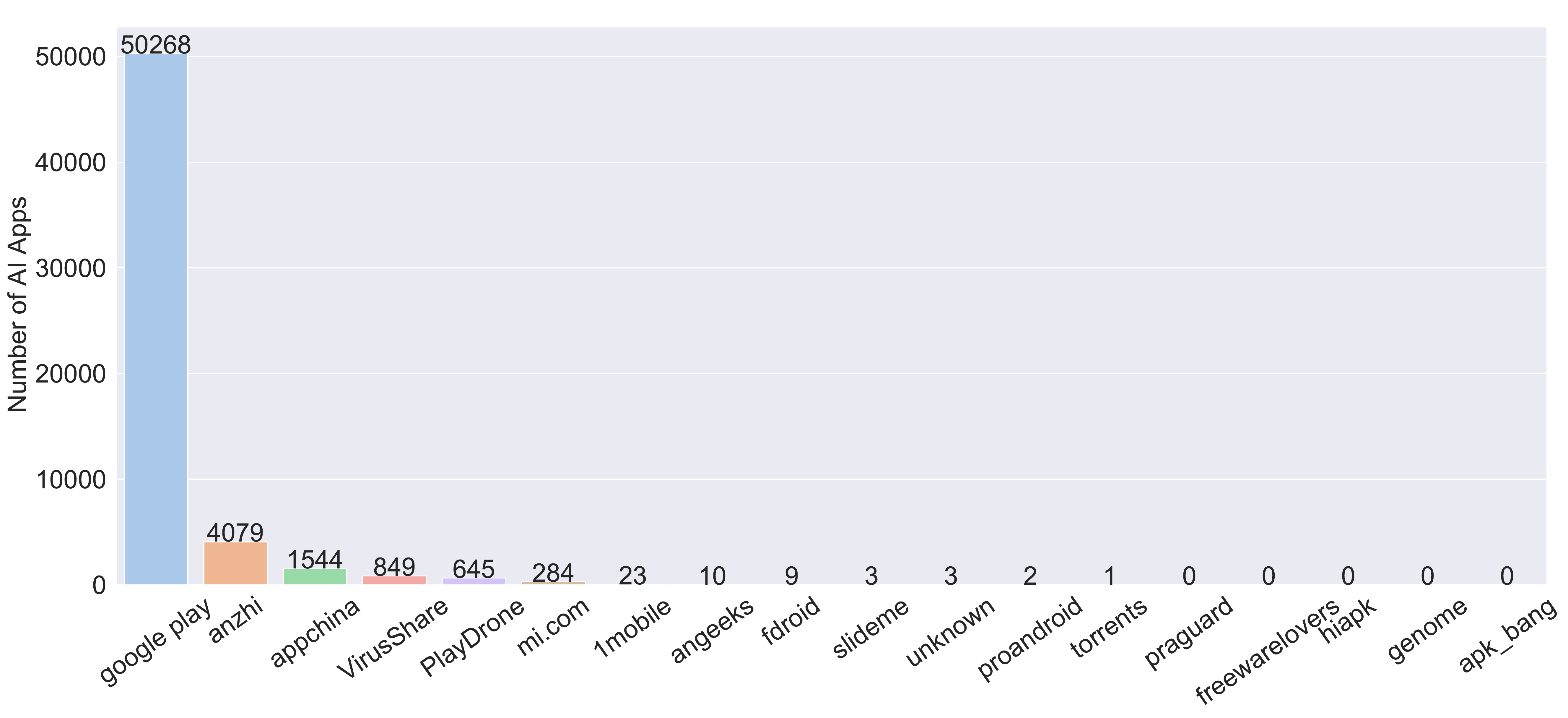}
	\caption{Number of AI apps in each market.}
	\label{fig:aiapps-market}
\end{figure}

Figure \ref{fig:aiapps-market} presents the experimental results for \textbf{RQ-1.2}, showcasing the market distributions for AI app releases. The X-axis displays the names of the markets, and the Y-axis indicates the number of AI apps released by each market. We see that Google Play has the highest number of AI apps published, with a total of 50,268 AI applications, which is 12 times more than the second highest, Anzhi, which released 4,079 AI apps. Following Anzhi are AppChina, VirusSharing, and PlayingDrones, which released 1,544, 849, and 645 AI apps, respectively.

\vspace{0.1cm}
\noindent\colorbox{gray!20}{{\parbox{1.0\linewidth}{
\textbf{Finding 2:} Currently, the market with the highest number of AI app releases is Google Play, significantly surpassing other markets. Other markets that have released over 1000 AI apps include Anzhi and AppChina. 
}}}
\vspace{0.1cm}

Table~\ref{tab:top10_companies_of_aiapps} presents the experimental results of \textbf{RQ-1.3}, showing the main categories of AI apps developed by the top-10 AI app providers. From left to right, the columns are the provider name, the number of AI apps each provider developed, and the main categories of these apps. We see that the top-10 AI app providers developed more AI apps in the Food \& Drink, Business, and Education categories. Specifically, 60\% of the companies published AI apps in Food \& Drink, 50\% released Business-oriented AI apps, and 40\% published Education-oriented AI apps.

\vspace{0.1cm}
\noindent\colorbox{gray!20}{{\parbox{1.0\linewidth}{
\textbf{Finding 3:} Top-10 AI app providers developed more AI apps in the Food \& Drink, Business, and Education categories. 
}}}

\vspace{0.1cm}

\begin{table}[htpb]
\caption{Top 10 Providers of AI apps}
\label{tab:top10_companies_of_aiapps}
\centering
\scalebox{0.55}
{
\begin{tabular}{lcl}
\toprule
\textbf{Provider name (Top 10)} & \textbf{Number of AI app} & \textbf{Category} \\ 
\midrule
Delivery Direto by Kekanto & 353 & \begin{tabular} [c]{@{}l@{}} \textcolor{orange}{Food \& Drink}, \textcolor{blue}{Business}, Shopping \end{tabular} \\ 
\midrule
CloudFaces & 129 & \begin{tabular} [c]{@{}l@{}} \textcolor{orange}{Food \& Drink}, \textcolor{blue}
{Business}, \textcolor{red}{Education}, \\ Beauty, Social, Entertainment, Travel \& Local,\\ Medical, Sports, Health \& Fitness \end{tabular} \\ 
\midrule
Tom McLeod Software & 102 & Productivity \\ 
\midrule
Laboratory X & 96 & \textcolor{orange}{Food \& Drink}, Lifestyle \\ 
\midrule
Klass Apps, Inc. & 87 & \textcolor{red}{Education}, Entertainment \\ 
\midrule
Core-apps & 85 & Books \& Reference \\ 
\midrule
Tara Blooms Private Ltd & 85 & \begin{tabular} [c]{@{}l@{}} \textcolor{orange}{Food \& Drink}, \textcolor{blue}{Business}, \textcolor{red}{Education},\\ Finance, Beauty, Travel \& Local, \\ Lifestyle, Shopping, Medical, \\ Auto \& Vehicles, Events \end{tabular} \\ 
\midrule
Tiffin Tom Ltd & 75 & \textcolor{orange}{Food \& Drink} \\ 
\midrule
InnoShop Co & 70 & \begin{tabular} [c]{@{}l@{}}  \textcolor{orange}{Food \& Drink}, \textcolor{blue}{Business}, \textcolor{red}{Education},\\  Entertainment, Beauty, Shopping \end{tabular} \\ 
\midrule
Via Transportation Inc. & 70 & \begin{tabular} [c]{@{}l@{}} \textcolor{blue}{Business}, Maps \& Navigation, Travel \& Local,\end{tabular} \\ 
\bottomrule
\end{tabular}
}
\end{table}

\begin{table}[htpb]
\caption{Categories of AI apps}
\label{tab:categories_ai_apps}
\centering
\scalebox{0.8}
{
\begin{tabular}{llllll}
\toprule
\multicolumn{2}{l}{\textbf{Category}} & \textbf{Ratio} & \textbf{Count} & \textbf{Avg Installs} & \textbf{Avg Score} \\ 
\midrule
\multicolumn{2}{l}{Finance} & 7.28\% & 2845 & 143337 & 4.1 \\ 
\multicolumn{2}{l}{Photography} & 4.75\% & 807 & 1639865 & 3.7 \\ 
\multicolumn{2}{l}{Productivity} & 3.80\% & 1752 & 2969649 & 3.8 \\ 
\multicolumn{2}{l}{Auto \& Vehicles} & 3.68\% & 374 & 18443 & 3.6 \\ 
\multicolumn{2}{l}{Food \& Drink} & 3.27\% & 1322 & 53698 & 4.0 \\ 
\multicolumn{2}{l}{Libraries \& Demo} & 3.13\% & 84 & 7535 & 3.6 \\ 
\multicolumn{2}{l}{Parenting} & 3.11\% & 62 & 57605 & 3.6 \\ 
\multicolumn{2}{l}{Business} & 3.01\% & 2398 & 19958 & 3.8 \\ 
\multicolumn{2}{l}{Beauty} & 2.88\% & 174 & 35702 & 3.7 \\ 
\multicolumn{2}{l}{Shopping} & 2.54\% & 1102 & 264894 & 4.0 \\ 
\multicolumn{2}{l}{Medical} & 2.39\% & 426 & 9105 & 3.7 \\ 
\multicolumn{2}{l}{Events} & 2.38\% & 158 & 739 & 4.7 \\ 
\multicolumn{2}{l}{Communication} & 2.34\% & 512 & 2011779 & 3.7 \\ 
\multicolumn{2}{l}{House \& Home} & 2.20\% & 173 & 12817 & 3.6 \\ 
\multicolumn{2}{l}{Tools} & 2.13\% & 1710 & 219064 & 3.6 \\ 
\multicolumn{2}{l}{Travel \& Local} & 2.13\% & 764 & 38989 & 3.8 \\ 
\multicolumn{2}{l}{Art \& Design} & 2.05\% & 185 & 20539 & 3.4 \\ 
\multicolumn{2}{l}{Social} & 1.83\% & 165 & 78650 & 3.7 \\ 
\multicolumn{2}{l}{Health \& Fitness} & 1.79\% & 865 & 26313 & 3.6 \\ 
\multicolumn{2}{l}{Lifestyle} & 1.68\% & 985 & 50755 & 3.6 \\ 
\multicolumn{2}{l}{Maps \& Navigation} & 1.63\% & 237 & 163105 & 3.7 \\ 
\multicolumn{2}{l}{Entertainment} & 1.63\% & 970 & 48776 & 3.5 \\ 
\multicolumn{2}{l}{Sports} & 1.58\% & 389 & 11748 & 3.7 \\ 
\multicolumn{2}{l}{Video Players \& Editors} & 1.34\% & 85 & 2128589 & 3.7 \\ 
\multicolumn{2}{l}{Education} & 1.30\% & 1920 & 20266 & 3.8 \\ 
\multicolumn{2}{l}{Comics} & 1.28\% & 10 & 1476 & 4.4 \\ 
\multicolumn{2}{l}{Weather} & 0.98\% & 39 & 13449 & 4.0 \\ 
\multicolumn{2}{l}{News \& Magazines} & 0.95\% & 211 & 8531 & 4.3 \\ 
\multicolumn{2}{l}{Books \& Reference} & 0.52\% & 299 & 10549 & 4.1 \\ 
\multicolumn{2}{l}{Music \& Audio} & 0.28\% & 161 & 36611 & 4.1 \\ 
\multicolumn{2}{l}{Dating} & 0.22\% & 1 & 50 & 0.0 \\ 
\multicolumn{2}{l}{Personalization} & 0.21\% & 79 & 2717519 & 3.8 \\ 
\multicolumn{1}{l}{\multirow{14}{*}{Game}} & Trivia & 0.43\% & 27 & 1546 & 3.7 \\ 
\multicolumn{1}{l}{} & Word & 0.33\% & 17 & 765 & 2.9 \\ 
\multicolumn{1}{l}{} & Strategy & 0.22\% & 9 & 5557578 & 2.9 \\ 
\multicolumn{1}{l}{} & Card & 0.99\% & 39 & 44559 & 3.8 \\ 
\multicolumn{1}{l}{} & Music & 0.64\% & 25 & 82 & 1.0 \\ 
\multicolumn{1}{l}{} & Board & 0.63\% & 34 & 8130 & 4.1 \\ 
\multicolumn{1}{l}{} & Casual & 0.26\% & 72 & 36145 & 3.7 \\ 
\multicolumn{1}{l}{} & Puzzle & 0.21\% & 61 & 97574 & 4.1 \\ 
\multicolumn{1}{l}{} & Role Playing & 0.20\% & 7 & 1436601 & 4.2 \\ 
\multicolumn{1}{l}{} & Simulation & 0.17\% & 19 & 326916 & 3.6 \\ 
\multicolumn{1}{l}{} & Adventure & 0.16\% & 16 & 64839 & 4.0 \\ 
\multicolumn{1}{l}{} & Action & 0.15\% & 16 & 641959 & 3.9 \\ 
\multicolumn{1}{l}{} & Racing & 0.10\% & 6 & 3500191 & 4.4 \\ 
\multicolumn{1}{l}{} & Arcade & 0.09\% & 27 & 1872319 & 3.2 \\ 
\bottomrule
\end{tabular}
}
\end{table}

The experimental results of \textbf{RQ-1.4} are presented in Table~\ref{tab:categories_ai_apps}, which shows the popularity of AI apps across different categories. In Table~\ref{tab:categories_ai_apps}, we present the number of AI apps in each category, their proportions (the ratio of AI apps to all apps), average installations, and average scores. Based on the analysis of Table~\ref{tab:categories_ai_apps}, we obtained the following findings (i.e., finding 4 to finding 7):

In Table \ref{tab:categories_ai_apps}, the column "Ratio" refers to the proportion of AI apps to all apps within each category. The data in the table is sorted in descending order based on this ratio. We see that the financial field has the highest proportion of AI apps, accounting for 7.28\%. Other categories with a high proportion of AI apps are Photography, Productivity, Auto \& Vehicles, Food \& Drink, Libraries \& Demo, and Parenting, accounting for 4.75\%, 3.80\%, 3.68\%, 3.27\%, 3.13\%, and 3.11\%, respectively.

\vspace{0.1cm}
\noindent\colorbox{gray!20}{{\parbox{1.0\linewidth}{
\textbf{Finding 4:} The financial field has the highest proportion of AI apps, followed by Photography, Productivity, Auto \& Vehicles, Food \& Drink, Libraries \& Demo, and Parenting. 
}}}
\vspace{0.1cm}

In Table \ref{tab:categories_ai_apps}, the column "Count" demonstrates the number of published AI apps in each category. We see that the categories Finance and Business released the highest number of AI apps, with 2845 and 2398 apps, respectively. This is followed by Education, Productivity, and Tools, with 1920, 1752, and 1322 AI apps released, respectively. In contrast, some areas have relatively few AI apps released, including Dating, Racing, Role-Playing, and Strategy games. 
\par 
\vspace{0.1cm}
\noindent\colorbox{gray!20}{{\parbox{1.0\linewidth}{
\textbf{Finding 5:} The categories Finance and Business released the highest number of AI apps, followed by Education and Tools. 
}}}
\vspace{0.1cm}

In Table~\ref{tab:categories_ai_apps}, the column "Avg Installs" demonstrates the average installs of AI apps in each category. We see that AI apps in the Strategy Games category have the highest number of installs, with 5,557,578 installs on average. This is followed by the Racing and Productivity categories, with 3,500,191 and 2,969,649 installs, respectively. 

\par 
\vspace{0.1cm}
\noindent\colorbox{gray!20}{{\parbox{1.0\linewidth}{
\textbf{Finding 6:} AI apps in the Strategy Games category have the highest number of installs, followed by the Racing and Productivity categories. 
}}}
\vspace{0.1cm}

In Table~\ref{tab:categories_ai_apps}, the column "Avg Score" shows the average score of AI apps in different categories, ranging from 0 to 5. We see that the categories with the highest average scores are Racing Games and Comics, both obtaining a score of 4.4. Additionally, 91.3\% of the categories have an average score above 3.0, and 87.0\% of the categories have an average score above 3.5. 
\par 
\vspace{0.1cm}
\noindent\colorbox{gray!20}{{\parbox{1.0\linewidth}{
\textbf{Finding 7:} AI apps in the Racing Games and Comics categories have the highest average scores. 91.3\% of the categories have an average score above 3.0, and 87.0\% of the categories have an average score above 3.5. 
}}}
\vspace{0.1cm}

\subsection{RQ2: Update status of published AI apps}
\label{subsec:RQ2}
\subsubsection{Objectives}

We aim to investigate the updates of AI apps, including the update frequency of AI apps, the correlation between app updates and embedded model updates, and the correlation between app updates and framework updates. Analyzing the update frequency of AI apps can reveal how often AI apps are maintained. Investigating the correlation between AI app updates and AI model updates can contribute to understanding the efficiency of developers in adopting new AI technologies. Understanding this relationship between AI app updates and AI framework updates can guide developers in selecting suitable frameworks. Specifically, we analyzed the updates of AI apps through three sub-questions: 

\begin{itemize}[leftmargin=*] 
    \item \textbf{RQ-2.1} How fast are the apps on the market updated?
    \item \textbf{RQ-2.2} To what extent do AI app updates accompany AI model updates? 
    \item \textbf{RQ-2.3} Which AI framework transitions typically occur when AI apps are updated? 
\end{itemize}

\subsubsection{Experimental Methodology} We conduct the following three experiments to answer the sub-questions.
\begin{itemize}[leftmargin=*]
    \item \textbf{Experiment RQ-2.1 (AI app update frequency)} To investigate the update frequency of the collected AI apps, we obtained the update information for each AI app from AndroZoo. In AndroZoo, applications under the same package name are different versions of the original app. We leveraged these package names to find out how many times each AI app has been updated.
    \item \textbf{Experiment RQ-2.2 (The correlation between app updates and model updates)} First, based on RQ-2.1, for each AI app, we identified its previous historical version apps. For each AI app and its historical versions, we downloaded the related application APKs from AndroZoo and decompiled them to obtain the internal files. We designed a program that can automatically extract the embedded models of each app and apply a hash function to calculate hash values for the extracted models. By comparing the hash values of the models in the original AI apps with those in their historical versions, we can determine whether the models have been updated. 
    \item \textbf{Experiment RQ-2.3 (The correlation between app updates and framework updates.)} Similarly, based on RQ-2.1, for each AI app, we identified its previous historical version apps. Then, we fed each AI app and its historical versions into AI Discriminator to obtain their applied AI frameworks and observed the changes in AI frameworks when the AI apps were updated. 
\end{itemize}

\subsubsection{Results} 
\begin{figure}[htpb]
	\centering
	\includegraphics[width=0.4\textwidth]{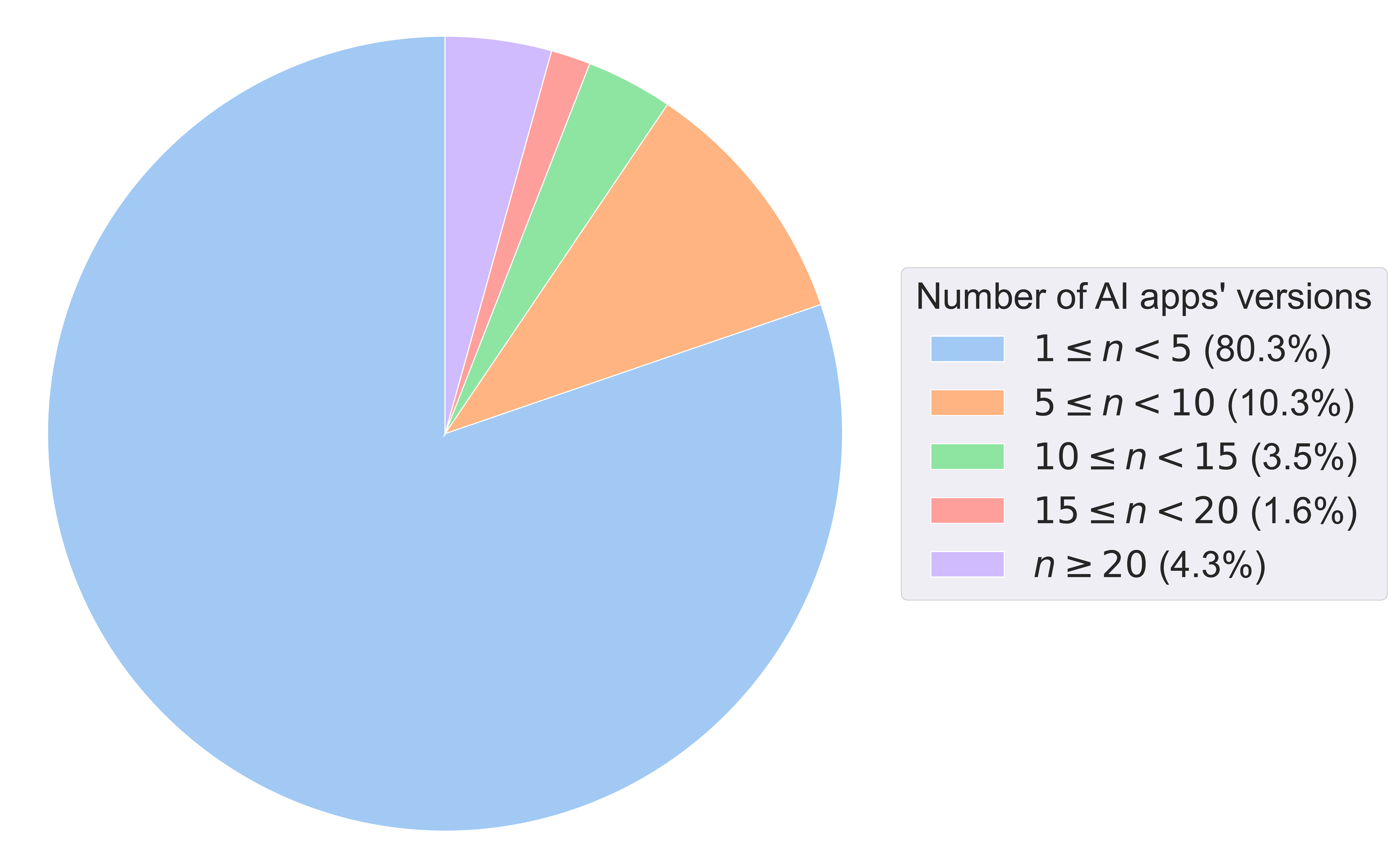}
	\caption{Distribution of historical version of AI apps.}
	\label{fig:aiapps_versions}
\end{figure}

Figure \ref{fig:aiapps_versions} presents the experimental results of \textbf{RQ-2.1}, showcasing the update frequency of the collected AI apps. The $n$ represents the number of updates. We separate all the collected AI apps into five groups based on the number of updates and present the proportion of AI apps in each category. In Figure~\ref{fig:aiapps_versions}, we see that the majority of AI apps (80.3\%) updated no more than 5 times. 10.3\% of AI apps are updated between 5 to 10 times. Only 4.3\% of AI apps are updated more than 20 times.

\vspace{0.1cm}
\noindent\colorbox{gray!20}{{\parbox{1.0\linewidth}{
\textbf{Finding 8:} The majority of AI apps (80.3\%) updated 0-5 times.
}}}
\vspace{0.1cm}

In response to \textbf{RQ-2.2}, we performed data analysis. Out of the 23,466 AI apps collected, 4,818 AI apps were updated, of which 2,225 had their AI models updated. Therefore, based on Formula~\ref{con:RQ2.2}, the proportion of model updates among the updated AI apps is 46.18\%. 
\begin{equation}
\label{con:RQ2.2}
\text { Ratio }=\frac{\text { \#model updates}}{\text { \#AI app updates }}=\frac{2225}{4818}=46.18 \%
\end{equation}

\noindent where \#model updates refer to the number of updated AI apps whose models were also updated. \#AI app updates refer to the number of updated AI apps. 

\vspace{0.1cm}
\noindent\colorbox{gray!20}{{\parbox{1.0\linewidth}{
\textbf{Finding 9:} Among all the updated AI apps, the proportion of AI apps whose models were also updated is 46.18\%. 
}}}
\vspace{0.1cm}

\begin{figure}[htpb]
	\centering
	\includegraphics[width=0.5\textwidth]{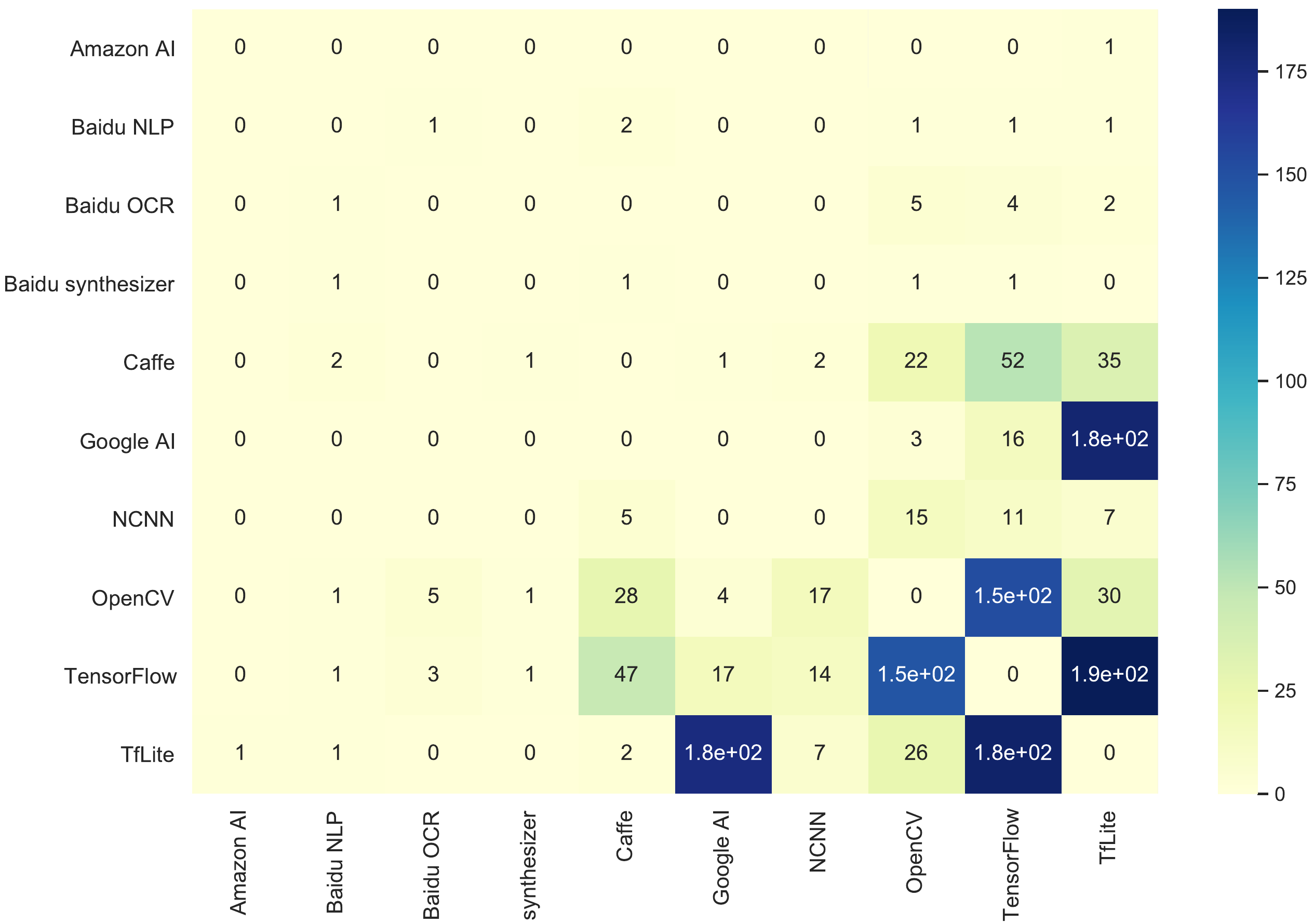}
	\caption{AI Frameworks change when AI apps are updated.}
	\label{fig:ai_frameworks_change}
\end{figure}

The experimental results of \textbf{RQ-2.3} are presented in Figure~\ref{fig:ai_frameworks_change}, which demonstrates the updates of AI frameworks when AI apps are updated. The X-axis refers to the AI framework after updates, and the Y-axis represents the frameworks before updates. The numbers in the grid indicate the number of transitions from the Y-axis framework to the X-axis framework, with darker colors indicating more frequent transitions. In Figure~\ref{fig:ai_frameworks_change}, we see that the top two most frequent framework transitions are TensorFlow $\rightarrow$ TFLite and TFLite $\rightarrow$ TensorFlow. Additionally, two other frequently occurring transitions are Google AI $\rightarrow$ TFLite and TFLite $\rightarrow$ Google AI.

\vspace{0.1cm}
\noindent\colorbox{gray!20}{{\parbox{1.0\linewidth}{
\textbf{Finding 10:} The most frequent framework transitions are TensorFlow $\rightarrow$ TFLite, TFLite $\rightarrow$ TensorFlow, Google AI $\rightarrow$ TFLite and TFLite $\rightarrow$ Google AI. 
}}}
\vspace{0.1cm}

\section{Framework and Model Analysis}
\label{sec:frame}

\subsection{RQ3: Framework popularity analysis of published AI apps}
\label{subsec:RQ3}

\subsubsection{Objectives} AI frameworks are utilized to build and deploy AI models (\cite{abadi2016tensorflow}). Some popular AI frameworks include TensorFlow, PyTorch, etc. In this research question, we investigate the popularity of different AI frameworks, the prevalence of single-framework and multi-framework AI applications, and the popularity trends of mainstream AI frameworks in recent years. We present our motivation for investigating these three aspects as follows. 

Understanding which AI frameworks are popular can help developers select more suitable frameworks among the many available AI frameworks. Investigating the usage condition of single-framework and multi-framework AI systems can help understand the prevalence of different development strategies. Understanding the popularity of mainstream AI frameworks can provide insights into the popularity trends of AI frameworks, specifically which AI frameworks are gradually being phased out and which ones are on the rise. Specifically, we analyzed AI frameworks based on three sub-questions:  

\begin{itemize}[leftmargin=*] 
    \item \textbf{RQ-3.1} How do on-device ML, on-device DL, and AI service apps differ in terms of usage and size?
    \item \textbf{RQ-3.2} What is the proportion and prevalence of AI apps using single and multiple AI frameworks? 
    \item \textbf{RQ-3.3} Among the mainstream AI frameworks, which frameworks have been the most popular in recent years?
\end{itemize}

\subsubsection{Experimental methodology} We conducted the following three experiments to answer the sub-questions above.

\begin{itemize}[leftmargin=*]
    \item \textbf{Experiment RQ-3.1 (Analysis on on-device ML, on-device DL, and AI service apps in terms of usage and size)} Existing AI frameworks can be broadly divided into three categories: on-device ML apps, on-device DL apps, and AI service apps. On-device ML apps refer to AI apps that perform inference using classical ML models deployed on mobile devices. On-device DL apps use DL models deployed on mobile devices for inference. AI service apps use cloud-based ML/DL services for inference. Here, we analyze these three types of AI apps from the following perspectives: application quantity and application size. 1) We calculated the number of AI apps in each category and the proportion of AI apps within each category. 2) We obtained the size information for each AI app and analyzed the size distribution of AI apps within each category. Moreover, we further categorized the AI frameworks and model format into six main categories, namely: 1) Traditional Deep Learning Frameworks, 2) Lite Deep Learning Frameworks, 3) Classical Machine Learning Frameworks, 4) Computer Vision Frameworks, 5) General Cloud AI Frameworks, and 6) Natural Language Processing Frameworks. In Table~\ref{tab:ai_frameworks}, we present the category to which each framework/model format belongs. In the following, we introduce each category and provide corresponding examples. 
    
\begin{itemize}[leftmargin=*]
   \item \textbf{Traditional Deep Learning Frameworks} are typically used for training and deploying deep neural networks. Typical examples include TensorFlow and PyTorch.
   \item \textbf{Lite Deep Learning Frameworks} are mainly used for deploying deep learning models on mobile devices. They are typically lightweight and can run efficiently on hardware with limited computational power. Typical examples include TensorFlow Lite and NCNN.
   \item \textbf{Classical Machine Learning Frameworks} are mainly used for implementing traditional machine learning algorithms, such as decision trees, Support Vector Machines (SVMs), k-nearest Neighbors (KNN), etc. Typical frameworks include Sklearn-porter and WEKA.
   \item \textbf{Computer Vision Frameworks} are specifically designed for image processing and computer vision tasks. OpenCV is a commonly used computer vision framework.
   \item \textbf{General Cloud AI Frameworks} are provided by cloud service providers and are designed to leverage the powerful computational capabilities and large-scale data processing capabilities of cloud computing for training and deploying AI models. Typical examples include Google AI Platform, Amazon AI, and Microsoft Azure AI.
   \item \textbf{Natural Language Processing Frameworks} focus on processing and analyzing natural language data, including tasks like semantic analysis. Typical examples include Baidu NLP.
\end{itemize}

    \item \textbf{Experiment RQ-3.2 (Single-framework and multi-framework AI apps)} We investigate the usage of single and multiple frameworks in AI apps. Single-framework AI apps refer to AI apps that use only one AI framework, while multi-framework AI apps use two or more frameworks. In RQ2.3 (cf. Section~\ref{subsec:RQ2}), we have obtained the framework information for each AI app. Based on this information, we ranked the frameworks/framework combinations according to their popularity and identified the top 10 most prevalent frameworks/framework combinations. 
    \item \textbf{Experiment RQ-3.3 (Mainstream AI frameworks)} We investigated the popularity of 16 mainstream frameworks in recent years. We counted the usage frequency of the mainstream AI frameworks for each time period.     
\end{itemize}

\subsubsection{Results}

\begin{figure}[htpb]
	\centering
	\includegraphics[width=0.4\textwidth]{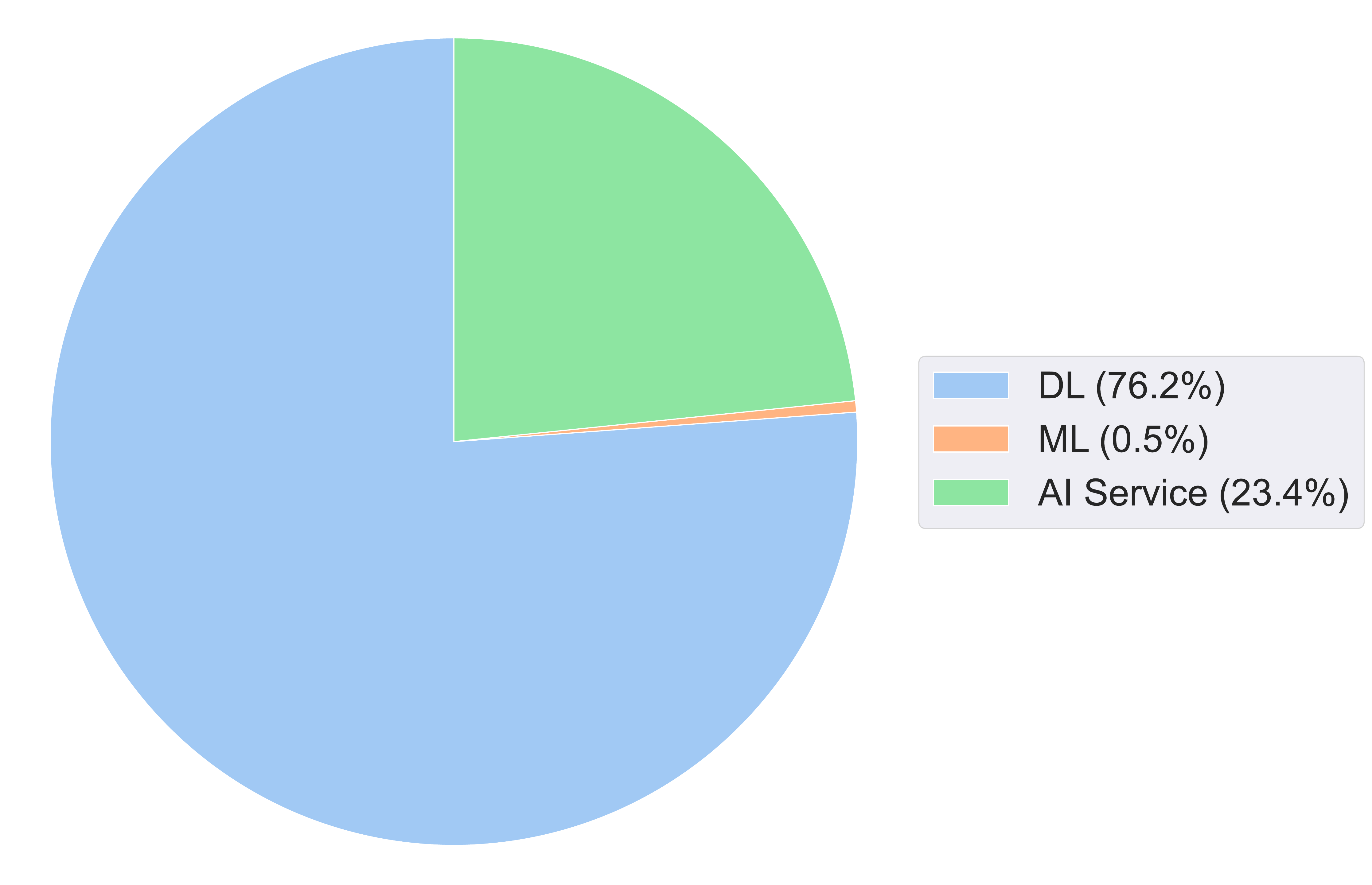}
	\caption{Ratio of AI apps in DL, ML and AI Service.}
	\label{fig:aiapps_ratio_dl_ml_service}
\end{figure}

\begin{figure}[htpb]
	\centering
	\includegraphics[width=0.45\textwidth]{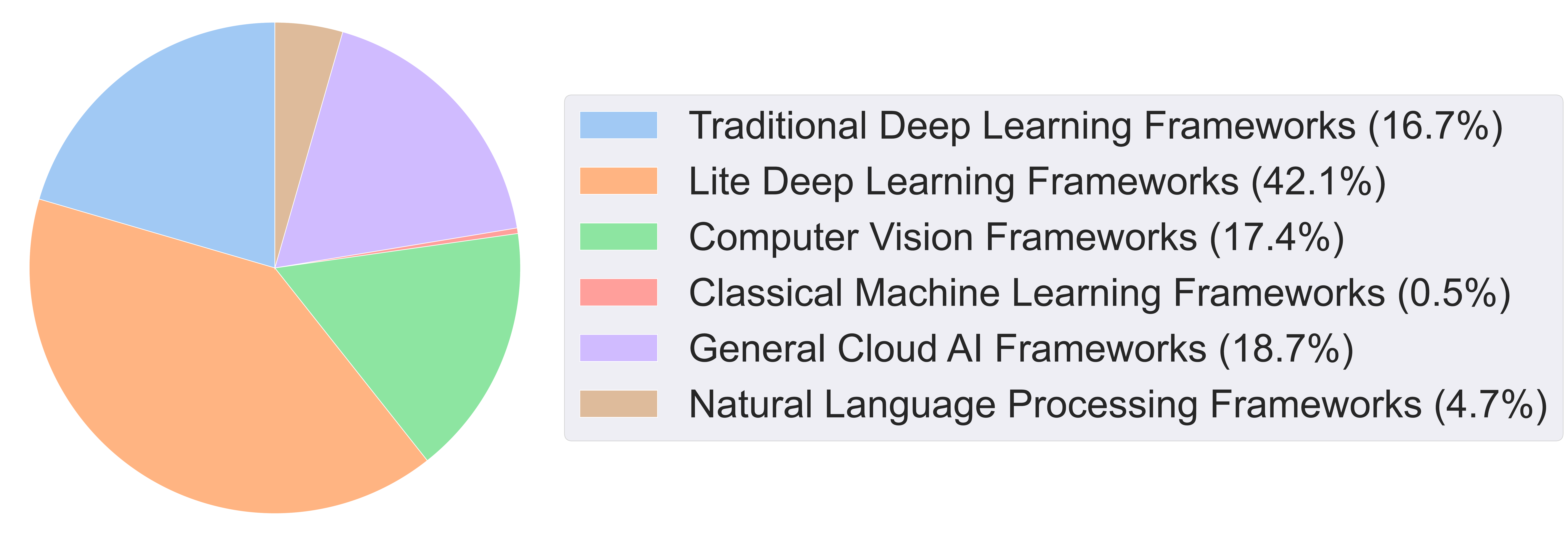}
	\caption{Ratio of AI apps across different framework categories.}
	\label{fig:aiapps_ratio_framework_specific}
\end{figure}

\begin{figure}[ht]
        \centering

        \includegraphics[width=5.0cm]{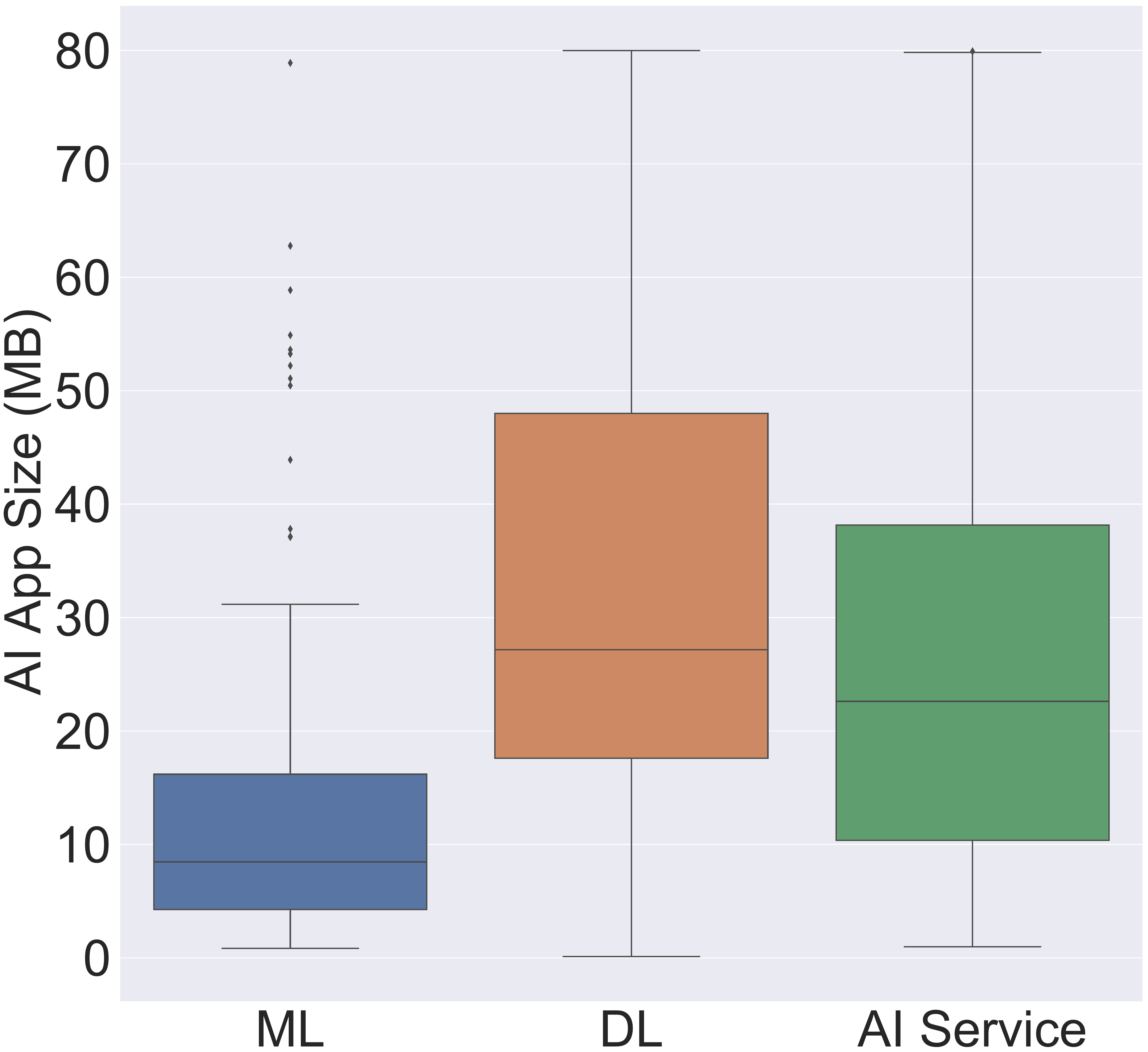}
        \label{aiapps_size_ml_dl_service}

    \caption{AI app size in DL, ML and AI Service}
    \label{fig:aiapp_lib_size_dl_ml_service}
    \end{figure}

The experimental results of \textbf{RQ-3.1} are presented in Figure~\ref{fig:aiapps_ratio_dl_ml_service}, Figure~\ref{fig:aiapps_ratio_framework_specific}, and Figure~\ref{fig:aiapp_lib_size_dl_ml_service}. Figure~\ref{fig:aiapps_ratio_dl_ml_service} presents the proportions of the three types of AI apps: on-device ML, on-device DL, and AI service. From Figure~\ref{fig:aiapp_lib_size_dl_ml_service}, we see that on-device DL apps have the highest proportion (76.2\%). Next are AI service-based apps, accounting for 23.4\%. Finally, on-device ML apps have the lowest proportion, accounting for only 0.5\%. 

Figure~\ref{fig:aiapps_ratio_framework_specific} presents a more detailed breakdown of the collected AI apps. As shown in the figure, Lite Deep Learning Frameworks have the highest proportion, accounting for 42.1\%. General Cloud AI Frameworks rank second, accounting for 18.7\% of the collected AI apps. Next are Computer Vision Frameworks and Traditional Deep Learning Frameworks, which account for 17.4\% and 16.7\%, respectively. To conclude, Lite Deep Learning Frameworks, General Cloud AI Frameworks, Computer Vision Frameworks, and Traditional Deep Learning Frameworks are the frameworks with the highest proportions among the collected AI apps.  

\vspace{0.1cm}
\noindent\colorbox{gray!20}{{\parbox{1.0\linewidth}{
\textbf{Finding 11:} AI apps supported by on-device DL techniques accounted for the highest proportion. Lite Deep Learning Frameworks, General Cloud AI Frameworks, Computer Vision Frameworks, and Traditional Deep Learning Frameworks are the frameworks with the highest proportions among the collected AI apps.} }}
\vspace{0.1cm}

Figure~\ref{fig:aiapp_lib_size_dl_ml_service} shows the size distribution of three different categories of AI apps. We see that on-device DL apps have the highest median, approximately 27 MB. The size distribution mainly ranges from 20 MB to 48 MB. AI service apps have the second-highest median, around 23 MB, with a main range of 10 MB$\sim$40 MB. On-device ML apps have the lowest median size, around 8 MB, with a main size range of approximately 5 MB$\sim$16 MB.

\vspace{0.1cm}
\noindent\colorbox{gray!20}{{\parbox{1.0\linewidth}{
\textbf{Finding 12:} On-device DL apps have a relatively larger size compared to AI service apps and on-device ML apps. }}}
\vspace{0.1cm}
\par

\begin{figure}[!ht]
    % \centering
    \flushleft
    \subfigure[Ratio of AI apps using AI frameworks]{\label{subfig:ratio}
    \includegraphics[scale=0.109]{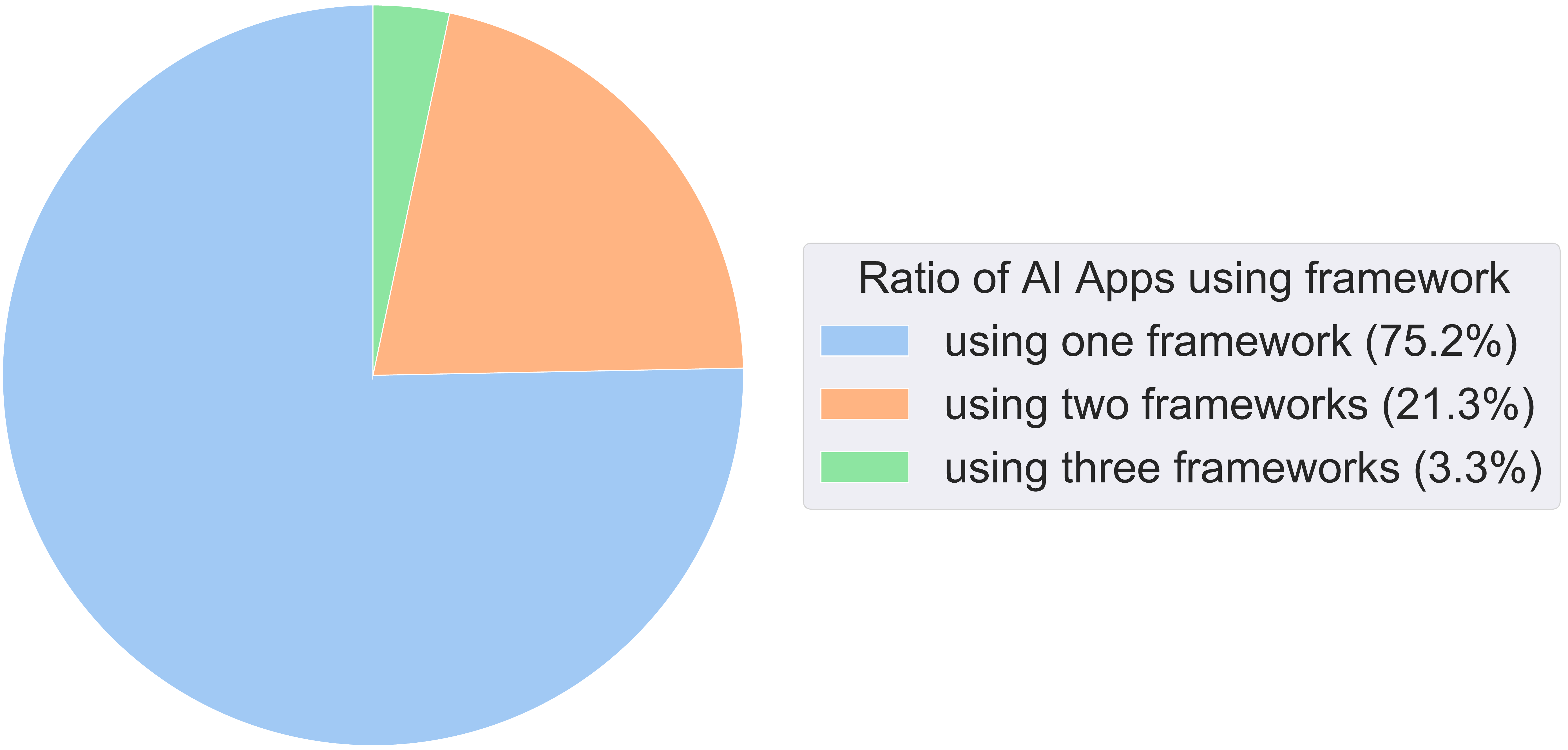}%
    }
    \subfigure[Ratio of AI apps using one framework]{\label{subfig:using_one}
    \includegraphics[scale=0.109]{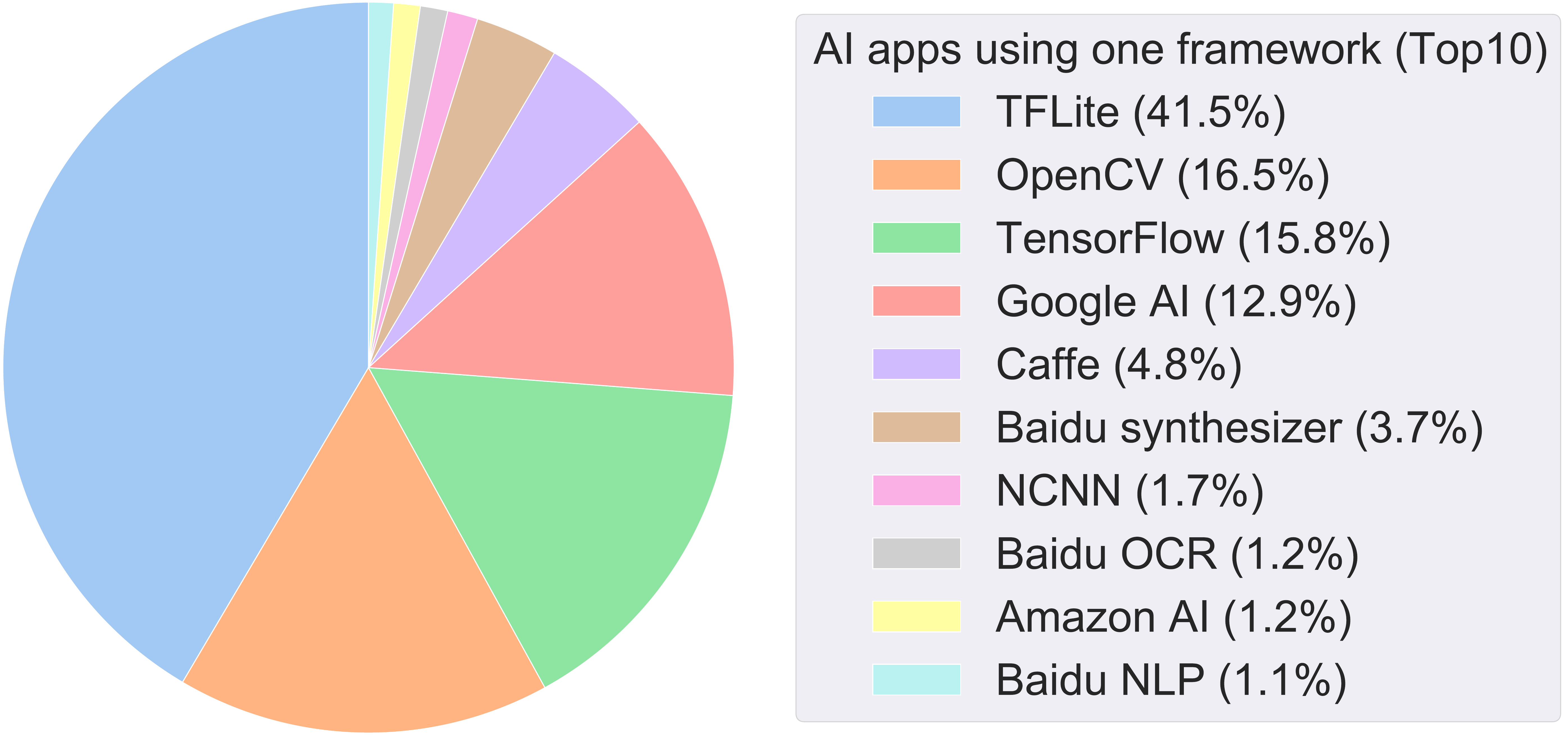}%
    }
    \subfigure[Ratio of AI apps using two framework]{\label{subfig:using_two}
    \includegraphics[scale=0.109]{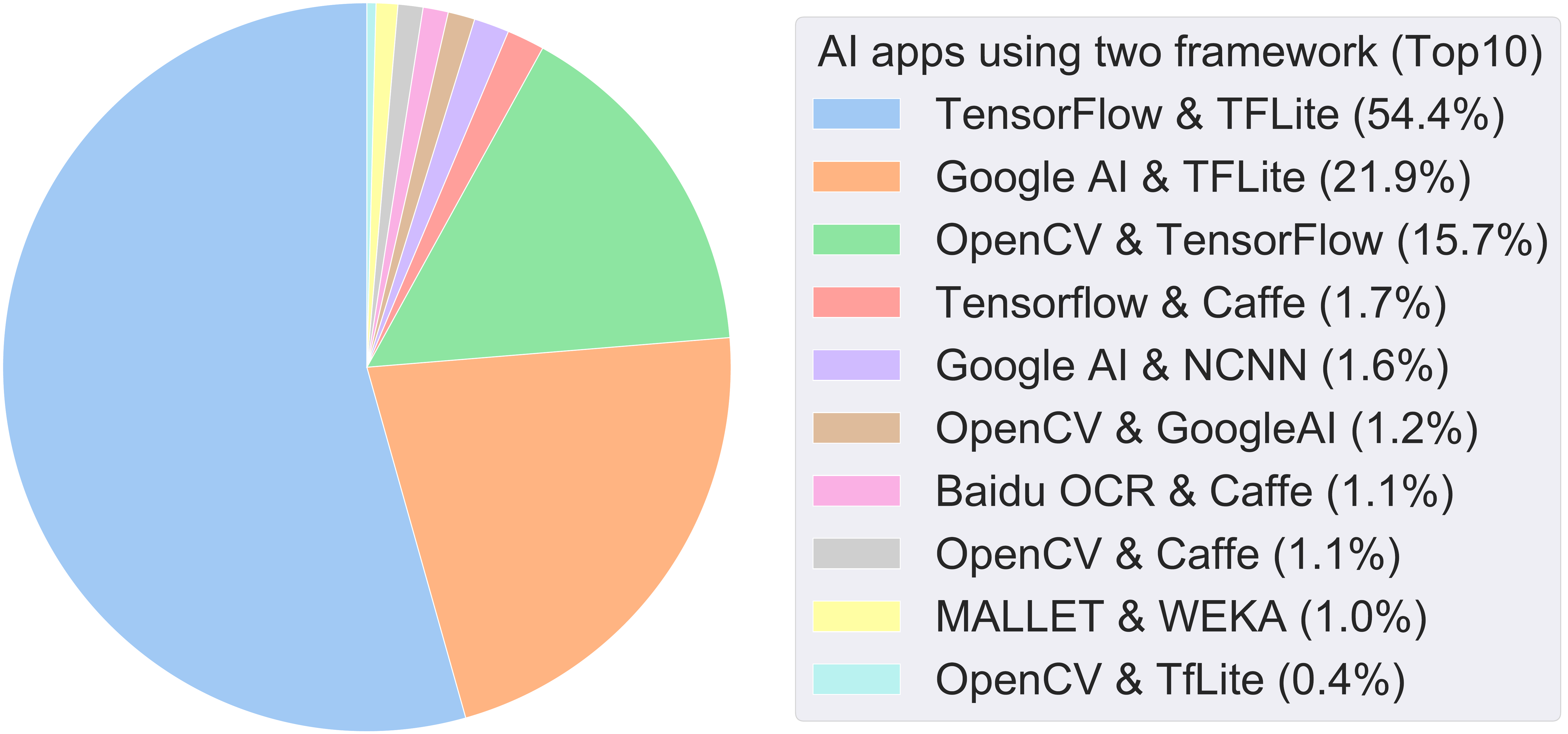}%
    }
    \subfigure[Ratio of AI apps using three framework]{\label{subfig:using_three}
    \includegraphics[scale=0.109]{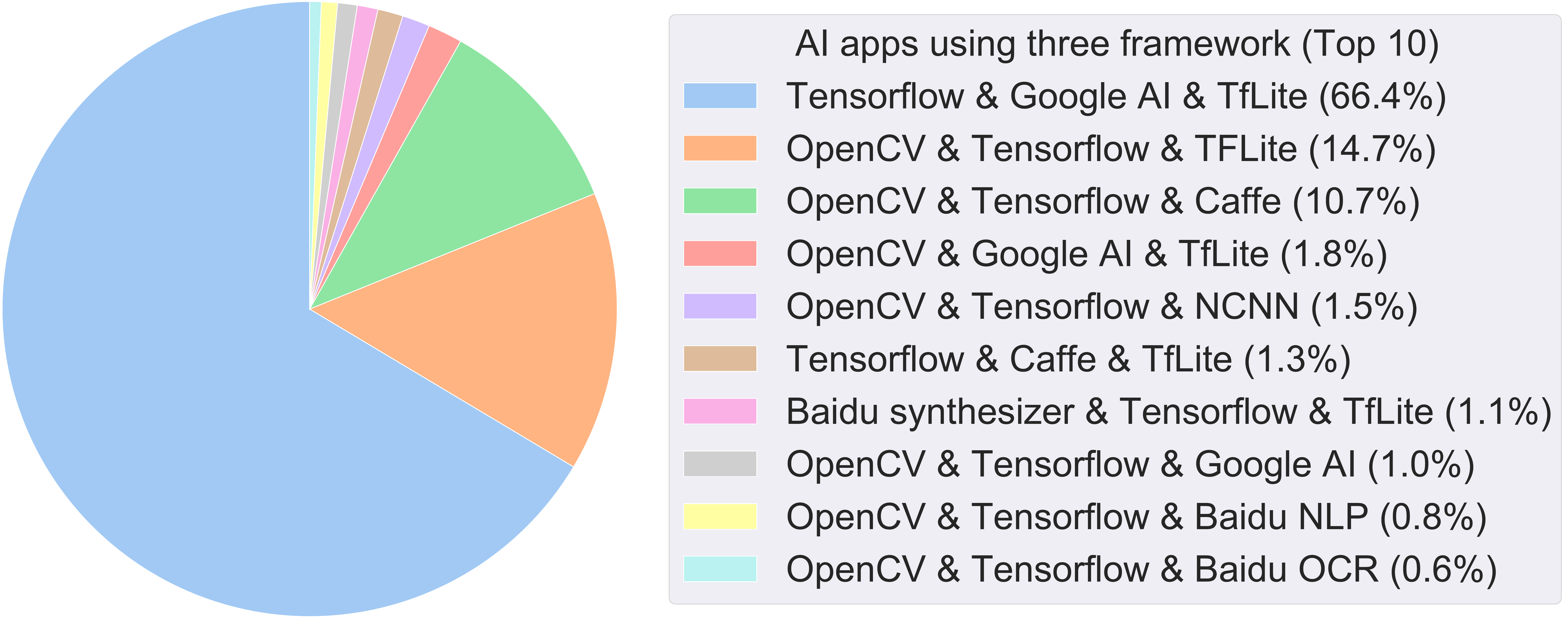}%
    }
    \caption{Distribution of AI apps using AI frameworks.}
    \label{fig:distribution_frameworks}
\end{figure}

The experimental results of \textbf{RQ-3.2} are presented in Figure~\ref{fig:distribution_frameworks}, which shows the usage of single framework and multiple frameworks in the collected AI apps. 

Figure~\ref{subfig:ratio} shows the proportion of collected AI apps using one AI framework, two AI frameworks, and three AI frameworks. We can see that AI apps using one AI framework account for the largest proportion (75.2\%). AI apps using two frameworks account for the second largest proportion, at 21.3\%. AI apps using three frameworks account for the smallest proportion, at 3.3\%.  

\vspace{0.1cm}
\noindent\colorbox{gray!20}{{\parbox{1.0\linewidth}{
\textbf{Finding 13:} Among all the collected AI apps, single-framework AI apps account for the highest proportion. }}}
\vspace{0.1cm}

Figure~\ref{subfig:using_one} exhibits which AI frameworks are more popular among AI apps using a single framework. We can see that AI apps using the AI framework TFLite account for the highest proportion, at 41.5\%. This is followed by OpenCV, TensorFlow, and Google AI, making up 16.5\%, 15.8\%, and 12.9\%, respectively.

\vspace{0.1cm}
\noindent\colorbox{gray!20}{{\parbox{1.0\linewidth}{
\textbf{Finding 14:} Among AI apps using a single framework, apps using the TFLite framework have the highest proportion, at 41.5\%. }}}
\vspace{0.1cm}

Figures~\ref{subfig:using_two} and Figure~\ref{subfig:using_three} show the prevalence of different AI framework combinations. Specifically, Figure~\ref{subfig:using_two} displays the most popular two-framework combinations, while Figure~\ref{subfig:using_three} shows the most popular three-framework combinations. In Figure~\ref{subfig:using_two}, we see that the combination of TensorFlow \& TFLite is the most popular. About 54.4\% of the collected two-framework AI apps use this combination. This is followed by Google AI \& TFLite and OpenCV \& TensorFlow, accounting for 21.9\% and 15.7\%, respectively. Regarding AI apps using three frameworks, the combination of TensorFlow \& Google AI \& TFLite is the most popular, accounting for 66.4\%. This is followed by OpenCV \& TensorFlow \& TFLite and OpenCV \& TensorFlow \& Caffe, accounting for 14.7\% and 10.7\%, respectively.

\vspace{0.1cm}
\noindent\colorbox{gray!20}{{\parbox{1.0\linewidth}{
\textbf{Finding 15:} Among AI apps using two frameworks, apps using the combination of TensorFlow \& TFLite are the most popular, accounting for 41.5\%. Among AI apps using three frameworks, apps using the combination of TensorFlow \& Google AI \& TFLite are the most popular, accounting for 66.4\%. }}}
\vspace{0.1cm}

\begin{figure*}[htpb]
	\centering
	\includegraphics[width=0.8\textwidth]{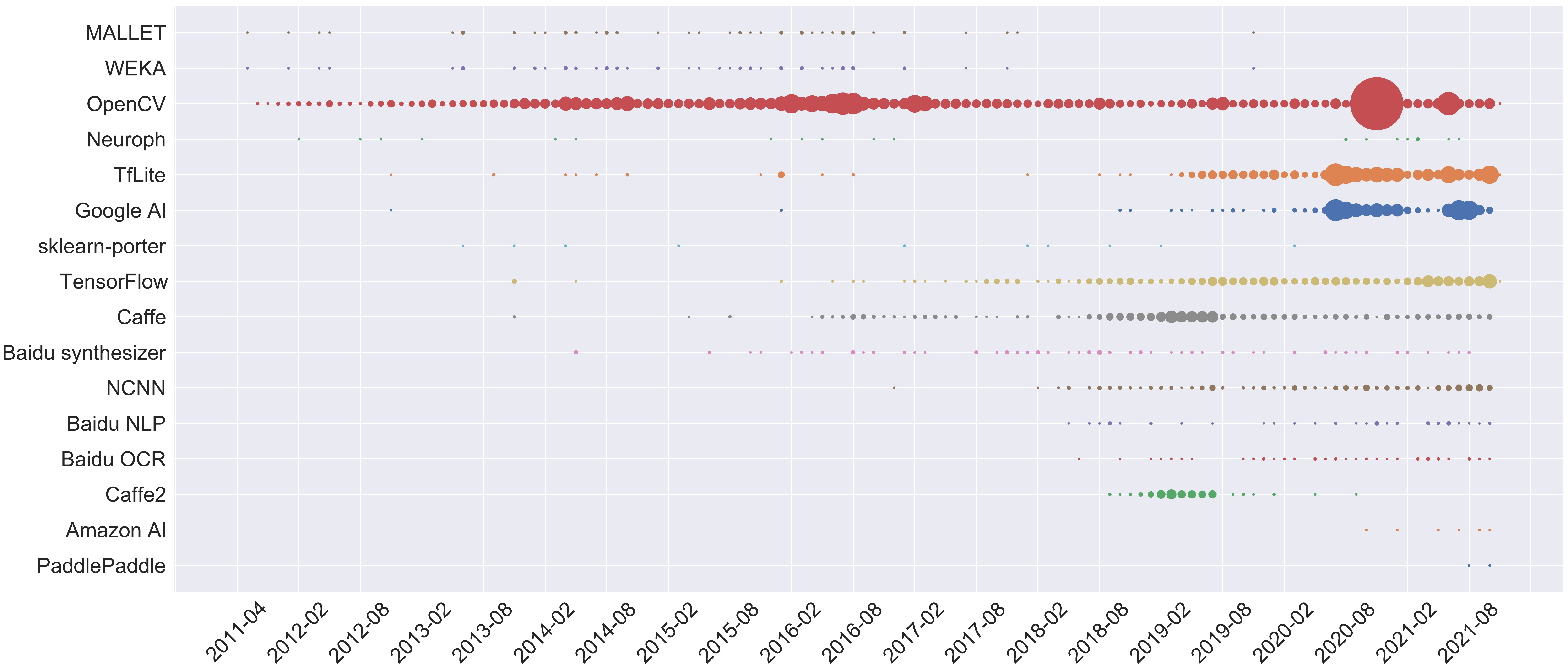}
	\caption{The usage of AI framework each year.}
	\label{fig:aiapps_gantt_framework_new}
\end{figure*}
\

The experimental results of \textbf{RQ-3.3} are demonstrated in Figure~\ref{fig:aiapps_gantt_framework_new}, which presents the popularity of 16 different AI frameworks from 2011 to 2021. The X-axis represents the year, and the Y-axis represents the AI frameworks. In Figure~\ref{fig:aiapps_gantt_framework_new}, the size of the points indicates the usage frequency of AI frameworks applied in each time slice by the newly released AI apps. The larger the point, the more frequently the framework was used in that year. From Figure~\ref{fig:aiapps_gantt_framework_new}, we can see that OpenCV has the largest point areas, indicating it has been consistently popular. On the other hand, TFLite and Google AI have rapidly become prevalent since 2020. Additionally, TensorFlow was consistently used between 2018 and 2021. The experimental results show that from the perspective of usage frequency, OpenCV, TFLite, Google AI, and TensorFlow are relatively popular compared to other AI frameworks.

\vspace{0.1cm}
\noindent\colorbox{gray!20}{{\parbox{1.0\linewidth}{
\textbf{Finding 16:} From the perspective of usage frequency, some relatively more popular AI frameworks are OpenCV, TFLite, Google AI, and TensorFlow. 
}}}
\vspace{0.1cm}

\subsection{RQ4: Model protection status of published AI apps}
\label{subsec:RQ4}

\subsubsection{Objectives} Sufficient model protection can prevent the core intellectual properties of AI apps from being stolen by malicious competitors. However, the boom in on-device AI apps increases the likelihood of model leaks as models are deployed directly on the client side (\cite {deng2022understanding}). In this research question, we investigate the model protection condition in AI apps. We conducted our study from two aspects: 1) We examined the use of open-source models in AI apps, as open-source models potentially pose higher security risks. 2) We investigated the encryption level of models embedded in AI apps, as strong encryption can significantly prevent AI app models from being stolen. 

The motivation for investigating the above two issues is twofold: 1) Open-source models can potentially pose security risks. Investigating the use of open-source models in AI apps can, to some extent, shed light on the state of AI model protection. 2) Investigating model encryption conditions can help understand to what extent on-device AI models are protected. We summarize the above two considerations into the following two sub-questions.

\begin{itemize}[leftmargin=*]
    \item \textbf{RQ-4.1} To what extent do public AI apps use open-source models to perform AI tasks?
    \item \textbf{RQ-4.2} Are models embedded in published AI applications well-encrypted?
\end{itemize}

\subsubsection{Experimental methodology} 
We conduct two corresponding experiments to answer the above two sub-questions. \par 

\begin{itemize}[leftmargin=*]
\item \textbf{Experiment for RQ-4.1 (Usage of open-source models in public AI apps)} We conducted experiments to calculate the ratio of AI apps directly using some common open-source models among the collected AI apps. In the first step, we collected 90 open-source mobile models from the TFLite Hub (\cite{tflitehub}), a repository that provides reusable ML models and calculated their hashes. TFLite was chosen as the source for open-source models because it is one of the most widely used framework codebases for deploying machine learning models on mobile devices. In the second step, we calculated the hashes of our collected 23,466 AI app models. For each AI app model, we matched its hash with the collection of open-source model hashes. If the hash of a model in an AI app matched any of the hashes in the open-source model collection, we considered that the AI app used open-source models. Using this method, we calculated the proportion of the AI apps that utilized common open-source models.  \par 

\item \textbf{Experiment for RQ-4.2 (Model encryption status of AI apps)} To study the encryption status of collected AI app models, we leverage the idea of the standard entropy test (\cite{sun2021mind}) to determine whether a given AI model is encrypted. The standard entropy test assesses whether an AI model is encrypted by calculating its entropy. According to its principle, encrypted AI models typically exhibit high entropy values. Following the existing study (\cite{sun2021mind}), we set the entropy threshold for encryption at 7.99. If an AI model file's entropy exceeds this value, we consider the AI model encrypted. 
\end{itemize}

Moreover, to prove the effectiveness of this approach, we conducted the following evaluation: First, we applied the standard entropy test approach with the threshold of 7.99 to obtain a set of AI models considered encrypted by this method. We randomly selected 50 models from this set and attempted to open them manually using the model viewer Netron. We found that all the models considered encrypted could not be opened. Next, we tried to use the official API of these AI models to load these models. Generally, unencrypted models can be successfully loaded. We also found that none of the models could be opened. Through these methods, we concluded that the standard entropy test with a threshold of 7.99 is a reasonable and effective method for verifying whether models are encrypted.

\subsubsection{Results} 

The experimental results of \textbf{RQ-4.1} are presented in Table~\ref{tab:cloned_model} and Table~\ref{tab:public-models-company}. Among the 23,466 AI app models collected, we found that 175 apps utilize open-source models by simply renaming them, accounting for 0.7\%. Since malicious attackers can easily obtain the input and output shapes of the open-source models, directly using open-source models exposes AI apps to security threats. Table~\ref{tab:cloned_model} shows some examples of directly renamed and used open-source AI models. The table from left to right presents the model name, model tasks, and company-provided renames. Table~\ref{tab:public-models-company} presents ten prevalent published models used by providers. From left to right, the table shows the model names, model execution tasks, and providers using these models. We see that the mainstream execution tasks of these models are image classification and image object detection.

\begin{table}[htbp]
\caption{Example of public models renamed in AI apps}
\label{tab:cloned_model}
\centering
\scalebox{0.6}
{
\begin{tabular}{lll}
\toprule
\multicolumn{1}{c}{\textbf{Public models}} & \multicolumn{1}{c}{\textbf{Task}} & \multicolumn{1}{c}{\textbf{Renamed models in AI apps}} \\ 
\midrule
\begin{tabular}[c]{@{}l@{}}magenta-arbitrary-image-styli-\\ zation-v1-256-int8-transfer-1.tflite\end{tabular} & Image-style-transfer & \begin{tabular}[c]{@{}l@{}}art-photo-384.tflite, \\ transfer-model.tflite,\\ style-transfer-quantized-384.tflite,\\ style-transfer.tflite\end{tabular} \\ 
\midrule
\begin{tabular}[c]{@{}l@{}}lite-model-aiy-vision-cla-\\ssifier-insects-V1-3.tflite\end{tabular} & Image-classification & \begin{tabular}[c]{@{}l@{}}insects-C.tflite,\\ aiy-classifier-natural-world-inse-\\ cts-V1-2-quantized-input-ui-\\ nt8-85018f9a4c0110bd69f70-\\be107f7d2207124c301-model-with\\-metadata.tflite\end{tabular} \\ 
\midrule
\begin{tabular}[c]{@{}l@{}}lite-model-ssd-mobil-\\ enet-v1-1-metadata-2.tflite\end{tabular} & Image Object Detection & detect.tflite \\ 
\bottomrule
\end{tabular}
}
\end{table}

\vspace{0.1cm}
\noindent\colorbox{gray!20}{{\parbox{1.0\linewidth}{
\textbf{Finding 17:} Among the collected AI apps, 175 utilized open-source AI models. The mainstream execution tasks of the ten prevalent open-source AI models are image classification and image object detection.
}}}
\vspace{0.1cm}

\begin{table}[htbp]
\caption{Examples of public mobile models used in provider}
\label{tab:public-models-company}
\centering
\scalebox{0.6}
{
\begin{tabular}{lll}
\toprule
\multicolumn{1}{c}{\textbf{public mobile models}} & \multicolumn{1}{c}{\textbf{Task}} & \multicolumn{1}{c}{\textbf{Providers}} \\ 
\midrule
efficientnet-lite0-int8-2.tflite & Image-classification & \begin{tabular}[c]{@{}l@{}}Shopping Deals \& Specials, \\ edobo\end{tabular} \\ 
\midrule
efficientnet-lite4-int8-2.tflite & Image Classification & Dave Bennett \\ 
\midrule
\begin{tabular}[c]{@{}l@{}}lite-model-aiy-vision-\\ classifier-birds-V1-3.tflite\end{tabular} & Image Classification & \begin{tabular}[c]{@{}l@{}}DSM Services, \\ Vanchel\end{tabular} \\ 
\midrule
\begin{tabular}[c]{@{}l@{}}lite-model-aiy-vision-\\ classifier-food-V1-1.tflite\end{tabular} & Image Classification & AG Apps Co \\ 
\midrule
\begin{tabular}[c]{@{}l@{}}mobilenet-v2-1.0-224-\\ 1-metadata-1.tflite\end{tabular} & Image Classification & Memorizer \\ 
\midrule
\begin{tabular}[c]{@{}l@{}}lite-model-cropnet-classifier-\\ cassava-disease-V1-1.tflite\end{tabular} & Image Classification & Solomon Nsumba \\ 
\midrule
\begin{tabular}[c]{@{}l@{}}lite-model-object-detection-\\ mobile-object-labeler-v1-1.tflite\end{tabular} & Image Classification & Glitter Technology Ventures LLC \\ 
\midrule
\begin{tabular}[c]{@{}l@{}}lite-model-ssd-mobilenet-\\ v1-1-metadata-2.tflite\end{tabular} & Image Object Detection & \begin{tabular}[c]{@{}l@{}}A La Carte Media Inc., \\ Apptastic Mobile, \\ Bridgewiz Engineering, \\ DistinctView, \\ Farid Ahmad Ahmadyar, \\ FavLabs,\\ LazyDroid, \\ MKK Games, \\  MLPJ DROID, \\ PlatineX TDC, \\ Polycents, \\ RajAppStudio, \\ SANE Tech, \\ Sifie Apps, \\ Sparkling India, \\ TeamLease EdTech Ltd., \\ Yusuf Suhair\end{tabular} \\ 
\midrule
\begin{tabular}[c]{@{}l@{}}object-detection-mobile-object-\\ localizer-v1-1-default-1.tflite\end{tabular} & Image Object Detection & \begin{tabular}[c]{@{}l@{}}FRUCT, \\ Nexart TechnoSolutions Pvt Ltd\end{tabular} \\ 
\midrule
lite-model-cartoongan-int8-1.tflite & Image Style Transfer & \begin{tabular}[c]{@{}l@{}}Dan Group, Dotsquares, \\ Pixel Force Pvt Ltd\end{tabular} \\ 
\bottomrule
\end{tabular}
}
\end{table}

\begin{figure}[htpb]
	\centering
	\includegraphics[width=0.45\textwidth]{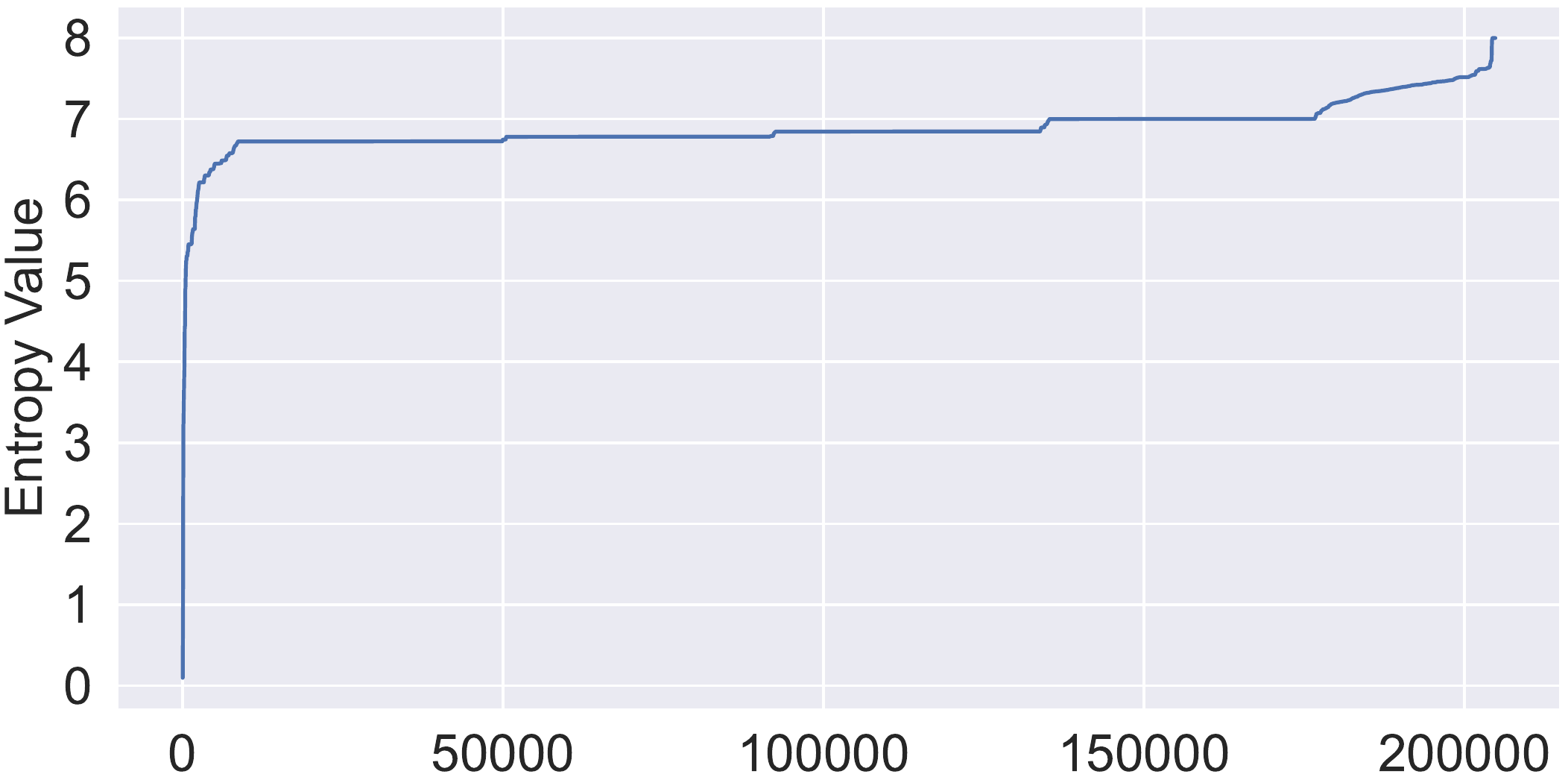}
	\caption{Model entropy of AI apps.}
	\label{fig:model_entropy}
\end{figure}

The experimental results of \textbf{RQ-4.2} are presented in Figure~\ref{fig:model_entropy}, which shows the encryption status of the collected AI app models. As mentioned in the experimental design of RQ-4.2, we used the standard entropy test (\cite{sun2021mind}) to determine whether an AI model is encrypted. If an AI model's entropy exceeds 7.99, this model is considered encrypted. In Figure~\ref{fig:model_entropy}, the Y-axis represents the entropy value of a model, while the X-axis represents the index of each AI model. All AI models are sorted by their entropy values from left to right. From Figure~\ref{fig:model_entropy}, we can see that the number of AI models with entropy values exceeding 7.99 is small. After manual verification, we found that the number of models considered to be encrypted is 520, accounting for 0.25\%. \par 
\vspace{0.1cm}
\noindent\colorbox{gray!20}{{\parbox{1.0\linewidth}{
\textbf{Finding 18:} Among all the collected AI models, the number of models considered to be encrypted is 520, accounting for only 0.25\%.
}}}
\vspace{0.1cm}

\section{User Analysis}
\label{sec:users}
\subsection{RQ5: User privacy protection of published AI apps}
\label{subsec:RQ5}

\subsubsection{Objectives}

Protecting user privacy in AI apps is crucial to maintaining user trust. Failure to adequately protect user data can lead to security breaches, misuse of personal information, and reputational damage for providers. In this research question, we investigate the state of user privacy protection in published AI apps. Analyzing the state of user privacy protection can contribute to identifying potential privacy protection vulnerabilities where improvements are needed. 

Specifically, we analyzed the state of user privacy protection in AI apps from the following perspectives:

\begin{itemize}[leftmargin=*]
    \item \textbf{RQ-5.1} To what extent is user privacy protected in AI apps?
    \item \textbf{RQ-5.2} What are the main privacy concerns of users regarding current AI apps?
\end{itemize}

\subsubsection{Experimental methodology}
We conducted the following two experiments to answer the sub-questions above. 

\begin{itemize}[leftmargin=*]
    \item \textbf{Experiment RQ-5.1 (Analysis on user privacy protection)}
    In the first step, we construct a privacy keyword list containing common user privacy terms (e.g., name, email, address, and birth). Then, we crawl the privacy policy data of 6016 AI apps from the Google Play market and match it with the pre-built privacy keyword list. If the crawl data includes keywords from the privacy list, we consider that the corresponding AI apps can access private data with respect to these keywords. 
    \item \textbf{Experiment RQ-5.2 (User concerns on current AI app privacy protection)}
    First, we filtered privacy-related comments using privacy-related keywords (such as ``privacy'') from all the reviews of collected AI apps. Next, we manually analyzed and summarized the collected privacy-related user comments, aiming to identify the core concerns users have about AI apps' privacy protection. Based on the manual analysis, we identified several core concerns, which are: 1) Privacy infringement and data misuse; 2) Lack of transparency in privacy policies; 3) Third-party data sharing; 4) Privacy protection features; 5) The conflict between privacy and user experience. Finally, we summarized and analyzed each of these concerns. 

\end{itemize}

\subsubsection{Results} 

\begin{table}[!htbp]
\caption{Private data of AI apps accessing}
\label{tab:private_data}
\centering
\scalebox{0.52}
{
\begin{tabular}{ccc|ccc|ccc}
\toprule
\multicolumn{3}{c|}{\textbf{Business}} & \multicolumn{3}{c|}{\textbf{Finance}} & \multicolumn{3}{c}{\textbf{Education}} \\ 
\midrule
\multicolumn{1}{c|}{privacy} & \multicolumn{1}{c|}{Count} & Ratio & \multicolumn{1}{c|}{privacy} & \multicolumn{1}{c|}{Count} & Ratio & \multicolumn{1}{c|}{privacy} & \multicolumn{1}{c|}{Count} & Ratio \\ 
\multicolumn{1}{c|}{name} & \multicolumn{1}{c|}{492} & 0.60 & \multicolumn{1}{c|}{name} & \multicolumn{1}{c|}{410} & 0.52 & \multicolumn{1}{c|}{name} & \multicolumn{1}{c|}{333} & 0.61 \\ 
\multicolumn{1}{c|}{address} & \multicolumn{1}{c|}{421} & 0.51 & \multicolumn{1}{c|}{account} & \multicolumn{1}{c|}{314} & 0.39 & \multicolumn{1}{c|}{address} & \multicolumn{1}{c|}{292} & 0.53 \\ 
\multicolumn{1}{c|}{email} & \multicolumn{1}{c|}{405} & 0.50 & \multicolumn{1}{c|}{location} & \multicolumn{1}{c|}{311} & 0.39 & \multicolumn{1}{c|}{email} & \multicolumn{1}{c|}{250} & 0.45 \\ 
\multicolumn{1}{c|}{location} & \multicolumn {1}{c|}{363} & 0.44 & \multicolumn{1}{c|}{email} & \multicolumn{1}{c|}{305} & 0.38 & \multicolumn{1}{c|}{location} & \multicolumn{1}{c|}{201} & 0.36 \\ 
\multicolumn{1}{c|}{image} & \multicolumn{1}{c|}{335} & 0.41 & \multicolumn{1}{c|}{address} & \multicolumn{1}{c|}{303} & 0.38 & \multicolumn{1}{c|}{image} & \multicolumn{1}{c|}{195} & 0.35 \\ 
\multicolumn{1}{c|}{account} & \multicolumn{1}{c|}{302} & 0.37 & \multicolumn{1}{c|}{image} & \multicolumn{1}{c|}{287} & 0.36 & \multicolumn{1}{c|}{account} & \multicolumn{1}{c|}{138} & 0.25 \\ 
\multicolumn{1}{c|}{country} & \multicolumn{1}{c|}{211} & 0.26 & \multicolumn{1}{c|}{phone number} & \multicolumn{1}{c|}{166} & 0.21 & \multicolumn{1}{c|}{video} & \multicolumn{1}{c|}{104} & 0.19 \\ 
\multicolumn{1}{c|}{phone number} & \multicolumn{1}{c|}{203} & 0.25 & \multicolumn{1}{c|}{credit card} & \multicolumn{1}{c|}{114} & 0.14 & \multicolumn{1}{c|}{credit card} & \multicolumn{1}{c|}{96} & 0.17 \\ 
\multicolumn{1}{c|}{video} & \multicolumn{1}{c|}{196} & 0.24 & \multicolumn{1}{c|}{country} & \multicolumn{1}{c|}{112} & 0.14 & \multicolumn{1}{c|}{phone number} & \multicolumn{1}{c|}{88} & 0.16 \\ 
\multicolumn{1}{c|}{interaction} & \multicolumn{1}{c|}{151} & 0.18 & \multicolumn{1}{c|}{video} & \multicolumn{1}{c|}{110} & 0.13 & \multicolumn{1}{c|}{device name} & \multicolumn{1}{c|}{87} & 0.15 \\ 
\multicolumn{1}{c|}{credit card} & \multicolumn{1}{c|}{141} & 0.17 & \multicolumn{1}{c|}{interaction} & \multicolumn{1}{c|}{110} & 0.13 & \multicolumn{1}{c|}{country} & \multicolumn{1}{c|}{81} & 0.14 \\ 
\multicolumn{1}{c|}{photo} & \multicolumn{1}{c|}{131} & 0.16 & \multicolumn{1}{c|}{birth} & \multicolumn{1}{c|}{85} & 0.10 & \multicolumn{1}{c|}{interaction} & \multicolumn{1}{c|}{80} & 0.14 \\ 
\multicolumn{1}{c|}{birth} & \multicolumn{1}{c|}{123} & 0.15 & \multicolumn{1}{c|}{audio} & \multicolumn{1}{c|}{71} & 0.09 & \multicolumn{1}{c|}{photo} & \multicolumn{1}{c|}{57} & 0.10 \\ 
\multicolumn{1}{c|}{device name} & \multicolumn{1}{c|}{88} & 0.10 & \multicolumn{1}{c|}{photo} & \multicolumn{1}{c|}{48} & 0.06 & \multicolumn{1}{c|}{gender} & \multicolumn{1}{c|}{51} & 0.09 \\ 
\multicolumn{1}{c|}{audio} & \multicolumn{1}{c|}{84} & 0.10 & \multicolumn{1}{c|}{gender} & \multicolumn{1}{c|}{44} & 0.05 & \multicolumn{1}{c|}{audio} & \multicolumn{1}{c|}{40} & 0.07 \\ 
\multicolumn{1}{c|}{gender} & \multicolumn{1}{c|}{68} & 0.08 & \multicolumn{1}{c|}{device name} & \multicolumn{1}{c|}{28} & 0.03 & \multicolumn{1}{c|}{birth} & \multicolumn{1}{c|}{33} & 0.06 \\ 
\multicolumn{1}{c|}{device ID} & \multicolumn{1}{c|}{28} & 0.03 & \multicolumn{1}{c|}{user ID} & \multicolumn{1}{c|}{15} & 0.01 & \multicolumn{1}{c|}{device ID} & \multicolumn{1}{c|}{15} & 0.02 \\ 
\multicolumn{1}{c|}{user ID} & \multicolumn{1}{c|}{21} & 0.02 & \multicolumn{1}{c|}{device ID} & \multicolumn{1}{c|}{9} & 0.01 & \multicolumn{1}{c|}{user ID} & \multicolumn{1}{c|}{10} & 0.01 \\ 
\multicolumn{1}{c|}{browsing history} & \multicolumn{1}{c|}{15} & 0.01 & \multicolumn{1}{c|}{browsing history} & \multicolumn{1}{c|}{6} & 0.00 & \multicolumn{1}{c|}{browsing history} & \multicolumn{1}{c|}{6} & 0.01 \\ 
\multicolumn{1}{c|}{serach history} & \multicolumn{1}{c|}{6} & 0.00 & \multicolumn{1}{c|}{search history} & \multicolumn{1}{c|}{4} & 0.00 & \multicolumn{1}{c|}{IMEI number} & \multicolumn{1}{c|}{2} & 0.00 \\ 
\multicolumn{1}{c|}{IMEI number} & \multicolumn{1}{c|}{2} & 0.00 & \multicolumn{1}{c|}{IMEI number} & \multicolumn{1}{c|}{2} & 0.00 & \multicolumn{1}{c|}{Android ID} & \multicolumn{1}{c|}{2} & 0.00 \\ 
\multicolumn{1}{c|}{Android ID} & \multicolumn{1}{c|}{1} & 0.00 & \multicolumn{1}{c|}{frequency of use} & \multicolumn{1}{c|}{1} & 0.00 & \multicolumn{1}{c|}{frequency of use} & \multicolumn{1}{c|}{2} & 0.00 \\ 
\multicolumn{1}{c|}{frequency of use} & \multicolumn{1}{c|}{0} & 0.00 & \multicolumn{1}{c|}{Android ID} & \multicolumn{1}{c|}{0} & 0.00 & \multicolumn{1}{c|}{search history} & \multicolumn{1}{c|}{1} & 0.00 \\ 
\midrule
\multicolumn{3}{c|}{\textbf{Productivity}} & \multicolumn{3}{c|}{\textbf{Tools}} & \multicolumn{3}{c}{\textbf{Shopping}} \\ 
\midrule
\multicolumn{1}{c|}{privacy} & \multicolumn{1}{c|}{Count} & Ratio & \multicolumn{1}{c|}{privacy} & \multicolumn{1}{c|}{Count} & Ratio & \multicolumn{1}{c|}{privacy} & \multicolumn{1}{c|}{Count} & Ratio \\ 
\multicolumn{1}{c|}{name} & \multicolumn{1}{c|}{371} & 0.72 & \multicolumn{1}{c|}{name} & \multicolumn{1}{c|}{289} & 0.66 & \multicolumn{1}{c|}{name} & \multicolumn{1}{c|}{173} & 0.63 \\ 
\multicolumn{1}{c|}{address} & \multicolumn{1}{c|}{328} & 0.64 & \multicolumn{1}{c|}{address} & \multicolumn{1}{c|}{241} & 0.55 & \multicolumn{1}{c|}{email} & \multicolumn{1}{c|}{155} & 0.56 \\ 
\multicolumn{1}{c|}{email} & \multicolumn{1}{c|}{324} & 0.63 & \multicolumn{1}{c|}{email} & \multicolumn{1}{c|}{225} & 0.52 & \multicolumn{1}{c|}{address} & \multicolumn{1}{c|}{151} & 0.55 \\ 
\multicolumn{1}{c|}{location} & \multicolumn{1}{c|}{167} & 0.32 & \multicolumn{1}{c|}{location} & \multicolumn{1}{c|}{185} & 0.42 & \multicolumn{1}{c|}{location} & \multicolumn{1}{c|}{142} & 0.51 \\ 
\multicolumn{1}{c|}{image} & \multicolumn{1}{c|}{154} & 0.30 & \multicolumn{1}{c|}{image} & \multicolumn{1}{c|}{185} & 0.42 & \multicolumn{1}{c|}{account} & \multicolumn{1}{c|}{122} & 0.44 \\ 
\multicolumn{1}{c|}{account} & \multicolumn{1}{c|}{147} & 0.28 & \multicolumn{1}{c|}{account} & \multicolumn{1}{c|}{152} & 0.35 & \multicolumn{1}{c|}{image} & \multicolumn{1}{c|}{118} & 0.43 \\ 
\multicolumn{1}{c|}{country} & \multicolumn{1}{c|}{109} & 0.21 & \multicolumn{1}{c|}{phone number} & \multicolumn{1}{c|}{104} & 0.24 & \multicolumn{1}{c|}{country} & \multicolumn{1}{c|}{81} & 0.29 \\ 
\multicolumn{1}{c|}{phone number} & \multicolumn{1}{c|}{91} & 0.17 & \multicolumn{1}{c|}{country} & \multicolumn{1}{c|}{95} & 0.21 & \multicolumn{1}{c|}{phone number} & \multicolumn{1}{c|}{71} & 0.25 \\ 
\multicolumn{1}{c|}{interaction} & \multicolumn{1}{c|}{72} & 0.14 & \multicolumn{1}{c|}{video} & \multicolumn{1}{c|}{90} & 0.20 & \multicolumn{1}{c|}{video} & \multicolumn{1}{c|}{52} & 0.18 \\ 
\multicolumn{1}{c|}{video} & \multicolumn{1}{c|}{69} & 0.13 & \multicolumn{1}{c|}{device name} & \multicolumn{1}{c|}{74} & 0.17 & \multicolumn{1}{c|}{interaction} & \multicolumn{1}{c|}{47} & 0.17 \\ 
\multicolumn{1}{c|}{photo} & \multicolumn{1}{c|}{56} & 0.11 & \multicolumn{1}{c|}{photo} & \multicolumn{1}{c|}{71} & 0.16 & \multicolumn{1}{c|}{credit card} & \multicolumn{1}{c|}{47} & 0.17 \\ 
\multicolumn{1}{c|}{device name} & \multicolumn{1}{c|}{52} & 0.10 & \multicolumn{1}{c|}{interaction} & \multicolumn{1}{c|}{60} & 0.13 & \multicolumn{1}{c|}{gender} & \multicolumn{1}{c|}{32} & 0.11 \\ 
\multicolumn{1}{c|}{credit card} & \multicolumn{1}{c|}{47} & 0.09 & \multicolumn{1}{c|}{audio} & \multicolumn{1}{c|}{51} & 0.11 & \multicolumn{1}{c|}{photo} & \multicolumn{1}{c|}{32} & 0.11 \\ 
\multicolumn{1}{c|}{audio} & \multicolumn{1}{c|}{44} & 0.08 & \multicolumn{1}{c|}{credit card} & \multicolumn{1}{c|}{46} & 0.10 & \multicolumn{1}{c|}{birth} & \multicolumn{1}{c|}{29} & 0.10 \\ 
\multicolumn{1}{c|}{gender} & \multicolumn{1}{c|}{25} & 0.04 & \multicolumn{1}{c|}{gender} & \multicolumn{1}{c|}{28} & 0.06 & \multicolumn{1}{c|}{device name} & \multicolumn{1}{c|}{29} & 0.10 \\ 
\multicolumn{1}{c|}{birth} & \multicolumn{1}{c|}{22} & 0.04 & \multicolumn{1}{c|}{device ID} & \multicolumn{1}{c|}{26} & 0.06 & \multicolumn{1}{c|}{audio} & \multicolumn{1}{c|}{27} & 0.09 \\ 
\multicolumn{1}{c|}{device ID} & \multicolumn{1}{c|}{16} & 0.03 & \multicolumn{1}{c|}{birth} & \multicolumn{1}{c|}{19} & 0.04 & \multicolumn{1}{c|}{device ID} & \multicolumn{1}{c|}{20} & 0.07 \\ 
\multicolumn{1}{c|}{browsing history} & \multicolumn{1}{c|}{7} & 0.01 & \multicolumn{1}{c|}{user ID} & \multicolumn{1}{c|}{14} & 0.03 & \multicolumn{1}{c|}{user ID} & \multicolumn{1}{c|}{7} & 0.02 \\ 
\multicolumn{1}{c|}{user ID} & \multicolumn{1}{c|}{7} & 0.01 & \multicolumn{1}{c|}{Android ID} & \multicolumn{1}{c|}{8} & 0.01 & \multicolumn{1}{c|}{search history} & \multicolumn{1}{c|}{3} & 0.01 \\ 
\multicolumn{1}{c|}{search history} & \multicolumn{1}{c|}{4} & 0.00 & \multicolumn{1}{c|}{IMEI number} & \multicolumn{1}{c|}{6} & 0.01 & \multicolumn{1}{c|}{browsing history} & \multicolumn{1}{c|}{3} & 0.01 \\ 
\multicolumn{1}{c|}{Android ID} & \multicolumn{1}{c|}{2} & 0.00 & \multicolumn{1}{c|}{frequency of use} & \multicolumn{1}{c|}{6} & 0.01 & \multicolumn{1}{c|}{IMEI number} & \multicolumn{1}{c|}{1} & 0.00 \\ 
\multicolumn{1}{c|}{IMEI number} & \multicolumn{1}{c|}{1} & 0.00 & \multicolumn{1}{c|}{browsing history} & \multicolumn{1}{c|}{4} & 0.00 & \multicolumn{1}{c|}{frequency of use} & \multicolumn{1}{c|}{1} & 0.00 \\ 
\multicolumn{1}{c|}{frequency of use} & \multicolumn{1}{c|}{0} & 0.00 & \multicolumn{1}{c|}{search history} & \multicolumn{1}{c|}{2} & 0.00 & \multicolumn{1}{c|}{Android ID} & \multicolumn{1}{c|}{0} & 0.00 \\ 
\bottomrule
\end{tabular}
}
\end{table}

\begin{table*}[!htbp]
\caption{Privacy concerns highlighted in user comments}
\label{tab:privacy_concerns}
\centering
\scalebox{0.72}
{
\begin{tabular}{lll}
\toprule
\multicolumn{1}{c}{\textbf{Main Privacy Concerns}} & \multicolumn{1}{c}{\textbf{Details}} & \multicolumn{1}{c}{\textbf{User Comment Examples}} \\
\midrule
Privacy Invasion \& Data Misuse & \begin{tabular}[c]{@{}l@{}}Users are concerned about apps requesting \\ excessive permissions unrelated to their core functions, \\ and unauthorized access to device resources.\end{tabular} & \begin{tabular}[c]{@{}l@{}}“To use video chat with a doctor, \\ I had to agree to invasive permissions like accessing my \\camera, microphone, calendar, and more.”
\\“This app accessed my microphone in the background \\even after I revoked permission. Who knows how long it's been\\ recording me? Major invasion of privacy!”\end{tabular} \\
\midrule
Lack of Transparency in Privacy Policies & \begin{tabular}[c]{@{}l@{}}Privacy policies are hard to access or understand, \\ with changes made without user notification.\end{tabular} & \begin{tabular}[c]{@{}l@{}}“If I can't read their privacy agreements, I can't use the app.”\\ “The privacy policy is not acceptable. \\Because it might be changed without informing the users, \\as it mentions in the roles of the application! ”\end{tabular} \\
\midrule
Third-Party Data Sharing & \begin{tabular}[c]{@{}l@{}}Apps shares data with third parties, \\ especially without explicit consent.\end{tabular} & \begin{tabular}[c]{@{}l@{}}“App worked well. Until today when I was asked if would agree\\ to a third party privacy advertising policy. I chose option\\ 'No thank you' but instead of being allowed to proceed,\\ it merely removed the option I had chose \\and would allow me to continue in app until I chose to agree.” \\“This app collects WAY too much data than should be needed to\\ use the service, this app also collects it to use toward advertising.”\end{tabular} \\
\midrule
Privacy Protection Features & \begin{tabular}[c]{@{}l@{}}Some users appreciate apps that offer privacy\\ protection, like encrypted communication.\end{tabular} & \begin{tabular}[c]{@{}l@{}}“This app actually respects your privacy.”\\ “Connect to people without disclosing privacy.”\end{tabular} \\
\midrule
Privacy vs. User Experience & \begin{tabular}[c]{@{}l@{}}Users report a trade-off between privacy protection \\ and user experience, with some apps limiting functionality \\ unless privacy terms are accepted.\end{tabular} & \begin{tabular}[c]{@{}l@{}}“Now I've disabled the annoying adds I now have to agree to \\a privacy policy every time I open the app??? Why can't \\it remember I have clicked accept?”\\ “Trying to sign up, but can't get past ticking the\\ terms and conditions and privacy policy box. No option \\to complete the action after ticking box.”\end{tabular} \\
\bottomrule
\end{tabular}
}
\end{table*}

Table~\ref{tab:private_data} presents the experimental results for \textbf{RQ5}. It shows the condition of user private data being accessed in six different categories of AI apps. The column "Count" indicates how many AI apps accessed this private content, and "Ratio" represents the proportion of AI apps that accessed this content. From Table~\ref{tab:private_data}, we can see that the most frequently accessed private attribute across all six categories is the user's name, indicating that this private attribute is the most commonly leaked to app providers. Other frequently accessed private attributes include address, email, and location. Across all six AI app categories, these private attributes are ranked in the top five in terms of the number and ratio of AI apps that accessed them. Moreover, we found that, for different categories of AI apps, the private attributes that are most easily leaked are similar (e.g., name, address, and email). 

\vspace{0.1cm}
\noindent\colorbox{gray!20}{{\parbox{1.0\linewidth}{
\textbf{Finding 19:} The most frequently accessed private attributes by the collected AI apps are name, address, email, and location. 

}}}
\vspace{0.1cm}

Table~\ref{tab:privacy_concerns} presents the privacy issues highlighted in user comments. The table categorizes these privacy issues into five main categories, providing detailed descriptions along with actual user comments as examples. Firstly, we see that privacy invasion and data misuse are among the top concerns. Users are worried about certain applications requesting excessive permissions that are unrelated to their core functions or even accessing device resources without authorization. This behavior makes users feel that their privacy is being violated, particularly when sensitive permissions such as camera and microphone access are accessed.

Secondly, users are worried about the lack of transparency in privacy policies. Many users find that privacy policies are difficult to access or understand, and they are concerned that these policies may be changed without notification. Another major concern for users is the issue of third-party data sharing. Users complain that some applications share their data with third parties without their explicit consent. Lastly, the trade-off between privacy and user experience is another significant issue. Users report that some applications. For example, users may be forced to accept privacy terms in order to continue using the app, and this compulsory approach has resulted in user dissatisfaction.

\vspace{0.1cm}
\noindent\colorbox{gray!20}{{\parbox{1.0\linewidth}{
\textbf{Finding 20:} The main privacy-related concerns raised by users include: privacy invasion and data misuse, lack of transparency in privacy policies, third-party data sharing, privacy protection features, and the balance between privacy and user experience.
}}}
\vspace{0.1cm}

\subsection{RQ6: Analysis of user reviews related to AI technology}
\label{subsec:RQ6}

\subsubsection{Objectives} 

User reviews are crucial for improving the AI techniques used in AI apps and enhancing user satisfaction. They can help identify potential issues with current AI techniques utilized in AI apps, contributing to the overall quality and reliability of the apps. In this research question, we investigate users' attitudes toward AI techniques in AI apps. Understanding users' attitudes can help developers identify current issues that users perceive in AI apps and create AI apps that better meet user expectations. 

\subsubsection{Experimental methodology}
In the first step, we conducted web crawling from the Google Play application market to retrieve available reviews associated with our collected AI apps. Since in our collected dataset of AI apps, 88.7\% of the AI apps belong to Google Play, we chose Google Play as the primary source for collecting reviews. In the second step, we constructed an AI-relevant technical keyword dictionary to filter out reviews that specifically focused on AI techniques. Based on the filtering, we obtained a collection of technical reviews (i.e., reviews regarding AI techniques). 

For the collected review data, we performed three types of analyses: \textbf{1) Overall sentiment analysis}, where we employed a Transformers-based model (\cite{wolf-etal-2020-transformers}) renowned for its effectiveness in sentiment classification tasks, achieving an impressive accuracy rate of 91.3\%. Each technical review was classified as either positive or negative by the Transformers-based model. Moreover, we conducted a deeper analysis of the reasons for the negative reviews. We manually checked all the negative reviews and categorized the reasons for the negativity. \textbf{2) Strengths and weaknesses analysis}, involving manual examination and summarization of the identified technical reviews' notable attributes and limitations; \textbf{3) Categorized sentiment analysis}, wherein we computed the ratio of positive and negative reviews for each app category, facilitating further exploration.

\subsubsection{Results} 

\begin{figure}[!htbp]
	\centering
	\includegraphics[width=0.35\textwidth]{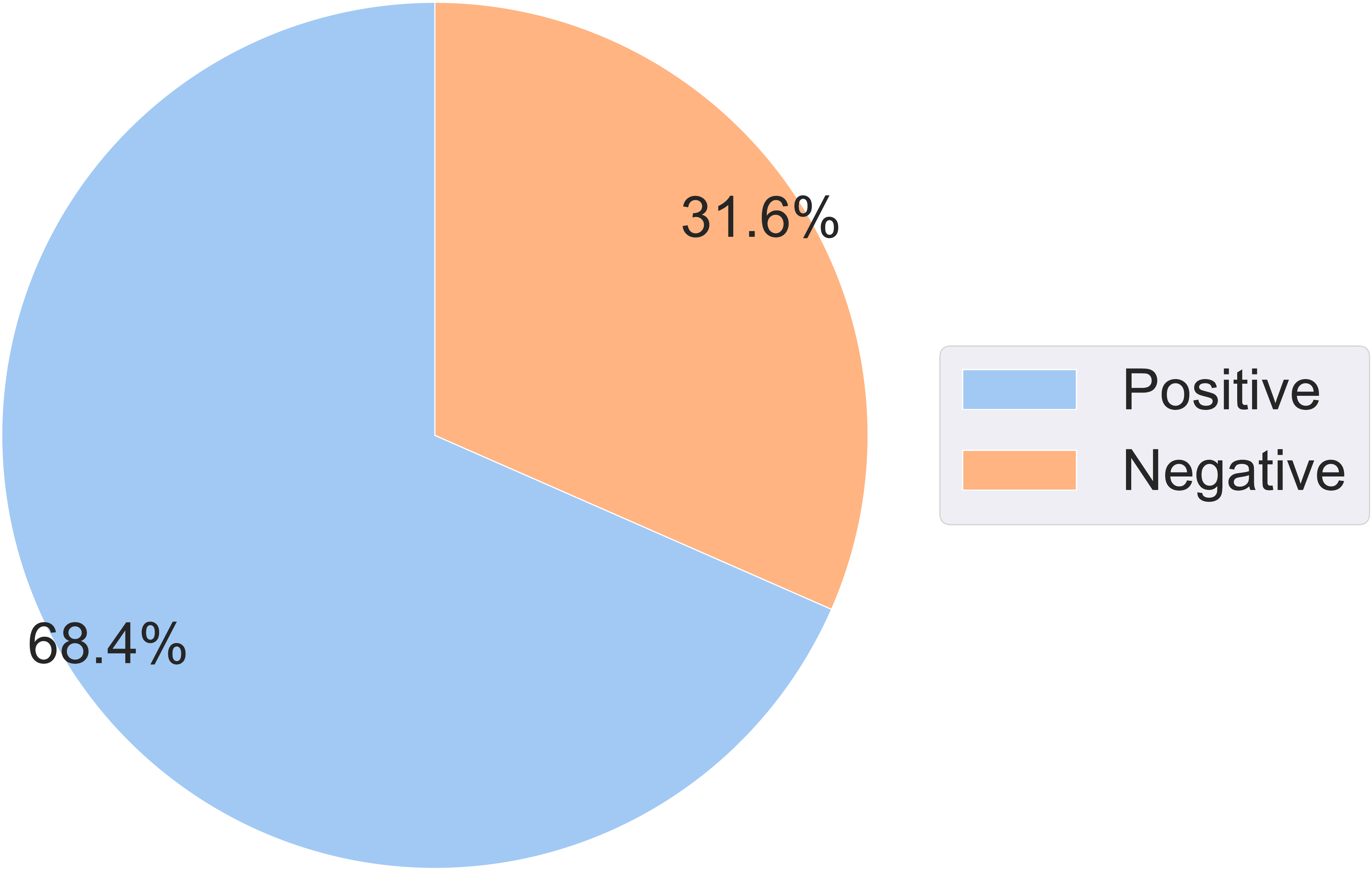}
	\caption{Sentiment analysis of technical reviews for AI apps: Proportion of positive and negative reviews}
	\label{fig:ratio_reviews}
\end{figure}

\begin{figure}[!htbp]
	\centering
	\includegraphics[width=0.47\textwidth]{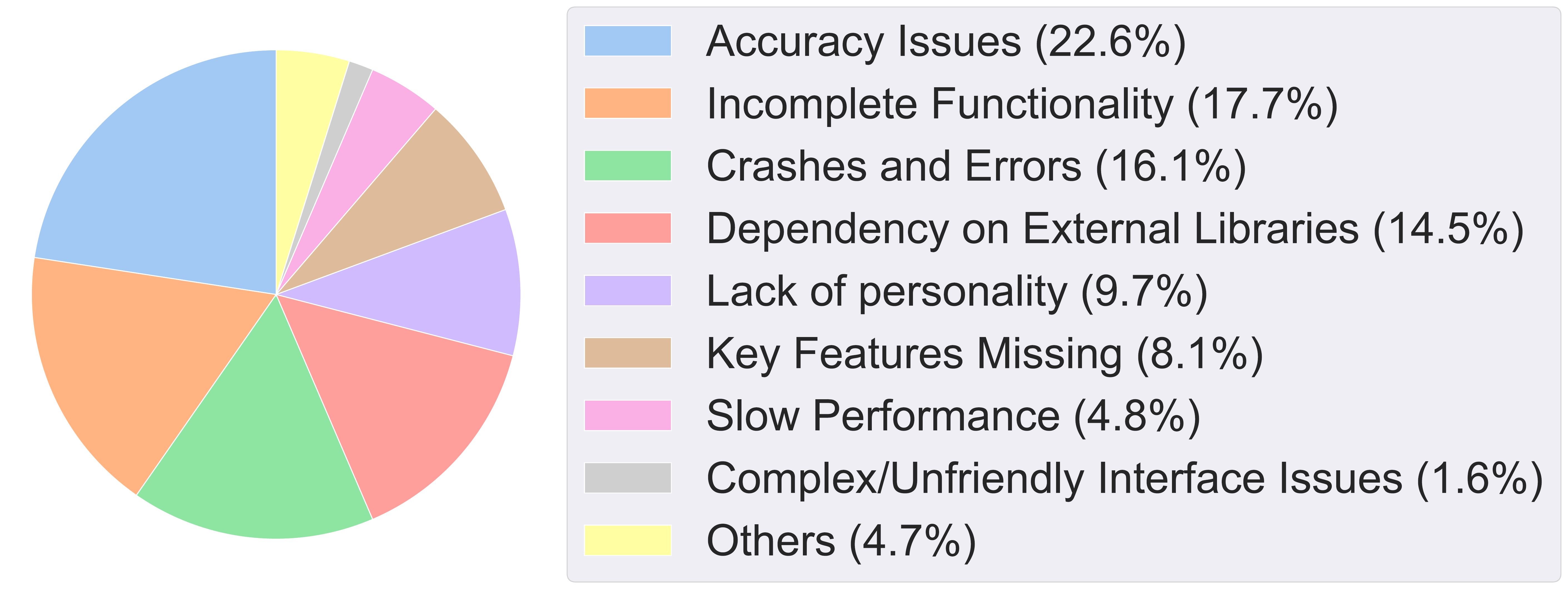}
	\caption{Analysis of causes for negative reviews of AI applications}
	\label{fig:issues_distribution}
\end{figure}

\begin{figure}[!htbp]
	\centering
	\includegraphics[width=0.48\textwidth]{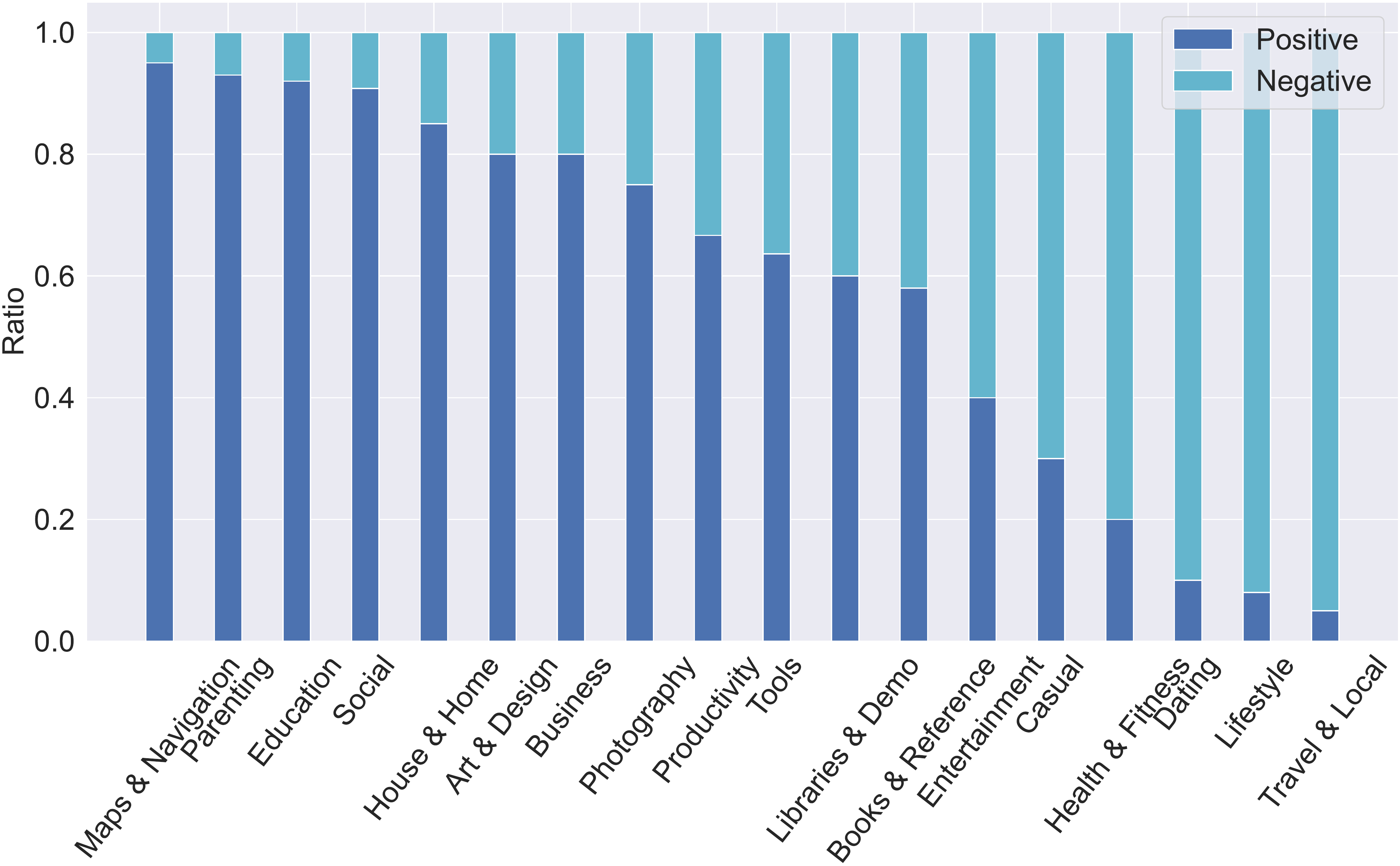}
	\caption{Sentiment analysis results on AI technical reviews across different categories}
	\label{fig:positive_nagetive}
\end{figure}

The experimental results for \textbf{RQ6} are presented in Figure~\ref{fig:ratio_reviews}, Figure~\ref{fig:issues_distribution}, Table~\ref{tab:negative_categories_analysis}, Table~\ref{tab:ai-reviews}, and Figure~\ref{fig:positive_nagetive}. Figure~\ref{fig:ratio_reviews} illustrates the proportion of positive and negative reviews among all the collected technical reviews using a pie chart. We can see that positive reviews account for 68.4\% of the total, while negative reviews constitute 31.6\%. Based on the experimental results, we find that for the techniques used in AI apps, the proportion of positive reviews is higher compared to negative reviews, indicating that users' attitudes towards the techniques in AI apps are relatively positive.

Figure~\ref{fig:issues_distribution} shows the main reasons causing negative reviews (as mentioned above, 31.6\% of reviews are negative). Below, we explain each of these main reasons.

\begin{itemize} [leftmargin=*] 
\item \textbf{Accuracy Issues} refer to the AI applications making inaccurate classifications, such as mistakenly identifying a hamburger as a hot dog.

\item \textbf{Incomplete Functionality} refers to the AI applications lacking some expected functionality. For instance, in the "Photography" category, a user mentioned that the face grouping feature did not work properly.

\item \textbf{Crashes and Errors} refer to instances where AI applications crash or generate error messages. For example, in the "Libraries \& Demo" category, a user mentioned, "When I click the Recognition Test Button, the app stops working."

\item \textbf{Dependency on External Libraries} refers to the application needing certain external libraries to function properly. For instance, in the "Video Players \& Editors" category, a user commented that the app required OpenCV Manager to run, a library that was not available in the app store.

\item \textbf{Lack of Personality} refers to the app's lack of personalized features, failing to meet users' needs for a personalized experience. For instance, in the "Education" category, a user commented that while the app was powerful, it lacked personalized elements during use.

\item \textbf{Key Features Missing} refers to the AI application lacking critical features compared to other similar apps. For example, in the "Finance" category, a user mentioned that the app lacked fingerprint authorization and NFC payment, which led them to consider other apps.

\item \textbf{Slow Performance} refers to the AI application running slowly, affecting the user experience. For example, in the "Libraries \& Demo" category, a user mentioned that the object detection speed of the app was not fast enough to meet real-time requirements.

\item \textbf{Complex/Unfriendly Interface Issues} refer to instances where the AI app has a complicated or user-unfriendly interface, making it difficult for users to operate. For example, in the "Productivity" category, users reported that the app's interface did not refresh after deleting a file, which made it challenging to use. 
\end{itemize}

From Figure~\ref{fig:issues_distribution}, we see that Accuracy Issues are the leading cause of negative user reviews, accounting for 22.6\% of all negative feedback. Following this, Incomplete Functionality accounts for 17.7\%, which is the second most significant cause. Crashes and Errors are the third major problem, comprising 16.1\% of the negative reviews. From the above analysis, we conclude that accuracy issues, incomplete functionality, and crashes and errors are the three main sources of negative reviews.

\vspace{0.1cm}
\noindent\colorbox{gray!20}{{\parbox{1.0\linewidth}{
\textbf{Finding 21:} AI app users' attitudes towards the techniques used in AI apps are relatively positive. Accuracy issues, incomplete functionality, and crashes and errors are the three main sources of negative reviews.} 
}}
\vspace{0.1cm}

Table~\ref{tab:ai-reviews} presents summaries for some example technical reviews across different AI app categories. We can see that, 73.7\% of reviews are completely positive, without pointing out any shortcomings. The remaining reviews are partly negative, highlighting advantages but pointing out issues. For example, for Travel \& Local-oriented AI apps, users emphasize the limitations in translating US English to Hebrew. Among the negative reviews, 75\% focus on accuracy issues (e.g., the prediction accuracy is insufficient), including in the Casual, Dating, Entertainment, and Photography categories of apps.

\vspace{0.1cm}
\noindent\colorbox{gray!20}{{\parbox{1.0\linewidth}{
\textbf{Finding 22:} Among the negative reviews of AI technology in AI apps, the most frequently reported issue by users is accuracy issues (e.g., the prediction accuracy is insufficient). 
}}}
\vspace{0.1cm}

Figure~\ref{fig:positive_nagetive} depicts the proportion of positive and negative technical reviews across different categories. Notably, in more than half of the categories, the proportion of positive exceeds that of negative reviews. For instance, in the categories of Maps \& Navigation, Parenting, Education, and Social, the ratio of positive reviews surpasses 90\%. However, certain categories exhibit a higher proportion of negative reviews, such as Book \& Reference, Entertainment, and Casual.

Table~\ref{tab:negative_categories_analysis} presents the causes of negative reviews across categories with higher negative sentiments. We see that, in the \textbf{Entertainment} category, users' negative reviews mainly focus on recognition accuracy issues, such as errors in object recognition. For example, identifying a door as a refrigerator or a desk as a microwave. In the \textbf{Casual} category, users' negative reviews mainly focus on inaccuracies in image classification, such as identifying a hamburger as a hotdog. Additionally, users suggested some functionality issues, such as adding confidence measures for algorithms. In the \textbf{Health \& Fitness} category, users were concerned about technical details, particularly regarding the machine learning modules used.

The negative reviews of the \textbf{Dating} category were mainly caused by algorithm accuracy and the lack of personalization. Negative feedback in the \textbf{Lifestyle} category focused on the dependency on external components, such as the requirement to install OpenCV Manager. In the \textbf{Travel \& Local} category, users mainly reported issues with voice recognition and translation, especially between American English and Hebrew. Lastly, negative reviews for \textbf{Books \& Reference} apps primarily focused on technical issues, such as crashes or failures to start the app. 

\vspace{0.1cm}
\noindent\colorbox{gray!20}{{\parbox{1.0\linewidth}{
\textbf{Finding 23:} In more than half of the AI app categories, the proportion of positive exceeds that of negative reviews. 
}}}
\vspace{0.1cm}

\begin{table*}
\caption{Causes of negative reviews across categories with higher negative sentiments}
\label{tab:negative_categories_analysis}
\centering
\scalebox{0.78}
{
\begin{tabular}{llll}
\toprule
\multicolumn{1}{c}{\textbf{Category}} & \multicolumn{1}{c}{\textbf{Issues}} & \multicolumn{1}{c}{\textbf{Detailed Description}} & \multicolumn{1}{c}{\textbf{Examples from Reviews}} \\
\midrule
Entertainment & Accuracy issues & \begin{tabular}[c]{@{}l@{}}Users mentioned poor object detection and recognition\\ accuracy, such as recognizing a door as a refrigerator or a desk\\ as a microwave, which obviously affects user experience.\end{tabular} & \begin{tabular}[c]{@{}l@{}}"Recognizing a door as a refrigerator, \\ a desk as a microwave"\end{tabular} \\
\midrule
Casual & \begin{tabular}[c]{@{}l@{}}Misclassification, \\ Functionality issues\end{tabular} & \begin{tabular}[c]{@{}l@{}}Users mentioned inaccurate image classification, \\ such as misclassifying a hamburger as a hotdog. \\ Users suggested adding confidence measures for algorithms\\ and using front-facing cameras,  which would improve\\ the app's practicality.\end{tabular} & \begin{tabular}[c]{@{}l@{}}"Misclassifying a hamburger as a hotdog, \\ lack of front-facing camera support"\end{tabular} \\
\midrule
Health \& Fitness & Technical details & \begin{tabular}[c]{@{}l@{}}Users inquired about the specific machine learning modules used \\ but did not receive satisfactory answers, \\ indicating a lack of technical transparency.\end{tabular} & \begin{tabular}[c]{@{}l@{}}"Questions about TensorFlow usage \\ not satisfactorily answered"\end{tabular} \\
\midrule
Dating & \begin{tabular}[c]{@{}l@{}}Algorithm accuracy, \\ Lack of personalization\end{tabular} & \begin{tabular}[c]{@{}l@{}}Users doubted the accuracy of facial recognition and \\ suggested adding user verification for algorithm accuracy. \\ Users expected more personalized and interactive features.\end{tabular} & \begin{tabular}[c]{@{}l@{}}"Facial recognition accuracy doubted, \\ lack of user verification"\end{tabular} \\
\midrule
Lifestyle & \begin{tabular}[c]{@{}l@{}}Dependency on\\ external components\end{tabular} & \begin{tabular}[c]{@{}l@{}}Users mentioned the need to install OpenCV Manager, \\ which could be difficult to find or install in some cases, \\ leading to the app not functioning properly\end{tabular} & \begin{tabular}[c]{@{}l@{}}"Difficulty finding or \\ installing OpenCV Manager"\end{tabular} \\
\midrule
Travel \& Local & \begin{tabular}[c]{@{}l@{}}Voice recognition and \\ translation issues\end{tabular} & \begin{tabular}[c]{@{}l@{}}Users reported that the app's voice model had difficulty with\\  American English and produced inaccurate translations, \\ especially from English to Hebrew. They noted that\\ other similar voice assistants (like Google and Alexa)\\ did not have these issues.\end{tabular} & \begin{tabular}[c]{@{}l@{}}"Poor recognition of American English, \\ inaccurate Hebrew translations"\end{tabular} \\
\midrule
Books \& Reference & \begin{tabular}[c]{@{}l@{}}Technical issues\end{tabular} & \begin{tabular}[c]{@{}l@{}}Some users reported crashes or failure to start, \\ such as inability to download necessary opencv packages. \end{tabular} & \begin{tabular}[c]{@{}l@{}}"Inability to download opencv packages"\end{tabular} \\
\bottomrule
\end{tabular}
}
\end{table*}

\begin{table}[ht]
\caption{Summaries of technically-relevant reviews of AI apps}
\label{tab:ai-reviews}
\centering
\scalebox{0.7}
{
\begin{tabular}{l|l}
\toprule
\textbf{Category} & \multicolumn{1}{c}{\textbf{AI Review summary}} \\ \hline
Art \& Design & \begin{tabular}[c]{@{}l@{}}Good application to create paintings from photos and\\ powerful application created using machine learning.\end{tabular} \\ \hline
Books \& Reference & Very good scikit-learn documentation. \\ \hline
Business & The scanning is good. \\ \hline
Casual & Useful but not always accurate. \\ \hline
Dating & \begin{tabular}[c]{@{}l@{}}It would be better for machine learning if the \\ users could verify its accuracy.\end{tabular} \\ \hline
Education & \begin{tabular}[c]{@{}l@{}}It covers all machine learning techniques and helps us \\ understand the basics of machine learning, which is excellent.\end{tabular} \\ \hline
Entertainment & \begin{tabular}[c]{@{}l@{}}It is an interesting app to learn about machine learning, \\ but the accuracy is not good.\end{tabular} \\ \hline
Health \& Fitness & \begin{tabular}[c]{@{}l@{}}Which machine learning Module have you used \\ for your predictive analysis.\end{tabular} \\ \hline
House \& Home & It contains object detection and is very excellent. \\ \hline
Libraries \& Demo & Great app, very helpful for exploring deep learning. \\ \hline
Lifestyle & Wants to install OpenCV Manager. \\ \hline
Maps \& Navigation & \begin{tabular}[c]{@{}l@{}}Use AI and deep learning to detect farm plots. \\ A very useful and light app.\end{tabular} \\ \hline
Parenting & \begin{tabular}[c]{@{}l@{}}Use machine learning to detect explicit \\ images on Reddit or Google.\end{tabular} \\ \hline
Photography & Object detection is not very accurate, but it was fun. \\ \hline
Productivity & Very good text recognition app. \\ \hline
Social & \begin{tabular}[c]{@{}l@{}}It can be used in various advanced scenes and balance modes, \\ such as using voice to invoke the camera.\end{tabular} \\ \hline
Tools & \begin{tabular}[c]{@{}l@{}}Text Recognition is an exciting and fantastic application \\ with outstanding new features.\end{tabular} \\ \hline
Travel \& Local & \begin{tabular}[c]{@{}l@{}}Spanish translation is perfect but abysmal results \\ for translating US English to Hebrew.\end{tabular} \\
\bottomrule
\end{tabular}
}
\end{table}

\section{Discussion}
\label{sec:discussion}

\subsection{Challenges and Opportunities}
In this section, we discuss the challenges and opportunities of deploying AI technologies to mobile applications.   \par 

\subsubsection{Challenges}
In the following, we elaborate on the challenges of deploying AI technologies to mobile applications from two perspectives: on-device deployment and on-cloud deployment.

One of the primary challenges in on-device AI deployment is the limitation of computational resources. Mobile devices have limited computational power, battery life, and storage capacity. Running AI models on such constrained devices can lead to performance issues, such as slow processing times, thus impacting the user experience. Another crucial challenge lies in model protection. As the models are deployed locally, the risk of model theft increases, which can result in intellectual property loss. In RQ4 of our empirical study (cf. Section~\ref{subsec:RQ4}), we show that several embedded models in AI apps are not encrypted, indicating the risk of model theft.

One of the primary challenges in on-cloud AI deployment is data privacy and security. AI systems typically require access to large amounts of user data, which can include sensitive information. Since AI models are deployed in the cloud, user data is stored on remote servers managed by third-party providers, leading to potential privacy and security issues. Another critical challenge lies in network dependency. Cloud-based processing requires a stable and high-speed internet connection. Poor connectivity can lead to a degraded user experience.

Some common challenges for both on-device and on-cloud deployment include model accuracy issues. In RQ6 of our empirical study, we find that 80\% of negative reviews highlight model accuracy issues, considering that the predictions of the models are not accurate enough. Another critical challenge is user privacy concerns. Our experimental results in RQ5 demonstrate that most AI apps can access several crucial attributes of sensitive user data (e.g., name, address, email, images, etc.), which increases the risk of user-sensitive information leaks. 

\subsubsection{Opportunities}
Deploying AI technologies in mobile applications offers numerous opportunities. First, AI techniques enhance the user experience through personalization. For instance, AI-driven personal assistants can learn from individual interactions to better predict user needs and offer more relevant assistance. Moreover, in some crucial fields like health monitoring and financial services, AI technologies can enable mobile applications to process and interpret complex data in real time. For example, mobile health apps can utilize AI technologies to provide real-time analysis of health data, preventing users from potential health issues before they become critical. 

\subsection{Future Directions}
Though AI techniques have demonstrated significant potential in enhancing mobile applications, several challenges remain due to the unique constraints and requirements of mobile environments. In this section, we suggest five future directions for AI in mobile applications.

\begin{itemize} [leftmargin=*]
\item \textbf{Exploring LLM-Driven AI Applications} With the growing capabilities of large language models (LLMs), there is a significant opportunity to integrate LLM models into mobile applications to enhance user interaction and provide more advanced features. By integrating LLMs, mobile apps can offer more personalized experiences, enabling sophisticated features like real-time language translation and intelligent content generation. Exploring LLM-driven AI applications is a valuable future direction.
\item \textbf{User Privacy Protection} As mobile applications increasingly rely on AI to process personal data, ensuring privacy and security becomes crucial (\cite{zhang2019deep}). This raises future research regarding ensuring data privacy and security without compromising the performance and usability of AI models. 
\item \textbf{AI Model Security} AI models are vulnerable to adversarial attacks, where small, deliberate perturbations to the input data can result in incorrect outputs. This vulnerability can severely affect the reliability of AI applications, particularly in critical fields like healthcare, finance, and autonomous driving. How to protect AI models from such attacks could be a future research direction. 
\item \textbf{Optimization of Lightweight Models} AI on mobile platforms requires balancing performance with limited computational and battery resources. While techniques like model compression and pruning strive to achieve this, maintaining optimal performance with minimal reduction in accuracy remains a challenge. This raises the question of how to design AI models that deliver high accuracy while being computationally lightweight on mobile devices.
\item \textbf{Real-time Processing} The demand for real-time AI processing on mobile devices is growing, particularly in applications like augmented reality, real-time translation, and autonomous navigation (\cite{battineni2024exploring}, \cite{omar2024systematic}). Achieving real-time performance with limited hardware resources is challenging. How to enable real-time AI capabilities while maintaining high accuracy and low power consumption is a crucial question to address.
\end{itemize}

\subsection{Recommendations for AI App Developers, Users and R\&D}
Based on the findings from the experimental results (cf. Sections~\ref{subsec:RQ1} to Section~\ref{subsec:RQ6}), we discuss the concrete recommendations for AI application developers, AI users, and AI R\&D.

For AI app developers: \textbf{1) Encrypt AI Models} Finding 18 reported that among all the collected AI models, only 520 models, accounting for just 0.25\%, are considered to be encrypted. Therefore, we recommend developers prioritize encrypting models, especially those handling sensitive tasks, to protect against potential breaches. \textbf{2) Enhance User Experience with Accurate AI Models} Finding 21 reported the accuracy problem of AI models in AI applications. Specifically, accuracy issues are the most commonly reported problem in negative user reviews. We recommend that developers focus on improving the predictive accuracy of AI applications to enhance the user experience. \textbf{3) Optimize AI App Size and Performance} Finding 12 indicated that AI apps supported by deep learning tend to be larger in size. This insight suggests developers to consider focusing on optimizing app size to reduce download time and improve user experience.

For AI app users: \textbf{1) Provide Feedback to Improve AI Apps} User comments can provide valuable insights for improving AI apps. For example, Finding 21 mentions that accuracy issues, incomplete functionality, and crashes and errors are the three main sources of negative reviews, which can help developers identify the shortcomings of current AI apps and make targeted improvements. Therefore, we recommend that users provide more suggestions to help developers better improve AI apps. \textbf{2) Manage Privacy Settings Proactively} Finding 19 indicates that AI applications frequently access sensitive user data, such as names, addresses, emails, and locations. Therefore, we recommend that users regularly review and manage the privacy settings of AI applications. This can effectively reduce the risk of potential data breaches and protect users' privacy.
   
For AI R\&D: \textbf{1) Analyze High-Scoring AI Apps} According to Finding 7, AI applications in the Racing Games and Comics categories have the highest average scores. We recommend analyzing the factors contributing to the high ratings of these applications and applying these successful strategies to the development of AI applications in other categories to enhance overall app quality and user ratings. \textbf{2) Continued Focus on User Feedback} Findings 20, 21, and 22 highlight user feedback on current AI apps. We recommend that AI R\&D teams continue to focus on user feedback, particularly on negative feedback, and use it as a crucial reference for optimizing applications. This approach will further enhance user experience and satisfaction.

\subsection{Threats to Validity} 

{\sc \em Threats to External Validity.}
The external threat of the study lies in the model protection measurement method we applied in RQ4. We adopted the standard entropy test approach to determine whether a model is encrypted, which assesses the encryption status of an AI model based on the model's entropy value. If the value exceeds the threshold, the model is considered encrypted. However, the accuracy of the method is not guaranteed. To mitigate this threat, we adopt the threshold value of 7.99 since it has been validated by the work of Sun \emph{et al.} Moreover, we validate the accuracy of the approach standard entropy test by manually checking. Specifically, we randomly selected 50 AI application models considered encrypted by the standard entropy test and manually verified whether they were actually protected through two steps. First, we tried to open them using the Netron viewer, but none of the models could be opened. Subsequently, we attempted to use the official API of these AI models to load them, but we also found that none of the models could be opened.  

{\sc \em Threats to Internal Validity.} 
A major threat to internal validity arises from the manual collection of keywords for identifying AI applications. We collected a set of AI-related keywords to distinguish AI apps from the mobile apps in AndroZoo. To mitigate this threat, we invested sufficient time in gathering the keywords by consulting extensive literature and related GitHub materials. Moreover, as some simple AI terms (e.g., LSTM, CNN) can also appear in non-AI words/phrases, we excluded them from the keyword dictionary to improve the accuracy of AI app identification. Additionally, recognizing that some AI apps do not rely on AI frameworks and thus may not be detected using framework keywords, we supplemented the keyword dictionary with package names of ML/DL algorithms.  
\section{Related Work}
\label{sec:relatedWork}

\subsection{Mobile Deep Learning}
Deploying deep learning (DL) techniques to mobile devices has shown remarkable benefits, including quick response time, network independence, and enhanced privacy protection, thus attracting much attention in recent studies (\cite{cheng2017survey, huang2022smart, he2018amc, xu2019first, Zhang_2018_CVPR, howard2017mobilenets}). \cite{cheng2017survey} conducted a comprehensive review of state-of-the-art techniques for compressing DNN models, including parameter pruning and quantization, low-rank factorization, compact convolutional filters, and knowledge distillation. \cite{he2018amc} proposed an effective model compression tool, AutoML, which utilized reinforcement learning to sample the design space and improve model compression quality. \cite{xu2019first} conducted the first large-scale study to explore the development progress of on-device deep learning and contributed valuable new findings. For example, early adopters of deep mobile learning are the top applications where embedded deep learning technology plays an important role. \cite{Zhang_2018_CVPR} introduced a computation-efficient CNN architecture, ShuffleNet, especially for mobile devices with minimal computing power. \cite{howard2017mobilenets} demonstrated MobileNets, a new model architecture for mobile and embedded vision applications that achieved significant performance compared to other popular models on ImageNet classification. 

\subsection{ML/DL as cloud services}
Unlike deploying ML/DL models directly to mobile devices, traditional computing paradigms prefer an online mode, where models are deployed on cloud platforms to perform training and inference. Under this mechanism, the mobile device sends data to the remote end and receives the prediction results. MLaaS (Machine Learning as a Service) (\cite{ribeiro2015mlaas}) is a prevalent cloud service that offers a suite of pre-built machine learning tools and capabilities, allowing users to perform data analysis and prediction without needing to deeply understand the principles of ML algorithms. \cite{yao2017complexity} reviewed the effectiveness of MLaaS systems ranging from fully automated, turnkey systems to fully customizable systems, observing that user control can affect ML task performance. \cite{shokri2017membership} empirically evaluated classification models trained by commercial MLaaS providers (e.g., Google and Amazon) from the perspective of model security, designing the first inference attack against ML models provided by Google Prediction API and Amazon.
\cite{tramer2016stealing} investigated the vulnerability of machine learning models offered by MLaaS providers to model extraction attacks. They demonstrated that adversaries with black-box access to these models can replicate them by making numerous queries to the prediction APIs, even when confidence scores are omitted. They underscore the insufficiency of removing confidence values as a protective measure and call for more robust security strategies for protecting these models.

\section{Conclusion}
\label{sec:conclusion}
In this paper, we conducted the most extensive empirical study on AI-driven applications, focusing on on-device ML apps, on-device DL apps, and AI service-supported apps. By analyzing 56,682 real-world AI applications identified from a pool of 7,259,232 mobile apps in the AndroZoo repository, we provide several key insights into the landscape of AI in mobile applications across three main perspectives: application analysis, framework and model analysis, and user analysis.
For example, from the \textbf{application analysis} perspective, we find that incorporating AI technology into applications has become a growing trend since 2018, with the Finance and Business categories releasing the highest number of AI apps. 
From the \textbf{framework and model analysis} perspective, we find that AI apps supported by on-device DL techniques accounted for the highest proportion, and TFLite is the most prevalent framework among single-framework AI apps. 
From the \textbf{user analysis} perspective, we find that the most frequently accessed private attributes by the collected AI apps are name, address, email, and location, and users generally have a positive attitude towards AI in apps, with accuracy issues being the most reported problem in negative reviews. 
Our detailed analysis offers insights into the prevalence, update practices, framework usage, and model protection in AI apps, guiding future AI app development and maintenance strategies. Moreover, by examining user privacy and attitudes, we highlight the importance of privacy protection in AI app development and offer insights into how users perceive current AI technologies utilized in AI apps. We provide a large-scale AI app dataset for further research, offering a valuable resource for the academic and developer communities.

\section{Declaration of competing interest}
The authors declare that they have no known competing financial interests or personal relationships that could have appeared to influence the work reported in this paper.

\section{Data Availability}
\label{Data Availability}

We make our data and scripts publicly available at \url{https://zenodo.org/records/12205325}.

\bibliographystyle{cas-model2-names}
\bibliography{reference}

\begin{thebibliography}{79}
\expandafter\ifx\csname natexlab\endcsname\relax\def\natexlab#1{#1}\fi
\providecommand{\url}[1]{\texttt{#1}}
\providecommand{\href}[2]{#2}
\providecommand{\path}[1]{#1}
\providecommand{\DOIprefix}{doi:}
\providecommand{\ArXivprefix}{arXiv:}
\providecommand{\URLprefix}{URL: }
\providecommand{\Pubmedprefix}{pmid:}
\providecommand{\doi}[1]{\href{http://dx.doi.org/#1}{\path{#1}}}
\providecommand{\Pubmed}[1]{\href{pmid:#1}{\path{#1}}}
\providecommand{\bibinfo}[2]{#2}
\ifx\xfnm\relax \def\xfnm[#1]{\unskip,\space#1}\fi
%Type = Inproceedings
\bibitem[{Abadi et~al.(2016)Abadi, Barham, Chen, Chen, Davis, Dean, Devin, Ghemawat, Irving, Isard et~al.}]{abadi2016tensorflow}
\bibinfo{author}{Abadi, M.}, \bibinfo{author}{Barham, P.}, \bibinfo{author}{Chen, J.}, \bibinfo{author}{Chen, Z.}, \bibinfo{author}{Davis, A.}, \bibinfo{author}{Dean, J.}, \bibinfo{author}{Devin, M.}, \bibinfo{author}{Ghemawat, S.}, \bibinfo{author}{Irving, G.}, \bibinfo{author}{Isard, M.}, et~al., \bibinfo{year}{2016}.
\newblock \bibinfo{title}{$\{$TensorFlow$\}$: a system for $\{$Large-Scale$\}$ machine learning}, in: \bibinfo{booktitle}{12th USENIX symposium on operating systems design and implementation (OSDI 16)}, pp. \bibinfo{pages}{265--283}.
%Type = Article
\bibitem[{AI(2023a)}]{alexaai}
\bibinfo{author}{AI, A.}, \bibinfo{year}{2023}a.
\newblock \bibinfo{title}{Alexa ai} \URLprefix \url{https://developer.amazon.com/en-US/alexa/.}
%Type = Article
\bibitem[{AI(2023b)}]{amazonai}
\bibinfo{author}{AI, A.}, \bibinfo{year}{2023}b.
\newblock \bibinfo{title}{Amazon ai} \URLprefix \url{https://aws.amazon.com/ai/.}
%Type = Article
\bibitem[{AI(2023c)}]{azureai}
\bibinfo{author}{AI, A.}, \bibinfo{year}{2023}c.
\newblock \bibinfo{title}{Azure ai} \URLprefix \url{https://azure.microsoft.com/.}
%Type = Article
\bibitem[{AI(2023d)}]{googleai}
\bibinfo{author}{AI, G.}, \bibinfo{year}{2023}d.
\newblock \bibinfo{title}{Google ai} \URLprefix \url{https://ai.google/.}
%Type = Article
\bibitem[{Ali(2023)}]{ali2023green}
\bibinfo{author}{Ali, A.H.}, \bibinfo{year}{2023}.
\newblock \bibinfo{title}{Green ai for sustainability: leveraging machine learning to drive a circular economy}.
\newblock \bibinfo{journal}{Babylonian Journal of Artificial Intelligence} \bibinfo{volume}{2023}, \bibinfo{pages}{15--16}.
%Type = Inproceedings
\bibitem[{Allix et~al.(2016)Allix, Bissyand{\'e}, Klein and Le~Traon}]{allix2016androzoo}
\bibinfo{author}{Allix, K.}, \bibinfo{author}{Bissyand{\'e}, T.F.}, \bibinfo{author}{Klein, J.}, \bibinfo{author}{Le~Traon, Y.}, \bibinfo{year}{2016}.
\newblock \bibinfo{title}{Androzoo: Collecting millions of android apps for the research community}, in: \bibinfo{booktitle}{2016 IEEE/ACM 13th Working Conference on Mining Software Repositories (MSR)}, \bibinfo{organization}{IEEE}. pp. \bibinfo{pages}{468--471}.
%Type = Article
\bibitem[{Amos et~al.(2016)Amos, Ludwiczuk, Satyanarayanan et~al.}]{amos2016openface}
\bibinfo{author}{Amos, B.}, \bibinfo{author}{Ludwiczuk, B.}, \bibinfo{author}{Satyanarayanan, M.}, et~al., \bibinfo{year}{2016}.
\newblock \bibinfo{title}{Openface: A general-purpose face recognition library with mobile applications}.
\newblock \bibinfo{journal}{CMU School of Computer Science} \bibinfo{volume}{6}, \bibinfo{pages}{20}.
%Type = Article
\bibitem[{APKTool(2023)}]{apktool}
\bibinfo{author}{APKTool}, \bibinfo{year}{2023}.
\newblock \bibinfo{title}{Apktool} \URLprefix \url{https://github.com/iBotPeaches/Apktool/.}
%Type = Article
\bibitem[{Battineni et~al.(2024)Battineni, Chintalapudi, Ricci, Ruocco and Amenta}]{battineni2024exploring}
\bibinfo{author}{Battineni, G.}, \bibinfo{author}{Chintalapudi, N.}, \bibinfo{author}{Ricci, G.}, \bibinfo{author}{Ruocco, C.}, \bibinfo{author}{Amenta, F.}, \bibinfo{year}{2024}.
\newblock \bibinfo{title}{Exploring the integration of artificial intelligence (ai) and augmented reality (ar) in maritime medicine}.
\newblock \bibinfo{journal}{Artificial Intelligence Review} \bibinfo{volume}{57}, \bibinfo{pages}{100}.
%Type = Inproceedings
\bibitem[{Bilyk et~al.(2020)Bilyk, Shapovalov, Shapovalov, Megalinska, Zhadan, Andruszkiewicz, Do{\l}ha{\'n}czuk-{\'S}r{\'o}dka and Antonenko}]{bilyk2020comparing}
\bibinfo{author}{Bilyk, Z.I.}, \bibinfo{author}{Shapovalov, Y.B.}, \bibinfo{author}{Shapovalov, V.B.}, \bibinfo{author}{Megalinska, A.P.}, \bibinfo{author}{Zhadan, S.O.}, \bibinfo{author}{Andruszkiewicz, F.}, \bibinfo{author}{Do{\l}ha{\'n}czuk-{\'S}r{\'o}dka, A.}, \bibinfo{author}{Antonenko, P.D.}, \bibinfo{year}{2020}.
\newblock \bibinfo{title}{Comparing google lens recognition accuracy with other plant recognition apps}, in: \bibinfo{booktitle}{Proceedings of the Symposium on Advances in Educational Technology, AET}.
%Type = Article
\bibitem[{Caffe(2023)}]{caffe}
\bibinfo{author}{Caffe}, \bibinfo{year}{2023}.
\newblock \bibinfo{title}{Caffe} \URLprefix \url{https://github.com/BVLC/caffe/.}
%Type = Article
\bibitem[{Caffe2(2023)}]{caffe2}
\bibinfo{author}{Caffe2}, \bibinfo{year}{2023}.
\newblock \bibinfo{title}{Caffe2} \URLprefix \url{https://github.com/facebookarchive/caffe2/.}
%Type = Article
\bibitem[{Chainer(2023)}]{chainer}
\bibinfo{author}{Chainer}, \bibinfo{year}{2023}.
\newblock \bibinfo{title}{Chainer} \URLprefix \url{https://github.com/chainer/chainer/.}
%Type = Inproceedings
\bibitem[{Chen et~al.(2021)Chen, Yao, Lou, Cao, Liu, Wang and Liu}]{chen2021empirical}
\bibinfo{author}{Chen, Z.}, \bibinfo{author}{Yao, H.}, \bibinfo{author}{Lou, Y.}, \bibinfo{author}{Cao, Y.}, \bibinfo{author}{Liu, Y.}, \bibinfo{author}{Wang, H.}, \bibinfo{author}{Liu, X.}, \bibinfo{year}{2021}.
\newblock \bibinfo{title}{An empirical study on deployment faults of deep learning based mobile applications}, in: \bibinfo{booktitle}{2021 IEEE/ACM 43rd International Conference on Software Engineering (ICSE)}, \bibinfo{organization}{IEEE}. pp. \bibinfo{pages}{674--685}.
%Type = Article
\bibitem[{Cheng et~al.(2017)Cheng, Wang, Zhou and Zhang}]{cheng2017survey}
\bibinfo{author}{Cheng, Y.}, \bibinfo{author}{Wang, D.}, \bibinfo{author}{Zhou, P.}, \bibinfo{author}{Zhang, T.}, \bibinfo{year}{2017}.
\newblock \bibinfo{title}{A survey of model compression and acceleration for deep neural networks}.
\newblock \bibinfo{journal}{arXiv preprint arXiv:1710.09282} .
%Type = Article
\bibitem[{CNNdroid(2023)}]{CNNdroid}
\bibinfo{author}{CNNdroid}, \bibinfo{year}{2023}.
\newblock \bibinfo{title}{Cnndroid} \URLprefix \url{https://github.com/ENCP/CNNdroid/.}
%Type = Article
\bibitem[{CNTK(2023)}]{cntk}
\bibinfo{author}{CNTK}, \bibinfo{year}{2023}.
\newblock \bibinfo{title}{Cntk} \URLprefix \url{https://github.com/microsoft/CNTK/.}
%Type = Article
\bibitem[{Dang et~al.(2024a)Dang, Li, Ma, Guo, Hu, Papadakis, Cordy and Traon}]{dang2024towards}
\bibinfo{author}{Dang, X.}, \bibinfo{author}{Li, Y.}, \bibinfo{author}{Ma, W.}, \bibinfo{author}{Guo, Y.}, \bibinfo{author}{Hu, Q.}, \bibinfo{author}{Papadakis, M.}, \bibinfo{author}{Cordy, M.}, \bibinfo{author}{Traon, Y.L.}, \bibinfo{year}{2024}a.
\newblock \bibinfo{title}{Towards exploring the limitations of test selection techniques on graph neural networks: An empirical study}.
\newblock \bibinfo{journal}{Empirical Software Engineering} \bibinfo{volume}{29}, \bibinfo{pages}{112}.
%Type = Article
\bibitem[{Dang et~al.(2023)Dang, Li, Papadakis, Klein, Bissyand{\'e} and Le~Traon}]{dang2023graphprior}
\bibinfo{author}{Dang, X.}, \bibinfo{author}{Li, Y.}, \bibinfo{author}{Papadakis, M.}, \bibinfo{author}{Klein, J.}, \bibinfo{author}{Bissyand{\'e}, T.F.}, \bibinfo{author}{Le~Traon, Y.}, \bibinfo{year}{2023}.
\newblock \bibinfo{title}{Graphprior: mutation-based test input prioritization for graph neural networks}.
\newblock \bibinfo{journal}{ACM Transactions on Software Engineering and Methodology} \bibinfo{volume}{33}, \bibinfo{pages}{1--40}.
%Type = Article
\bibitem[{Dang et~al.(2024b)Dang, Li, Papadakis, Klein, Bissyande{\'e} and Le~Traon}]{dang2024test}
\bibinfo{author}{Dang, X.}, \bibinfo{author}{Li, Y.}, \bibinfo{author}{Papadakis, M.}, \bibinfo{author}{Klein, J.}, \bibinfo{author}{Bissyande{\'e}, T.F.}, \bibinfo{author}{Le~Traon, Y.}, \bibinfo{year}{2024}b.
\newblock \bibinfo{title}{Test input prioritization for machine learning classifiers}.
\newblock \bibinfo{journal}{IEEE Transactions on Software Engineering} .
%Type = Article
\bibitem[{Datumbox(2023)}]{datumbox}
\bibinfo{author}{Datumbox}, \bibinfo{year}{2023}.
\newblock \bibinfo{title}{Datumbox} \URLprefix \url{https://github.com/datumbox/datumbox-framework/.}
%Type = Article
\bibitem[{David et~al.(2021)David, Duke, Jain, Janapa~Reddi, Jeffries, Li, Kreeger, Nappier, Natraj, Wang et~al.}]{tflite}
\bibinfo{author}{David, R.}, \bibinfo{author}{Duke, J.}, \bibinfo{author}{Jain, A.}, \bibinfo{author}{Janapa~Reddi, V.}, \bibinfo{author}{Jeffries, N.}, \bibinfo{author}{Li, J.}, \bibinfo{author}{Kreeger, N.}, \bibinfo{author}{Nappier, I.}, \bibinfo{author}{Natraj, M.}, \bibinfo{author}{Wang, T.}, et~al., \bibinfo{year}{2021}.
\newblock \bibinfo{title}{Tensorflow lite micro: Embedded machine learning for tinyml systems}.
\newblock \bibinfo{journal}{Proceedings of Machine Learning and Systems} \bibinfo{volume}{3}, \bibinfo{pages}{800--811}.
%Type = Article
\bibitem[{DeepLearning4J(2023)}]{deeplearning4j}
\bibinfo{author}{DeepLearning4J}, \bibinfo{year}{2023}.
\newblock \bibinfo{title}{Deeplearning4j} \URLprefix \url{https://github.com/eclipse/deeplearning4j/.}
%Type = Inproceedings
\bibitem[{Deng et~al.(2022)Deng, Chen, Meng, Zhang, Xu and Cheng}]{deng2022understanding}
\bibinfo{author}{Deng, Z.}, \bibinfo{author}{Chen, K.}, \bibinfo{author}{Meng, G.}, \bibinfo{author}{Zhang, X.}, \bibinfo{author}{Xu, K.}, \bibinfo{author}{Cheng, Y.}, \bibinfo{year}{2022}.
\newblock \bibinfo{title}{Understanding real-world threats to deep learning models in android apps}, in: \bibinfo{booktitle}{Proceedings of the 2022 ACM SIGSAC Conference on Computer and Communications Security}, pp. \bibinfo{pages}{785--799}.
%Type = Article
\bibitem[{Dospinescu and Popa(2016)}]{dospinescu2016face}
\bibinfo{author}{Dospinescu, O.}, \bibinfo{author}{Popa, I.}, \bibinfo{year}{2016}.
\newblock \bibinfo{title}{Face detection and face recognition in android mobile applications}.
\newblock \bibinfo{journal}{Informatica Economica} \bibinfo{volume}{20}, \bibinfo{pages}{20}.
%Type = Article
\bibitem[{FeatherCNN(2023)}]{feathercnn}
\bibinfo{author}{FeatherCNN}, \bibinfo{year}{2023}.
\newblock \bibinfo{title}{Feathercnn} \URLprefix \url{https://github.com/Tencent/FeatherCNN/.}
%Type = Article
\bibitem[{Framework(2023)}]{neuroph}
\bibinfo{author}{Framework, N.}, \bibinfo{year}{2023}.
\newblock \bibinfo{title}{Neuroph framework} \URLprefix \url{https://github.com/neuroph/NeurophFramework/.}
%Type = Article
\bibitem[{Gamble(2020)}]{gamble2020artificial}
\bibinfo{author}{Gamble, A.}, \bibinfo{year}{2020}.
\newblock \bibinfo{title}{Artificial intelligence and mobile apps for mental healthcare: a social informatics perspective}.
\newblock \bibinfo{journal}{Aslib Journal of Information Management} \bibinfo{volume}{72}, \bibinfo{pages}{509--523}.
%Type = Inproceedings
\bibitem[{He et~al.(2018)He, Lin, Liu, Wang, Li and Han}]{he2018amc}
\bibinfo{author}{He, Y.}, \bibinfo{author}{Lin, J.}, \bibinfo{author}{Liu, Z.}, \bibinfo{author}{Wang, H.}, \bibinfo{author}{Li, L.J.}, \bibinfo{author}{Han, S.}, \bibinfo{year}{2018}.
\newblock \bibinfo{title}{Amc: Automl for model compression and acceleration on mobile devices}, in: \bibinfo{booktitle}{Proceedings of the European conference on computer vision (ECCV)}, pp. \bibinfo{pages}{784--800}.
%Type = Article
\bibitem[{Hjelm{\aa}s and Low(2001)}]{hjelmaas2001face}
\bibinfo{author}{Hjelm{\aa}s, E.}, \bibinfo{author}{Low, B.K.}, \bibinfo{year}{2001}.
\newblock \bibinfo{title}{Face detection: A survey}.
\newblock \bibinfo{journal}{Computer vision and image understanding} \bibinfo{volume}{83}, \bibinfo{pages}{236--274}.
%Type = Article
\bibitem[{Howard et~al.(2017)Howard, Zhu, Chen, Kalenichenko, Wang, Weyand, Andreetto and Adam}]{howard2017mobilenets}
\bibinfo{author}{Howard, A.G.}, \bibinfo{author}{Zhu, M.}, \bibinfo{author}{Chen, B.}, \bibinfo{author}{Kalenichenko, D.}, \bibinfo{author}{Wang, W.}, \bibinfo{author}{Weyand, T.}, \bibinfo{author}{Andreetto, M.}, \bibinfo{author}{Adam, H.}, \bibinfo{year}{2017}.
\newblock \bibinfo{title}{Mobilenets: Efficient convolutional neural networks for mobile vision applications}.
\newblock \bibinfo{journal}{arXiv preprint arXiv:1704.04861} .
%Type = Article
\bibitem[{Huang and Chen(2022)}]{huang2022smart}
\bibinfo{author}{Huang, Y.}, \bibinfo{author}{Chen, C.}, \bibinfo{year}{2022}.
\newblock \bibinfo{title}{Smart app attack: hacking deep learning models in android apps}.
\newblock \bibinfo{journal}{IEEE Transactions on Information Forensics and Security} \bibinfo{volume}{17}, \bibinfo{pages}{1827--1840}.
%Type = Inproceedings
\bibitem[{Huang et~al.(2021)Huang, Hu and Chen}]{huang2021robustness}
\bibinfo{author}{Huang, Y.}, \bibinfo{author}{Hu, H.}, \bibinfo{author}{Chen, C.}, \bibinfo{year}{2021}.
\newblock \bibinfo{title}{Robustness of on-device models: Adversarial attack to deep learning models on android apps}, in: \bibinfo{booktitle}{2021 IEEE/ACM 43rd International Conference on Software Engineering: Software Engineering in Practice (ICSE-SEIP)}, \bibinfo{organization}{IEEE}. pp. \bibinfo{pages}{101--110}.
%Type = Article
\bibitem[{Hub(2023)}]{tflitehub}
\bibinfo{author}{Hub, T.L.}, \bibinfo{year}{2023}.
\newblock \bibinfo{title}{Tensorflow lite hub} \URLprefix \url{https://tfhub.dev/s?deployment-format=lite&subtype=module,placeholder/.}
%Type = Article
\bibitem[{Jones(2020)}]{jones2020plant}
\bibinfo{author}{Jones, H.G.}, \bibinfo{year}{2020}.
\newblock \bibinfo{title}{What plant is that? tests of automated image recognition apps for plant identification on plants from the british flora}.
\newblock \bibinfo{journal}{AoB Plants} \bibinfo{volume}{12}, \bibinfo{pages}{plaa052}.
%Type = Article
\bibitem[{Joshi and Dhakal(2021)}]{joshi2021predicting}
\bibinfo{author}{Joshi, R.D.}, \bibinfo{author}{Dhakal, C.K.}, \bibinfo{year}{2021}.
\newblock \bibinfo{title}{Predicting type 2 diabetes using logistic regression and machine learning approaches}.
\newblock \bibinfo{journal}{International journal of environmental research and public health} \bibinfo{volume}{18}, \bibinfo{pages}{7346}.
%Type = Article
\bibitem[{keras(2023)}]{keras}
\bibinfo{author}{keras}, \bibinfo{year}{2023}.
\newblock \bibinfo{title}{Keras} \URLprefix \url{https://github.com/keras-team/keras/.}
%Type = Article
\bibitem[{Kindylidi and Cabral(2021)}]{kindylidi2021sustainability}
\bibinfo{author}{Kindylidi, I.}, \bibinfo{author}{Cabral, T.S.}, \bibinfo{year}{2021}.
\newblock \bibinfo{title}{Sustainability of ai: The case of provision of information to consumers}.
\newblock \bibinfo{journal}{Sustainability} \bibinfo{volume}{13}, \bibinfo{pages}{12064}.
%Type = Article
\bibitem[{LaValley(2008)}]{lavalley2008logistic}
\bibinfo{author}{LaValley, M.P.}, \bibinfo{year}{2008}.
\newblock \bibinfo{title}{Logistic regression}.
\newblock \bibinfo{journal}{Circulation} \bibinfo{volume}{117}, \bibinfo{pages}{2395--2399}.
%Type = Article
\bibitem[{Li et~al.(2023)Li, Dang, Ma, Klein, Traon and Bissyand{\'e}}]{li2023test}
\bibinfo{author}{Li, Y.}, \bibinfo{author}{Dang, X.}, \bibinfo{author}{Ma, L.}, \bibinfo{author}{Klein, J.}, \bibinfo{author}{Traon, Y.L.}, \bibinfo{author}{Bissyand{\'e}, T.F.}, \bibinfo{year}{2023}.
\newblock \bibinfo{title}{Test input prioritization for 3d point clouds}.
\newblock \bibinfo{journal}{ACM Transactions on Software Engineering and Methodology} .
%Type = Article
\bibitem[{Li et~al.(2024)Li, Dang, Pian, Habib, Klein and Bissyand{\'e}}]{li2024test}
\bibinfo{author}{Li, Y.}, \bibinfo{author}{Dang, X.}, \bibinfo{author}{Pian, W.}, \bibinfo{author}{Habib, A.}, \bibinfo{author}{Klein, J.}, \bibinfo{author}{Bissyand{\'e}, T.}, \bibinfo{year}{2024}.
\newblock \bibinfo{title}{Test input prioritization for graph neural networks}.
\newblock \bibinfo{journal}{IEEE Transactions on Software Engineering} .
%Type = Inproceedings
\bibitem[{Li et~al.(2021)Li, Hua, Wang, Chen and Liu}]{li2021deeppayload}
\bibinfo{author}{Li, Y.}, \bibinfo{author}{Hua, J.}, \bibinfo{author}{Wang, H.}, \bibinfo{author}{Chen, C.}, \bibinfo{author}{Liu, Y.}, \bibinfo{year}{2021}.
\newblock \bibinfo{title}{Deeppayload: Black-box backdoor attack on deep learning models through neural payload injection}, in: \bibinfo{booktitle}{2021 IEEE/ACM 43rd International Conference on Software Engineering (ICSE)}, \bibinfo{organization}{IEEE}. pp. \bibinfo{pages}{263--274}.
%Type = Article
\bibitem[{Lite(2023)}]{paddlelite}
\bibinfo{author}{Lite, P.}, \bibinfo{year}{2023}.
\newblock \bibinfo{title}{Paddle lite} \URLprefix \url{https://github.com/PaddlePaddle/Paddle-Lite/.}
%Type = Article
\bibitem[{Locke et~al.(2021)Locke, Bashall, Al-Adely, Moore, Wilson and Kitchen}]{locke2021natural}
\bibinfo{author}{Locke, S.}, \bibinfo{author}{Bashall, A.}, \bibinfo{author}{Al-Adely, S.}, \bibinfo{author}{Moore, J.}, \bibinfo{author}{Wilson, A.}, \bibinfo{author}{Kitchen, G.B.}, \bibinfo{year}{2021}.
\newblock \bibinfo{title}{Natural language processing in medicine: a review}.
\newblock \bibinfo{journal}{Trends in Anaesthesia and Critical Care} \bibinfo{volume}{38}, \bibinfo{pages}{4--9}.
%Type = Article
\bibitem[{Lu et~al.(2015)Lu, Wu, Mao, Wang and Zhang}]{lu2015recommender}
\bibinfo{author}{Lu, J.}, \bibinfo{author}{Wu, D.}, \bibinfo{author}{Mao, M.}, \bibinfo{author}{Wang, W.}, \bibinfo{author}{Zhang, G.}, \bibinfo{year}{2015}.
\newblock \bibinfo{title}{Recommender system application developments: a survey}.
\newblock \bibinfo{journal}{Decision Support Systems} \bibinfo{volume}{74}, \bibinfo{pages}{12--32}.
%Type = Article
\bibitem[{MACE(2023)}]{mace}
\bibinfo{author}{MACE}, \bibinfo{year}{2023}.
\newblock \bibinfo{title}{Mace} \URLprefix \url{https://github.com/XiaoMi/mace/.}
%Type = Article
\bibitem[{MALLET(2023)}]{mallet}
\bibinfo{author}{MALLET}, \bibinfo{year}{2023}.
\newblock \bibinfo{title}{Mallet} \URLprefix \url{https://github.com/mimno/Mallet/.}
%Type = Article
\bibitem[{Matarneh et~al.(2017)Matarneh, Maksymova, Lyashenko and Belova}]{matarneh2017speech}
\bibinfo{author}{Matarneh, R.}, \bibinfo{author}{Maksymova, S.}, \bibinfo{author}{Lyashenko, V.}, \bibinfo{author}{Belova, N.}, \bibinfo{year}{2017}.
\newblock \bibinfo{title}{Speech recognition systems: A comparative review} .
%Type = Article
\bibitem[{Miner(2023)}]{rapidminer}
\bibinfo{author}{Miner, R.}, \bibinfo{year}{2023}.
\newblock \bibinfo{title}{Rapid miner} \URLprefix \url{https://rapidminer.com/.}
%Type = Article
\bibitem[{MLPACK(2023)}]{mlpack}
\bibinfo{author}{MLPACK}, \bibinfo{year}{2023}.
\newblock \bibinfo{title}{Mlpack} \URLprefix \url{https://github.com/mlpack/mlpack/.}
%Type = Unpublished
\bibitem[{Morawiec(2021)}]{skpodamo}
\bibinfo{author}{Morawiec, D.}, \bibinfo{year}{2021}.
\newblock \bibinfo{title}{sklearn-porter}.
\newblock \URLprefix \url{https://github.com/nok/sklearn-porter}. \bibinfo{note}{transpile trained scikit-learn estimators to C, Java, JavaScript and others}.
%Type = Article
\bibitem[{NCNN(2023)}]{ncnn}
\bibinfo{author}{NCNN}, \bibinfo{year}{2023}.
\newblock \bibinfo{title}{Ncnn} \URLprefix \url{https://github.com/Tencent/ncnn/.}
%Type = Article
\bibitem[{NLP(2023)}]{baidunlp}
\bibinfo{author}{NLP, B.}, \bibinfo{year}{2023}.
\newblock \bibinfo{title}{Baidu nlp} \URLprefix \url{https://ai.baidu.com/ai-doc/NLP/.}
%Type = Article
\bibitem[{OCR(2023)}]{baiduocr}
\bibinfo{author}{OCR, B.}, \bibinfo{year}{2023}.
\newblock \bibinfo{title}{Baidu ocr} \URLprefix \url{https://ai.baidu.com/ai-doc/OCR/.}
%Type = Inproceedings
\bibitem[{Omar and Salih(2024)}]{omar2024systematic}
\bibinfo{author}{Omar, L.I.}, \bibinfo{author}{Salih, A.A.}, \bibinfo{year}{2024}.
\newblock \bibinfo{title}{Systematic review of english/arabic machine translation postediting: Implications for ai application in translation research and pedagogy}, in: \bibinfo{booktitle}{Informatics}, \bibinfo{organization}{MDPI}. p.~\bibinfo{pages}{23}.
%Type = Article
\bibitem[{OpenCV(2023)}]{opencv}
\bibinfo{author}{OpenCV}, \bibinfo{year}{2023}.
\newblock \bibinfo{title}{Opencv} \URLprefix \url{https://github.com/opencv/opencv/.}
%Type = Article
\bibitem[{Paszke et~al.(2019)Paszke, Gross, Massa, Lerer, Bradbury, Chanan, Killeen, Lin, Gimelshein, Antiga et~al.}]{paszke2019pytorch}
\bibinfo{author}{Paszke, A.}, \bibinfo{author}{Gross, S.}, \bibinfo{author}{Massa, F.}, \bibinfo{author}{Lerer, A.}, \bibinfo{author}{Bradbury, J.}, \bibinfo{author}{Chanan, G.}, \bibinfo{author}{Killeen, T.}, \bibinfo{author}{Lin, Z.}, \bibinfo{author}{Gimelshein, N.}, \bibinfo{author}{Antiga, L.}, et~al., \bibinfo{year}{2019}.
\newblock \bibinfo{title}{Pytorch: An imperative style, high-performance deep learning library}.
\newblock \bibinfo{journal}{Advances in neural information processing systems} \bibinfo{volume}{32}, \bibinfo{pages}{8026--8037}.
%Type = Inproceedings
\bibitem[{Pham et~al.(2014)Pham, Bluche, Kermorvant and Louradour}]{pham2014dropout}
\bibinfo{author}{Pham, V.}, \bibinfo{author}{Bluche, T.}, \bibinfo{author}{Kermorvant, C.}, \bibinfo{author}{Louradour, J.}, \bibinfo{year}{2014}.
\newblock \bibinfo{title}{Dropout improves recurrent neural networks for handwriting recognition}, in: \bibinfo{booktitle}{2014 14th international conference on frontiers in handwriting recognition}, \bibinfo{organization}{IEEE}. pp. \bibinfo{pages}{285--290}.
%Type = Inproceedings
\bibitem[{Ribeiro et~al.(2015)Ribeiro, Grolinger and Capretz}]{ribeiro2015mlaas}
\bibinfo{author}{Ribeiro, M.}, \bibinfo{author}{Grolinger, K.}, \bibinfo{author}{Capretz, M.A.}, \bibinfo{year}{2015}.
\newblock \bibinfo{title}{Mlaas: Machine learning as a service}, in: \bibinfo{booktitle}{2015 IEEE 14th international conference on machine learning and applications (ICMLA)}, \bibinfo{organization}{IEEE}. pp. \bibinfo{pages}{896--902}.
%Type = Article
\bibitem[{Rokach and Maimon(2005)}]{rokach2005decision}
\bibinfo{author}{Rokach, L.}, \bibinfo{author}{Maimon, O.}, \bibinfo{year}{2005}.
\newblock \bibinfo{title}{Decision trees}.
\newblock \bibinfo{journal}{Data mining and knowledge discovery handbook} , \bibinfo{pages}{165--192}.
%Type = Article
\bibitem[{Searcher(2023)}]{ag}
\bibinfo{author}{Searcher, T.S.}, \bibinfo{year}{2023}.
\newblock \bibinfo{title}{The silver searcher} \URLprefix \url{https://github.com/ggreer/the_silver_searcher/.}
%Type = Article
\bibitem[{Shogun(2023)}]{shogun}
\bibinfo{author}{Shogun}, \bibinfo{year}{2023}.
\newblock \bibinfo{title}{Shogun} \URLprefix \url{https://github.com/shogun-toolbox/shogun/.}
%Type = Inproceedings
\bibitem[{Shokri et~al.(2017)Shokri, Stronati, Song and Shmatikov}]{shokri2017membership}
\bibinfo{author}{Shokri, R.}, \bibinfo{author}{Stronati, M.}, \bibinfo{author}{Song, C.}, \bibinfo{author}{Shmatikov, V.}, \bibinfo{year}{2017}.
\newblock \bibinfo{title}{Membership inference attacks against machine learning models}, in: \bibinfo{booktitle}{2017 IEEE symposium on security and privacy (SP)}, \bibinfo{organization}{IEEE}. pp. \bibinfo{pages}{3--18}.
%Type = Article
\bibitem[{SNPE(2023)}]{snpe}
\bibinfo{author}{SNPE}, \bibinfo{year}{2023}.
\newblock \bibinfo{title}{Snpe} \URLprefix \url{https://developer.qualcomm.com/software/qualcomm-neural-processing-sdk/.}
%Type = Inproceedings
\bibitem[{Sun et~al.(2021)Sun, Sun, Lu and Mislove}]{sun2021mind}
\bibinfo{author}{Sun, Z.}, \bibinfo{author}{Sun, R.}, \bibinfo{author}{Lu, L.}, \bibinfo{author}{Mislove, A.}, \bibinfo{year}{2021}.
\newblock \bibinfo{title}{Mind your weight (s): A large-scale study on insufficient machine learning model protection in mobile apps}, in: \bibinfo{booktitle}{30th $\{$USENIX$\}$ Security Symposium ($\{$USENIX$\}$ Security 21)}.
%Type = Article
\bibitem[{synthesizer(2023)}]{synthesizer}
\bibinfo{author}{synthesizer}, \bibinfo{year}{2023}.
\newblock \bibinfo{title}{Synthesizer} \URLprefix \url{https://ai.baidu.com/tech/speech/.}
%Type = Article
\bibitem[{Thakkar and Thakkar(2019)}]{thakkar2019introduction}
\bibinfo{author}{Thakkar, M.}, \bibinfo{author}{Thakkar, M.}, \bibinfo{year}{2019}.
\newblock \bibinfo{title}{Introduction to core ml framework}.
\newblock \bibinfo{journal}{Beginning Machine Learning in iOS: CoreML Framework} , \bibinfo{pages}{15--49}.
%Type = Inproceedings
\bibitem[{Tram{\`e}r et~al.(2016)Tram{\`e}r, Zhang, Juels, Reiter and Ristenpart}]{tramer2016stealing}
\bibinfo{author}{Tram{\`e}r, F.}, \bibinfo{author}{Zhang, F.}, \bibinfo{author}{Juels, A.}, \bibinfo{author}{Reiter, M.K.}, \bibinfo{author}{Ristenpart, T.}, \bibinfo{year}{2016}.
\newblock \bibinfo{title}{Stealing machine learning models via prediction $\{$APIs$\}$}, in: \bibinfo{booktitle}{25th USENIX security symposium (USENIX Security 16)}, pp. \bibinfo{pages}{601--618}.
%Type = Article
\bibitem[{TVM(2023)}]{tvm}
\bibinfo{author}{TVM}, \bibinfo{year}{2023}.
\newblock \bibinfo{title}{Tvm} \URLprefix \url{https://github.com/apache/tvm/.}
%Type = Inproceedings
\bibitem[{Wang et~al.(2015)Wang, Wang and Yeung}]{wang2015collaborative}
\bibinfo{author}{Wang, H.}, \bibinfo{author}{Wang, N.}, \bibinfo{author}{Yeung, D.Y.}, \bibinfo{year}{2015}.
\newblock \bibinfo{title}{Collaborative deep learning for recommender systems}, in: \bibinfo{booktitle}{Proceedings of the 21th ACM SIGKDD international conference on knowledge discovery and data mining}, pp. \bibinfo{pages}{1235--1244}.
%Type = Article
\bibitem[{WEKA(2023)}]{weka}
\bibinfo{author}{WEKA}, \bibinfo{year}{2023}.
\newblock \bibinfo{title}{Weka} \URLprefix \url{https://github.com/ishaanjav/Weka-ML-Face-Recognition/.}
%Type = Inproceedings
\bibitem[{Wolf et~al.(2020)Wolf, Debut, Sanh, Chaumond, Delangue, Moi, Cistac, Rault, Louf, Funtowicz, Davison, Shleifer, von Platen, Ma, Jernite, Plu, Xu, Scao, Gugger, Drame, Lhoest and Rush}]{wolf-etal-2020-transformers}
\bibinfo{author}{Wolf, T.}, \bibinfo{author}{Debut, L.}, \bibinfo{author}{Sanh, V.}, \bibinfo{author}{Chaumond, J.}, \bibinfo{author}{Delangue, C.}, \bibinfo{author}{Moi, A.}, \bibinfo{author}{Cistac, P.}, \bibinfo{author}{Rault, T.}, \bibinfo{author}{Louf, R.}, \bibinfo{author}{Funtowicz, M.}, \bibinfo{author}{Davison, J.}, \bibinfo{author}{Shleifer, S.}, \bibinfo{author}{von Platen, P.}, \bibinfo{author}{Ma, C.}, \bibinfo{author}{Jernite, Y.}, \bibinfo{author}{Plu, J.}, \bibinfo{author}{Xu, C.}, \bibinfo{author}{Scao, T.L.}, \bibinfo{author}{Gugger, S.}, \bibinfo{author}{Drame, M.}, \bibinfo{author}{Lhoest, Q.}, \bibinfo{author}{Rush, A.M.}, \bibinfo{year}{2020}.
\newblock \bibinfo{title}{Transformers: State-of-the-art natural language processing}, in: \bibinfo{booktitle}{Proceedings of the 2020 Conference on Empirical Methods in Natural Language Processing: System Demonstrations}, \bibinfo{publisher}{Association for Computational Linguistics}, \bibinfo{address}{Online}. pp. \bibinfo{pages}{38--45}.
\newblock \URLprefix \url{https://www.aclweb.org/anthology/2020.emnlp-demos.6}.
%Type = Article
\bibitem[{Xu and Tian(2015)}]{xu2015comprehensive}
\bibinfo{author}{Xu, D.}, \bibinfo{author}{Tian, Y.}, \bibinfo{year}{2015}.
\newblock \bibinfo{title}{A comprehensive survey of clustering algorithms}.
\newblock \bibinfo{journal}{Annals of data science} \bibinfo{volume}{2}, \bibinfo{pages}{165--193}.
%Type = Inproceedings
\bibitem[{Xu et~al.(2019)Xu, Liu, Liu, Lin, Liu and Liu}]{xu2019first}
\bibinfo{author}{Xu, M.}, \bibinfo{author}{Liu, J.}, \bibinfo{author}{Liu, Y.}, \bibinfo{author}{Lin, F.X.}, \bibinfo{author}{Liu, Y.}, \bibinfo{author}{Liu, X.}, \bibinfo{year}{2019}.
\newblock \bibinfo{title}{A first look at deep learning apps on smartphones}, in: \bibinfo{booktitle}{The World Wide Web Conference}, pp. \bibinfo{pages}{2125--2136}.
%Type = Inproceedings
\bibitem[{Yao et~al.(2017)Yao, Xiao, Wang, Viswanath, Zheng and Zhao}]{yao2017complexity}
\bibinfo{author}{Yao, Y.}, \bibinfo{author}{Xiao, Z.}, \bibinfo{author}{Wang, B.}, \bibinfo{author}{Viswanath, B.}, \bibinfo{author}{Zheng, H.}, \bibinfo{author}{Zhao, B.Y.}, \bibinfo{year}{2017}.
\newblock \bibinfo{title}{Complexity vs. performance: empirical analysis of machine learning as a service}, in: \bibinfo{booktitle}{Proceedings of the 2017 Internet Measurement Conference}, pp. \bibinfo{pages}{384--397}.
%Type = Article
\bibitem[{Zhang et~al.(2019)Zhang, Patras and Haddadi}]{zhang2019deep}
\bibinfo{author}{Zhang, C.}, \bibinfo{author}{Patras, P.}, \bibinfo{author}{Haddadi, H.}, \bibinfo{year}{2019}.
\newblock \bibinfo{title}{Deep learning in mobile and wireless networking: A survey}.
\newblock \bibinfo{journal}{IEEE Communications surveys \& tutorials} \bibinfo{volume}{21}, \bibinfo{pages}{2224--2287}.
%Type = Inproceedings
\bibitem[{Zhang et~al.(2018)Zhang, Zhou, Lin and Sun}]{Zhang_2018_CVPR}
\bibinfo{author}{Zhang, X.}, \bibinfo{author}{Zhou, X.}, \bibinfo{author}{Lin, M.}, \bibinfo{author}{Sun, J.}, \bibinfo{year}{2018}.
\newblock \bibinfo{title}{Shufflenet: An extremely efficient convolutional neural network for mobile devices}, in: \bibinfo{booktitle}{Proceedings of the IEEE Conference on Computer Vision and Pattern Recognition (CVPR)}.
%Type = Article
\bibitem[{Zhao et~al.(2017)Zhao, Wu and Chen}]{zhao2017android}
\bibinfo{author}{Zhao, N.}, \bibinfo{author}{Wu, M.}, \bibinfo{author}{Chen, J.}, \bibinfo{year}{2017}.
\newblock \bibinfo{title}{Android-based mobile educational platform for speech signal processing}.
\newblock \bibinfo{journal}{International Journal of Electrical Engineering Education} \bibinfo{volume}{54}, \bibinfo{pages}{3--16}.

\end{thebibliography}

\end{document}